\documentclass[final,3p,times,twocolumn]{elsarticle}

\usepackage{amssymb}
\usepackage{amsmath}

\journal{Chaos, Solitons \& Fractals}

\usepackage[colorlinks=true,urlcolor=blue,citecolor=blue,linkcolor=blue]{hyperref}
\usepackage{graphicx}
\usepackage{hyperref}
\usepackage{bm}
\usepackage{newfloat}
\usepackage{romannum}

\usepackage{algorithm}
\usepackage{algcompatible}

\usepackage{lipsum}
\usepackage{cuted}

\usepackage[mathlines,columnwise]{lineno}

\def\dbar{{\mathchar'26\mkern-12mu d}}

\newcommand{\RNum}[1]{\uppercase\expandafter{\romannumeral #1\relax}}

\begin{document}

\begin{frontmatter}



\title{Inferring the Langevin Equation with Uncertainty via Bayesian Neural Networks}


\author[label1]{Youngkyoung Bae}
\affiliation[label1]{Department of Physics and Astronomy & Center for Theoretical Physics, Seoul National University, Seoul 08826, Korea}

\author[label2]{Seungwoong Ha}
\affiliation[label2]{Santa Fe Institute, Santa Fe, NM 87501, USA}

\author[label3,label4]{Hawoong Jeong\corref{cor1}}
\affiliation[label3]{Department of Physics, Korea Advanced Institute of Science and Technology, Daejeon 34141, Korea}
\affiliation[label4]{Center for Complex Systems, Korea Advanced Institute of Science and Technology, Daejeon 34141, Korea}
\ead{hjeong@kaist.edu}
\cortext[cor1]{Corresponding author.}


\begin{abstract}
Pervasive across diverse domains, stochastic systems exhibit fluctuations in processes ranging from molecular dynamics to climate phenomena.
The Langevin equation has served as a common mathematical model for studying such systems, enabling predictions of their temporal evolution and analyses of thermodynamic quantities, including absorbed heat, work done on the system, and entropy production. 
However, inferring the Langevin equation from observed trajectories is a challenging problem, and assessing the uncertainty associated with the inferred equation has yet to be accomplished.
In this study, we present a comprehensive framework that employs Bayesian neural networks for inferring Langevin equations in both overdamped and underdamped regimes. 
Our framework first provides the drift force and diffusion matrix separately and then combines them to construct the Langevin equation. 
By providing a distribution of predictions instead of a single value, our approach allows us to assess prediction uncertainties, which can help prevent potential misunderstandings and erroneous decisions about the system.
We demonstrate the effectiveness of our framework in inferring Langevin equations for various scenarios including a neuron model and microscopic engine, highlighting its versatility and potential impact.
\end{abstract}








\end{frontmatter}



\section{Introduction}

\begin{figure*}[!ht]
    \includegraphics[width=\linewidth]{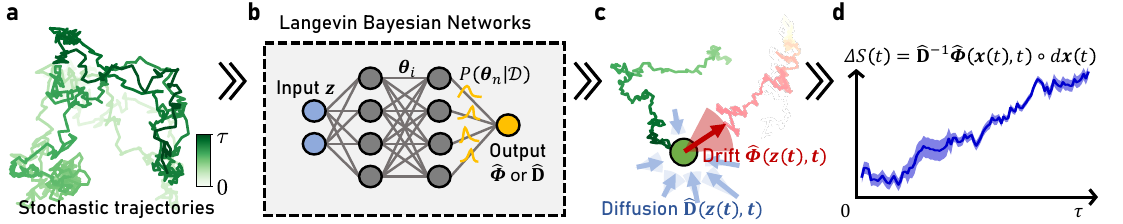}
    \vskip -0.1in
    \caption{
    Schematics of our framework to infer the Langevin dynamics via Langevin Bayesian networks (LBN).
    To build the Langevin equation from observed trajectories ({\bf a}), we feed the trajectories to LBN and obtain the inferred drift vector $\hat{\bm{\Phi}}(\bm{z}, t)$ and diffusion matrix $\hat{\bm{\mathsf{D}}}(\bm{z}, t)$ ({\bf b}).
    The inferred Langevin equation enables us to predict system evolution over time and analyze thermodynamic quantities ({\bf c} and {\bf d}).
    Here, the shaded areas represent the uncertainty of its prediction.
    }\label{fig1}
\end{figure*}

The dynamics of numerous complex systems are inherently stochastic due to fluctuation occurring at various levels. 
From molecular processes~\cite{ariga2018nonequilibrium, fang2019nonequilibrium, liu2020quantifying} to collective behaviors of multicellular organisms~\cite{zhang2010collective, hayakawa2020polar, bruckner2021learning}, soft matter and biological substances exhibit both long-term persistent motion and short-term random displacements by a combination of deterministic forces and thermal fluctuations of the environment.
Fluctuations originating from thermal or non-thermal sources also play an essential role in both the description and design of electronic devices~\cite{agudov2021stochastic, roldan2023variability, aifer2024thermodynamic}.
Zooming out to larger scales, phenomena like bird flocks~\cite{bialek2012statistical, Mora2016local}, fish schools~\cite{yates2009inherent, jhawar2020noise}, financial market prices~\cite{kanazawa2018kinetic}, and climate dynamics~\cite{lucarini2014mathematical} unfold in characteristic patterns of persistent, gradually changing trends with fluctuations stemming from fast and unresolved degrees of freedom like rapid interactions between constituents.
The Langevin equation stands as the prevalent mathematical framework for describing the evolution of such stochastic systems over time, incorporating a drift vector that captures persistent, deterministic trends and a stochastic force that captures environmental fluctuations~\cite{risken1996fokker, gardiner2004handbook}.
A profound understanding of the Langevin equation governing a specific system unlocks substantial advantages: prediction of system evolution over time~\cite{bruckner2019stochastic, bruckner2021learning, bruckner2023learning}, quantification of interactions between degrees of freedom, calibration of an optical trap for particle manipulation~\cite{garvrilov2014realtime}, and measurement of the stochastic nature of thermodynamic quantities like absorbed heat, work done on the system, or entropy production~\cite{sekimoto2010stochastic, seifert2012stochastic}.
Particularly, how dissipation or entropy production can be noninvasively measured from observed trajectories has recently garnered much attention~\cite{roldan2010estimating, li2019quantifying, liu2020quantifying, vu2020entropy, kim2020learning, lee2023multidimensional, kwon2023alpha}. 
However, inferring the Langevin equation itself from observed trajectories still remains challenging, especially for nonlinear and high-dimensional complex systems.

Various techniques to infer the Langevin equation from observed trajectories have been investigated, such as regular or adaptive binning~\cite{friedrich2011approaching, hoze2012heterogeneity, beheiry2015inferenceMAP, sungkaworn2017single}, kernel-based regression~\cite{lamouroux2009kernel-based}, projections onto predetermined basis functions~\cite{frishman2020learning, bruckner2020inferring}, machine learning~\cite{yu2022extracting, bishnoi2024brognet, gao2024learning}, Bayesian methods~\cite{ferretti2020building, bryan2022inferring, siler2023bayesian}, and a linear regression~\cite{garcia2018highperformance}.
Most approaches, though, have struggled to accurately infer the equation of high-dimensional and nonlinear systems due to the exponential increase in parameters with dimensionality and the difficulty in selecting appropriate functional forms for regression.
Additionally, although these methods can provide predictions, they often fall short in assessing the uncertainty of their predictions.
This limitation can lead to erroneous conclusions; for example, an inferred equation based on observed data for cell movements may yield incorrect behavioral predictions for unobserved or barely observed regions, which may encourage a fundamentally flawed understanding of the system~\cite{ghahramani2015probabilistic}.
Therefore, a comprehensive framework should be developed that enables accurate inference of the Langevin equation while also providing a robust assessment of prediction uncertainties.

In this paper, we introduce a practical framework called Langevin Bayesian networks (LBN) that enables the inference of the Langevin equation from observed trajectories without detailed knowledge of the system.
LBN facilitates precise inference of the drift field and diffusion matrix separately, while also providing a quantification of the associated uncertainties for each prediction by harnessing Bayesian neural networks (BNNs) and associated techniques. 
Additionally, we establish that the observed dynamics of an underdamped Langevin system deviate from the original underlying dynamics. Rigorous derivation of unbiased estimators to first order in $\Delta t$ for both overdamped and underdamped Langevin systems is presented.
LBN possesses several salient advantages: 1) no dependency on binning, 2) no assumptions of functional forms for the drift field and diffusion matrix, 3) availability of uncertainty estimates associated with its predictions, 4) ability to handle multidimensional trajectories and large datasets, 5) accessibility of the gradient of outputs, and 6) no leading-order bias for estimations.
These inherent strengths underscore the significance of LBN, offering a versatile and robust tool for researchers in diverse fields.
After describing our methods, we demonstrate the effectiveness of LBN through a series of illustrative examples in the following sections.

\section{Description of our methods}
\subsection{Problem setup}
We consider a general stochastic system whose states at time $t$, denoted by $\bm{z}$, are governed by the Langevin equation given by
\begin{equation}
\begin{aligned}
    \dot{\bm{z}}(t) = \bm{\Phi}(\bm{z}, t) + \sqrt{2\bm{\mathsf{D}}(\bm{z}, t)}\bm{\xi}(t),
\end{aligned}
\label{eq:GeneralLangevinEq}
\end{equation}
in the It\^{o} convention where $\bm{\xi}(t)$ is a Gaussian white noise vector that satisfies $\langle \xi_\mu \rangle = 0$ and $\langle \xi_\mu (t) \xi_\nu (t') \rangle = \delta(t-t') \delta_{\mu \nu}$, and $\langle \cdot \rangle$ denotes the ensemble average.
Here, $\bm{\Phi}(\bm{z}, t)$ is the drift vector and $\bm{\mathsf{D}}(\bm{z}, t)$ is the diffusion matrix.
These functions are defined by $\lim_{\Delta t \rightarrow 0} \frac{1}{\Delta t} \langle \bm{z}(t+\Delta t)-\bm{z}(t) \rangle|_{\bm{z}(t)}$ and $\lim_{\Delta t \rightarrow 0} \frac{1}{2\Delta t} \langle \left( \bm{z}(t+\Delta t)-\bm{z}(t) \right)^2 \rangle|_{\bm{z}(t)}$, respectively, commonly referred to as Kramers--Moyal coefficients of the Fokker--Planck equation~\cite{risken1996fokker}.
The state vector can be either $\bm{z} = \bm{x} \in \mathbb{R}^d$ or $\bm{z} = [\bm{x}^{\rm T}, \bm{v}^{\rm T}]^{\rm T} \in \mathbb{R}^{2d}$ with $\bm{v} \equiv \dot{\bm{x}}$ depending on the characteristic size and timescales of system. 
The first scenario is called the overdamped Langevin equation (OLE), and the second is called the underdamped Langevin equation (ULE).

Critical issues must be addressed to infer the OLE and ULE from observed trajectories.
In the OLE inference case, it is important to notice that the drift vector $\bm{\Phi}(\bm{x}, t)$ is often equivalent to a deterministic force $\bm{F}(\bm{x}, t)$ acting on the system, but $\bm{\Phi}(\bm{x}, t)$ differs when the diffusion matrix varies depending on the state (i.e., inhomogeneous diffusion). 
In such cases, $\bm{\Phi}(\bm{x}, t)=\bm{F}(\bm{x}, t) + \bm{\nabla}_{\bm{x}}\cdot \bm{\mathsf{D}}(\bm{x}, t)$ where $\bm{\nabla}_{\bm{x}}\cdot \bm{\mathsf{D}}(\bm{x}, t)$ is often called spurious or noise-dependent drift, which originates from the procedure of eliminating the velocity variable~\cite{sancho1982adiabatic, lau2007state-dependent, yang2013brownian, durang2015overdamped, volpe2016effective}.
Thus, to infer $\bm{F}(\bm{x}, t)$ acting on an inhomogeneously diffusing system, it is essential to calculate the gradient of the estimated diffusion matrix.
On the other hand, in the ULE inference case, a critical issue is that the velocity $\bm{v}$ cannot be directly measured but rather should be estimated from trajectories of $\bm{x}$. 
As we describe in the next section, the observed dynamics of estimated $\bm{v}$ is governed by a qualitatively different ULE from the true underlying ULE, requiring a leading-order correction to recover the true ULE regardless of the sampling intervals of time.
For instance, previous reports have shown that a constant rescaling factor should be multiplied by $3/2$ to retrieve the true friction coefficient $\gamma$ for a linear friction force $\bm{F}(\bm{v})=-\gamma \bm{v}$~\cite{lehle2015analyzing, pedersen2016how, ferretti2020building, bruckner2020inferring}.
Therefore, it is necessary to understand the observed ULE via terms of the true ULE as well as to derive unbiased estimators for accurate inference.

\subsection{Training and inference procedure}

Now, let us delve into the construction of datasets to feed LBN and the training process to learn Langevin equations.
A dataset $\mathcal{D}$ consists of $M$ independent stochastic trajectories with length $L$ [Fig.~\ref{fig1}(a)] as follows: $\mathcal{D} \equiv \left\{ \left\{ \left({\bm{z}}^{j}(t_i), \tilde{\bm{y}}^j(t_i)\right) \right\}_{i=1}^{L} \right\}_{j=1}^{M}$, where each element contains a pair of a state vector ${\bm{z}}^j(t_i)$ as an input and a target vector $\tilde{\bm{y}}^j(t_i)$ (called a label) with sampling times $t_i \equiv i\Delta t$ and the duration of a trajectory $\tau \equiv t_{L} = L \Delta t$.
We utilize three datasets for each example: a training dataset $\mathcal{D}_{\rm tr}$ for training LBN, a validation dataset $\mathcal{D}_{\rm val}$ for selecting the best model, and a test dataset $\mathcal{D}_{\rm te}$ for evaluating the performance of estimators. Here, $\mathcal{D}_{\rm te}$ is only used to demonstrate the performance of methods on unseen data and is not a requirement in practice.
Importantly, observed trajectories are not required to be stationary random processes or of substantial duration, especially when they can be observed multiple times. 
This adaptability is particularly relevant in experimental contexts where tracking a system for an extended period may be difficult due to factors such as blinking, photobleaching, or missed localization~\cite{manzo2015review}.
Note that $\tilde{\bm{y}}^j(t_i)$ is not a true drift vector or diffusion matrix but is assigned depending on the quantity we aim to infer, such as the drift vector or the diffusion matrix, as well as the regime of the system, such as whether it follows the OLE or ULE.
We specify $\tilde{\bm{y}}^j(t_i)$ in each scenario (hereafter, we skip the superscript indicating trajectory indices for simplicity).

To achieve probabilistic predictions and quantify prediction uncertainty, we employ a BNN architecture for inferring the Langevin equation [Fig.~\ref{fig1}(b)].
While ordinary neural networks yield fixed outputs on learned network parameter $\bm{\theta}$ after training, BNNs are built by introducing probabilistic components into the network, allowing them to provide not a single output but a distribution of possible outputs for a given input.
In our case, instead of learning a fixed value of $\bm{\theta}$ as in ordinary neural networks, LBN learns distributions $Q_{\bm{\phi}}(\bm{\theta})$ of $\bm{\theta}$ through variational inference~\cite{hinton1993keeping, blei2017variational} and Bayes-by-backdrop methods~\cite{blundell2015weight}.
The distribution $Q_{\bm{\phi}}(\bm{\theta})$, parameterized by trainable parameters $\bm{\phi}$, is obtained by minimizing the modified variational free energy $\mathcal{F}[Q_{\bm{\phi}}]$ (\ref{sec:ApendixA1}):
\begin{equation}
\begin{aligned}
    \mathcal{F}\left[ Q_{\bm{\phi}} \right] &\equiv \mathcal{L}^E(\bm{\phi}, \mathcal{D}) + \kappa \mathcal{L}^C(\bm{\phi}),
\end{aligned}
\label{eq:free_energy}
\end{equation}
with $0 \leq \kappa \leq 1$~\cite{liu2018advbnn, ashukha2020pitfalls, wenzel2020how} where the error loss $\mathcal{L}^E(\bm{\phi}, \mathcal{D})$ and the complexity loss $\mathcal{L}^C(\bm{\phi})$ are given by 
\begin{equation}
\begin{aligned}
    \mathcal{L}^E(\bm{\phi}, \mathcal{D}) &= \frac{1}{2}\left\langle \left\langle \left( \tilde{\bm{y}}(t) - \bm{y}_{\bm{\theta}}\left(\bm{z}, t \right) \right)^2 \right\rangle_{\bm{\theta}} \right\rangle_{\mathcal{D}} + \rm{const.},\\
    \mathcal{L}^C(\bm{\phi}) &= D_{\rm KL}\left[ Q_{\bm{\phi}}(\bm{\theta}) || P(\bm{\theta}) \right].
\end{aligned}
\label{eq:error_losses}
\end{equation}
Here, $\langle \cdot \rangle_{\mathcal{D}}$ denotes the average over $\mathcal{D}$ with $(\bm{z}(t), \tilde{\bm{y}}(t)) \in \mathcal{D}$, $\langle \cdot \rangle_{\bm{\theta}} \equiv \int Q_{\bm{\phi}} (\bm{\theta}) d\bm{\theta}$ is the marginalization over $\bm{\theta}$, $\bm{y}_{\bm{\theta}}(\bm{z}, t)$ is LBN's output for a given input $\bm{z}(t)$ and a sampled $\bm{\theta}$ from $Q_{\bm{\phi}}(\bm{\theta})$, and $D_{\rm KL}\left[ Q_{\bm{\phi}}(\bm{\theta}) || P(\bm{\theta})  \right]$ is the Kullback--Liebler divergence between $Q_{\bm{\phi}}(\bm{\theta})$ and the prior $P(\bm{\theta})$.
If we set $Q_{\bm{\phi}}(\bm{\theta})$ to be a fully factorized form of Gaussian distributions and $P(\bm{\theta})$ to be Gaussian, $D_{\rm KL}\left[ Q_{\bm{\phi}}(\bm{\theta}) || P(\bm{\theta})  \right]$ can be easily calculated (\ref{sec:ApendixA1}).
Because the exact calculation of marginalizing over $\bm{\theta}$ is infeasible in practice, $\langle \cdot \rangle_{\bm{\theta}}$ is calculated via Monte Carlo sampling from $Q_{\bm{\phi}}(\bm{\theta})$.
We also note that, to ensure the inferred diffusion matrix $\bm{\mathsf{D}}_{\bm{\theta}}$ is positive semi-definite and symmetric, mirroring the true diffusion matrix, LBN outputs a lower triangular matrix $\bm{\mathsf{L}}_{\bm{\theta}}$, which is then used to reconstruct $\bm{\mathsf{D}}_{\bm{\theta}}$ through $\bm{\mathsf{L}}_{\bm{\theta}}\bm{\mathsf{L}}_{\bm{\theta}}^{\rm T}$ (\ref{sec:ApendixB3}).

$\mathcal{F}[Q_{\bm{\phi}}]$ in Eq.~\eqref{eq:free_energy} consists of two terms, the error loss $\mathcal{L}^{E}$ and the complexity loss $\mathcal{L}^{C}$. Also called the negative log-likelihood, $\mathcal{L}^{E}$ is the mean squared error (MSE) between LBN's outputs and labels over $\bm{\theta}$ and $\mathcal{D}$.
Thus, estimators trained by minimizing $\mathcal{L}^{E}$ directly correspond to the maximum likelihood estimator and can be used to quantify how accurate the estimator's predictions are.
On the other hand, $\mathcal{L}^{C}$ is a statistical distance between the trained distribution $Q_{\bm{\phi}}(\bm{\theta})$ and the prior $P(\bm{\theta})$, which serves as a regularization to prevent overfitting and improve the generalization of neural networks.
Our first aim is to accurately infer the Langevin equation, and thus we introduce a prefactor $\kappa$ and modulate the influence of $\mathcal{L}^{C}$ by setting a small $\kappa$; a similar modification has also been used in Refs.~\cite{higgins2017betavae, liu2018advbnn, lee2023graddiv}.
Note that LBN is trained using $\mathcal{F}[Q_{\bm{\phi}}]$ on $\mathcal{D}_{\rm tr}$, and the final LBN is selected at the minimum point of $\mathcal{L}^E(\bm{\phi}, \mathcal{D}_{\rm val})$.
See Appendices~\ref{sec:ApendixA} and \ref{sec:ApendixB} for more background on BNNs and training details of LBN.

After completing the training process, prediction statistics can be obtained through an ensemble of predictions.
Given an input $\bm{z}(t)$, LBN provides a prediction $\bm{y}_{\bm{\theta}}(\bm{z}, t)$ by sampling $\bm{\theta}$ from $Q_{\bm{\phi}}(\bm{\theta})$ at each trial, so that all predictions for the given $\bm{z}(t)$ are different but form a certain distribution.
Collecting the predictions over multiple trials $N_{\bm{\theta}}$, we can obtain the ensemble of predictions $\{y_{\bm{\theta}^k}\}_{k=1}^{N_{\bm{\theta}}}$ and compute its average and variance as follows:
\begin{algorithm}[H]
    \begin{algorithmic}[1]
    \REQUIRE{The trained LBN for $\bm{y}$, dataset $\mathcal{D}$.}
    \LOOP
        \STATE Draw $\bm{\theta}^k \sim Q_{\bm{\phi}}(\bm{\theta})$.
        \State Compute $\bm{y}_{\bm{\theta}^k}(\bm{z}, t)$ where $\bm{z} \in \mathcal{D}$.
    \ENDLOOP
    \State Calculate from $\{\bm{y}_{\bm{\theta}^k} \}_{k=1}^{N_{\bm{\theta}}}$:
    \begin{equation}
    \begin{aligned}
        &\hat{\bm{y}}(\bm{z}, t) \equiv \left\langle \bm{y}_{\bm{\theta}}(\bm{z}, t) \right\rangle_{\bm{\theta}} = \frac{1}{N_{\bm{\theta}}}\sum_{k} \bm{y}_{\bm{\theta}^k}(\bm{z}, t)
        ,\\
        &\hat{\Sigma}_{\bm{y}} (\bm{z}, t) \equiv \frac{{\rm Var}[\bm{y}_{\bm{\theta}}(\bm{z}, t)]}{ \left\langle \bm{y}^2_{\bm{\theta}}(\bm{z}, t) \right\rangle_{\bm{\theta}}} = \frac{ \sum_{k} \left[ \hat{\bm{y}}(\bm{z}, t)-{\bm{y}}_{\bm{\theta}^k}(\bm{z}, t)\right]^2}{\sum_{k} \bm{y}^2_{\bm{\theta}^k}(\bm{z}, t)},
    \end{aligned}
    \label{eq:InferedDrift&Uncertainty}
    \end{equation}
    where $N_{\bm{\theta}}$ is the cardinality of a set $\{ \bm{\theta}^k \}_{k=1}^{N_{\bm{\theta}}}$.
    \end{algorithmic}
\caption{Inference procedure for $\bm{y}(\bm{z}, t)$ on $\mathcal{D}$}
\label{alg:inference}
\end{algorithm}
\noindent 
Here, $\hat{\Sigma}_{\bm{y}}(\bm{z}, t)$ is the relative uncertainty of predictions, where the uncertainty is quantified by ${\rm Var}[\bm{y}_{\bm{\theta}}(\bm{z}, t)] \equiv \langle (\hat{\bm{y}}(\bm{z}, t) - \bm{y}_{\bm{\theta}}(\bm{z}, t))^2 \rangle_{\bm{\theta}}$ as commonly used~\cite{jospin2022hands-on}.
${\rm Var}[\bm{y}_{\bm{\theta}}(\bm{z}, t)]$ indicates the extent of the spread of the predictions, thereby directly reflecting the uncertainty of the predictions. 
Note that ${\rm Var}[\bm{y}_{\bm{\theta}}(\bm{z}, t)]$ decreases as parameter uncertainty decreases with an increase in the number of training data.

\begin{figure*}[!t]
    \includegraphics[width=\linewidth]{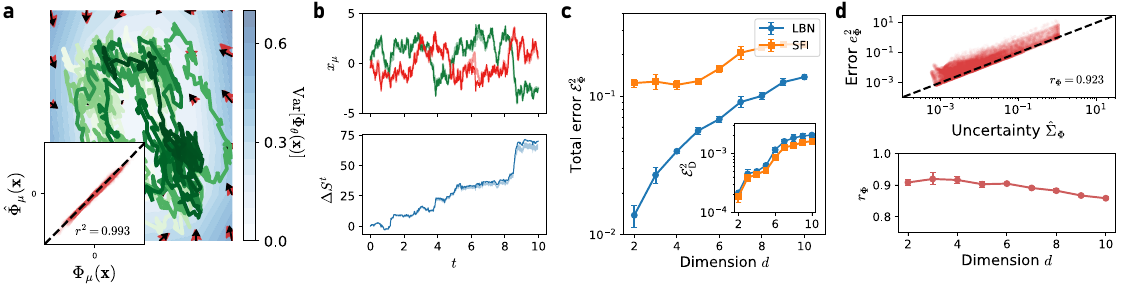}
    \vskip -0.1in
    \caption{
    Overdamped Langevin equation (OLE) inference for systems with a nonlinear force.
    (a) Example trajectory and the average of inferred drift fields $\hat{\bm{\Phi}}(\bm{x})$ (red arrows) for the $d=2$ case. The black arrows represent the true drift fields ${\bm{\Phi}}(\bm{x})$ and the colormap indicates the uncertainty of predictions. Inset: $\hat{{\Phi}}_\mu(\bm{x})$ vs. ${{\Phi}}_\mu(\bm{x})$ on $\mathcal{D}_{\rm te}$.
    (b) The generated trajectory (top) and the cumulative entropy production $\Delta S^t$ (bottom) along the trajectory are compared between the true model (solid lines) and the trained LBN (shaded area, behind the solid line) using the same random seed.
    (c) Total errors of LBN and stochastic force inference (SFI) for drift field ($\mathcal{E}^2_{\bm{\Phi}}$) and diffusion matrix ($\mathcal{E}^2_{\bm{\mathsf{D}}}$, inset) with increasing dimension $d$.
    (d) Pointwise errors for drift inference $e^2_{\Phi}$ vs. uncertainties of LBN $\hat{\Sigma}_{\Phi}$ for the $d=2$ case along a trajectory (top), and Pearson correlation coefficient between $e^2_{\Phi}$ and $\hat{\Sigma}_{\Phi}$ with $d$ ($r_{\bm{\Phi}}$, bottom).
    Error bars indicate the standard deviation of estimates from five independent trajectories and estimators. 
    }\label{fig2}
\end{figure*}

Importantly, we leverage the relationship between the pointwise error $e^2_{\bm{y}}(\bm{z}, t)$ and $\hat{\Sigma}_{\bm{y}}(\bm{z}, t)$ as follows (\ref{sec:ApendixC2}):
\begin{equation}
\begin{aligned}
    e_{\bm{y}}^2(\bm{z}, t) = \hat{{\Sigma}}_{\bm{y}} (\bm{z}, t)+ \frac{{\rm Bias}^2[\bm{y}_{\bm{\theta}}(\bm{z}, t)]}{\langle \bm{y}^2_{\bm{\theta}}(\bm{z}, t) \rangle_{\bm{\theta}}},
\end{aligned}
\label{eq:er_un_relation}
\end{equation}
where $e^2_{\bm{y}}(\bm{z}, t)$ is defined by
\begin{equation}
\begin{aligned}
    e^2_{\bm{y}}(\bm{z}, t) \equiv \frac{ \left\langle \left( \bm{y} (\bm{z}, t) - \bm{y}_{\bm{\theta}}(\bm{z}, t) \right)^2 \right\rangle_{\bm{\theta}}}{ \left\langle \bm{y}_{\bm{\theta}}^2 (\bm{z}, t) \right\rangle_{\bm{\theta}}},
\end{aligned}
\end{equation}
and ${\rm Bias}^2[\bm{y}_{\bm{\theta}}(\bm{z}, t)] \equiv \left[\bm{y}(\bm{z}, t) - \hat{\bm{y}}(\bm{z}, t)\right]^2$.
The total error $\mathcal{E}^2_{\bm{y}}$ in dataset $\mathcal{D}$ is defined by taking the average over $\mathcal{D}$ in both the denominator and numerator of $e^2_{\bm{y}}(\bm{z}, t)$, similarly with $\mathcal{L}^E$ in Eq.~\eqref{eq:error_losses}.
Note that ${\rm Bias}^2[\bm{y}_{\bm{\theta}}(\bm{z}, t)] \geq 0$, ensuring that $\hat{{\Sigma}}_{\bm{y}} (\bm{z}, t)$ gives a lower bound of $e_{\bm{y}}^2(\bm{z}, t)$.
Furthermore, since LBN is an unbiased estimator for the leading-order term, ${\rm Bias}^2[\bm{y}_{\bm{\theta}}(\bm{z}, t)]$ is minimized during the training so that $\hat{\Sigma}_{\bm{y}}(\bm{z}, t)$ and $e^2_{\bm{y}}(\bm{z}, t)$ become highly correlated.
In the next section, we empirically show the positive relationship between $e_{\bm{y}}^2(\bm{z}, t)$ and $\hat{\Sigma}_{\bm{y}} (\bm{z}, t)$ in various scenarios.

\section{Results}

We mention that all results here are assessed on $\mathcal{D}_{\rm te}$, a set of data not used during training, to demonstrate the generalization capability of LBN.
Simulation details and additional results are presented in \ref{sec:ApendixD}.

\subsection{Overdamped Langevin equation (OLE)}

In the context of a system with a timescale significantly longer than the velocity--velocity correlation time, the system can be described by the OLE by eliminating the velocity variable $\bm{v}$ from the ULE.
The discretized version of OLE is expressed as $\Delta \bm{x}(t_i)|_{\bm{x}(t_i)} = \bm{\Phi}(\bm{x}, t_i)\Delta t + \sqrt{2\bm{\mathsf{D}}(\bm{x}, t_i)}\Delta \bm{W}(t_i) + \bm{R}(t_i)$, where $\Delta \bm{x}(t_i)|_{\bm{x}(t_i)} \equiv \bm{x}(t_{i+1})-\bm{x}(t_i)$, $\bm{W}(t_i)$ is the Wiener process with zero mean and variance $\Delta t$, and $\bm{R}(t_i)$ is a remainder with $\langle \bm{R}(t_i) \rangle \sim \mathcal{O}(\Delta t^2)$.
This equation suggests that $\Delta \bm{x}(t_i)|_{\bm{x}(t_i)}$ can be considered as a sample from a Gaussian distribution:
\begin{equation}
\begin{aligned}
    \Delta \bm{x}(t_i)|_{\bm{x}(t_i)} \sim \mathcal{N}\left(\bm{\Phi}(\bm{x}, t_i)\Delta t, 2\bm{\mathsf{D}}(\bm{x}, t_i) \Delta t \right),
\end{aligned}
\label{eq:OLE_OnsagerMachlup}
\end{equation}
referred to as the Onsager--Machlup function~\cite{risken1996fokker}.
From Eq.~\eqref{eq:OLE_OnsagerMachlup}, it becomes evident that $\bm{\Phi}(\bm{x}, t_i)\Delta t$ and $2\bm{\mathsf{D}}(\bm{x}, t_i)\Delta t$ can be measured from the statistical moments $\langle \Delta {\bm{x}}(t_i) \rangle|_{\bm{x}(t_i)}$ and $\langle \Delta \bm{x}(t_i)\Delta \bm{x}(t_i)^{\rm T} \rangle|_{\bm{x}(t_i)}$, respectively.
Therefore, by assigning $\tilde{\bm{y}}(t_i)=\Delta \bm{x}(t_i)|_{\bm{x}(t_i)}/\Delta t$ and $\tilde{\bm{y}}(t_i)=\Delta \bm{x}(t_i)\Delta \bm{x}(t_i)^{\rm T}|_{\bm{x}(t_i)}/(2\Delta t)$ to construct datasets for separate drift and diffusion estimators, the LBN comprises the following unbiased OLE estimators to first order in $\Delta t$:
\begin{equation}
\begin{aligned}
    \hat{\bm{\Phi}}(\bm{x}, t) &= \frac{\langle \Delta \bm{x}(t) \rangle|_{\bm{x}(t)}}{\Delta t}, \\
    \hat{\bm{\mathsf{D}}}(\bm{x}, t) &= \frac{\langle \Delta \bm{x}(t)\Delta \bm{x}(t)^{\rm T}\rangle|_{\bm{x}(t)}}{2\Delta t}.
\end{aligned}
\label{eq:OLE_estimators}
\end{equation}
The detailed derivation of Eq.~\eqref{eq:OLE_estimators} is given in \ref{sec:ApendixC1}.

{\it Nonlinear force field} ---
To investigate the capability of our method to capture arbitrary nonlinear dynamics, we initially examine $d$-dimensional systems characterized by a nonlinear force and a diffusion matrix, specifically $\Phi_\mu(\bm{x}) = F_\mu(\bm{x}) = -\sum_{\nu}\mathsf{A}_{\mu \nu}x_\nu + \alpha x_\mu e^{-x_\mu^2}$ and $\mathsf{D}_{\mu\nu} = T\delta_{\mu\nu} - \sqrt{T} \delta_{\mu,\nu+1} - \sqrt{T} \delta_{\mu,\nu-1}$.
Given the homogeneity of $\bm{\mathsf{D}}$ in this system, LBN generates $\hat{\bm{\mathsf{D}}}$ state-independently with an inputless structure (see \ref{sec:ApendixB3}).
After training on a trajectory, LBN accurately learns both the drift field and the diffusion matrix, achieving $r^2=0.993$ and $r^2=0.999$ for the $d=2$ case, respectively [Fig.~\ref{fig2}(a)].
In addition, LBN not only offers precise drift inference for states outside the training range but also quantifies uncertainty ${\rm Var}[\bm{\Phi}_{\bm{\theta}}(\bm{x})]$ in this inference [colormap in Fig.~\ref{fig2}(a)].
This uncertainty estimation proves valuable for identifying uncertain regions and quantifying the extent of the uncertainty.

Demonstrating the practical utility of our trained LBN is important.
The key strengths of our method are its ability to perform precise inferences regardless of functional forms and its capacity to provide quantifiable uncertainty for these inferences.
Capitalizing on these dual capabilities, we can now leverage LBN to predict the system dynamics over time, analyze thermodynamic quantities such as entropy production $\Delta S(t)$ along the trajectory, and also track the uncertainties associated with these calculations [Fig.~\ref{fig2}(b)].
Unlike deterministic methods that offer a single prediction for a given state, by yielding multiple predictions LBN can construct a possible range of predictions; if a given state falls within (outside) the training range, the distribution of predictions will be narrow (wide).
In \ref{sec:ApendixC2}, we present how the uncertainty in the phase space varies with the length $L$ of the training trajectory, showing that the uncertainty diminishes as $L$ increases.

We apply our approach to this nonlinear system with increasing $d$ and compare its performance against stochastic force inference (SFI)~\cite{frishman2020learning}, a well-established OLE inference method (see \ref{sec:ApendixB1} for how we apply the SFI).
As illustrated in Fig.~\ref{fig2}(c), LBN consistently outperforms SFI in accurately inferring the drift field across the entire range of $d$. Additionally, the total error $\mathcal{E}^2_{\bm{\mathsf{D}}}$ for the diffusion matrix is not only remarkably small but also virtually indistinguishable between the two methods.
Convergence of the total error with increasing duration $\tau$ is depicted in \ref{sec:ApendixD2}.
Importantly, we confirm that $\hat{\Sigma}_{\bm{\Phi}} (\bm{x}(t))$ is highly correlated with $e^2_{\bm{\Phi}}(\bm{x}(t))$ [Fig.~\ref{fig2}(d)], as expected from Eq.~\eqref{eq:er_un_relation}.
This strong positive correlation remains robust across varying $d$, implying that the uncertainty provided by LBN can be used to estimate prediction errors. 
In particular, higher uncertainties guarantee higher errors because $e^2_{\bm{y}}$ must be larger than $\hat{{\Sigma}}_{\bm{y}} (\bm{x}, t)$ [Eq.~\eqref{eq:er_un_relation}].

\begin{figure*}[!ht]
    \includegraphics[width=\linewidth]{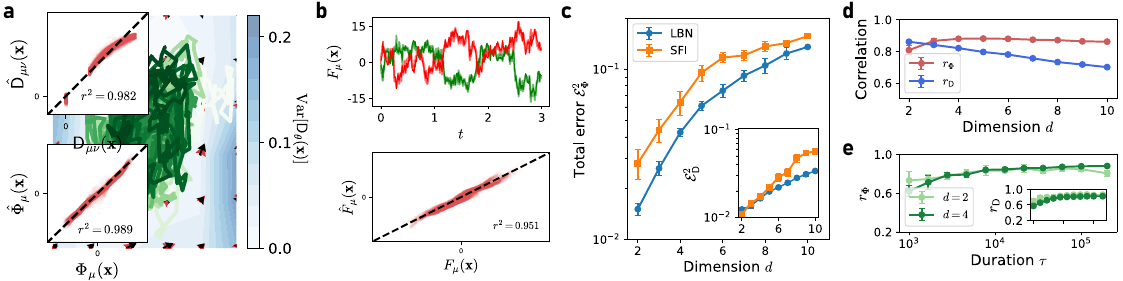}
    \vskip -0.1in
    \caption{
    OLE inference for systems with an inhomogeneous diffusion matrix.
    (a) Example trajectory and diagonal entries of the diffusion matrix (red arrows), $[\hat{\mathsf{D}}_{11}(\bm{x}), \hat{\mathsf{D}}_{22}(\bm{x})]^{\rm T}$, for the $d=2$ case. The black arrows represent the true values $[{\mathsf{D}}_{11}(\bm{x}), {\mathsf{D}}_{22}(\bm{x})]^{\rm T}$ and the colormap indicates the uncertainty of predictions. Insets: $\hat{{\Phi}}_{\mu}(\bm{x})$ vs. ${\Phi}_{\mu}(\bm{x})$ (top) and $\hat{\mathsf{D}}_{\mu\nu}(\bm{x})$ vs. ${\mathsf{D}}_{\mu\nu}(\bm{x})$ (bottom) on $\mathcal{D}_{\rm te}$.
    (b) Exact and inferred forces, $F_\mu(\bm{x})$ (solid line) and $\hat{F}_\mu(\bm{x})$ (shaded area, behind the solid line), along a trajectory (top) and a scatter plot between them (bottom).
    (c) Total errors of LBN and SFI for the drift field ($\mathcal{E}^2_{\bm{\Phi}}$) and diffusion matrix ($\mathcal{E}^2_{\bm{\mathsf{D}}}$, inset) with increasing dimension $d$. (d, e)
    Pearson correlation coefficient $r_{\bm{\Phi}}$ between $e^2_{\Phi}$ and $\hat{\Sigma}_{\Phi}$ ($r_{\bm{\mathsf{D}}}$ between $e^2_{\bm{\mathsf{D}}}$ and $\hat{\Sigma}_{\bm{\mathsf{D}}}$) with increasing $d$ (d) and the duration $\tau$ of a training trajectory for $d=2$ and $d=4$ (e).
    Error bars indicate the standard deviation of estimates from five independent trajectories and estimators. 
    }\label{fig3}
\end{figure*}

{\it Inhomogeneous diffusion matrix} ---
As the next scenario, we explore a $d$-dimensional system with an inhomogeneous diffusion matrix.
In the analysis of systems, while the assumption of a homogeneous diffusion matrix is commonly employed for convenience, such homogeneity is often rare in practical situations and can potentially lead to incorrect analyses.
For instance, in situations where particles are near a wall~\cite{brenner1961slow, lacon2001drift, volpe2010influence, volpe2016effective}, particles undergo an inhomogeneous diffusion and a spurious drift vector $\bm{\nabla}_{\bm{x}} \cdot \bm{\mathsf{D}}(\bm{x})$ arises, causing a discrepancy between $\bm{\Phi}(\bm{x})$ and $\bm{F}(\bm{x})$.
Therefore, for accurate inference of force field $\bm{F}(\bm{x})$, it is essential to address both the accurate inference of $\bm{\mathsf{D}}(\bm{x})$ and the calculation of $\bm{\nabla}_{\bm{x}} \cdot \hat{\bm{\mathsf{D}}}(\bm{x})$.
Here, we demonstrate the effectiveness of LBN in handling both aspects by applying it to an inhomogeneously diffusing system.

We consider a model characterized by a linear force $\bm{F}(\bm{x})=-\bm{\mathsf{A}}\bm{x}$ and an inhomogeneous diffusion matrix $\mathsf{D}_{\mu\nu}(\bm{x}) = T(1 + \alpha e^{-x_\mu^2/2})\delta_{\mu\nu}$, which induce the drift vector $\Phi_\mu(\bm{x})=-\sum_{\nu}\mathsf{A}_{\mu \nu}x_\nu - \alpha T x_\mu e^{-x_\mu^2/2}$.
As shown in Fig.~\ref{fig3}(a), LBN precisely infers the drift field and the inhomogeneous diffusion matrix in the $d=2$ case, achieving $r^2 = 0.979$ and $r^2=0.982$, respectively.
We present the uncertainty of $\hat{\bm{\mathsf{D}}}(\bm{x})$ as the colormap in Fig.~\ref{fig3}(a), which provides insight into how the uncertainty varies by region.
Having obtained precise estimators for the drift and diffusion, we can now proceed to infer the force $\bm{F}(\bm{x})$.
Importantly, our method easily handles the calculation of the gradient of $\hat{\bm{\mathsf{D}}}(\bm{x})$ through backpropagation.
By applying backpropagation to the diffusion estimator and combining it with the drift estimator, the force field can be inferred by
\begin{equation}
\begin{aligned}
    \bm{F}_{\bm{\theta}}(\bm{x}) = \bm{\Phi}_{\bm{\theta}}(\bm{x}) - \nabla_{\bm{x}} \cdot \bm{\mathsf{D}}_{\bm{\theta}}(\bm{x}).
\end{aligned}
\end{equation}
Note that network parameters $\bm{\theta}$ for both estimators are separately sampled from different variational distributions.
The ensemble of force estimators, $\{\bm{F}_{\bm{\theta}^k} \}_{k=1}^{N_{\bm{\theta}}}$, is calculated from two different ensembles, $\{\bm{\Phi}_{\bm{\theta}^k} \}_{k=1}^{N_{\bm{\theta}}}$ and $\{ \bm{\nabla}_{\bm{x}} \cdot \bm{\mathsf{D}}_{\bm{\theta}^k} \}_{k=1}^{N_{\bm{\theta}}}$.
We confirm that $\bm{F}(\bm{x})$ is successfully retrieved from this procedure with $r^2=0.942$ [Fig.~\ref{fig3}(b)].

We now evaluate the performance of LBN in comparison to SFI as well as investigate the correlation between $e^2_{\bm{y}}$ and $\hat{{\Sigma}}_{\bm{y}}$ as the dimensionality $d$ increases in the inhomogeneous diffusion model.
Fig.~\ref{fig3}(c) illustrates that LBN outperforms the SFI across all ranges of $d$ for both drift and diffusion inferences.
The convergence of $\mathcal{E}^2_{\bm{\Phi}}$ and $\mathcal{E}^2_{\bm{\mathsf{D}}}$ with increasing trajectory duration $\tau$ is depicted in \ref{sec:ApendixD3}.
The correlations for both estimators, denoted by $r_{\bm{\Phi}}$ and $r_{\bm{\mathsf{D}}}$, also remain consistently high as $d$ increases [Fig.~\ref{fig3}(d)].
Additionally, we explore $r_{\bm{\Phi}}$ and $r_{\bm{\mathsf{D}}}$ with varying $\tau$ [Fig.~\ref{fig3}(e)] and find that they consistently exhibit strong correlations, indicating that $\hat{{\Sigma}}_{\bm{y}}$ successfully represents $e^2_{\bm{\mathsf{D}}}$.

\begin{figure}[!t]
    \includegraphics[width=\linewidth]{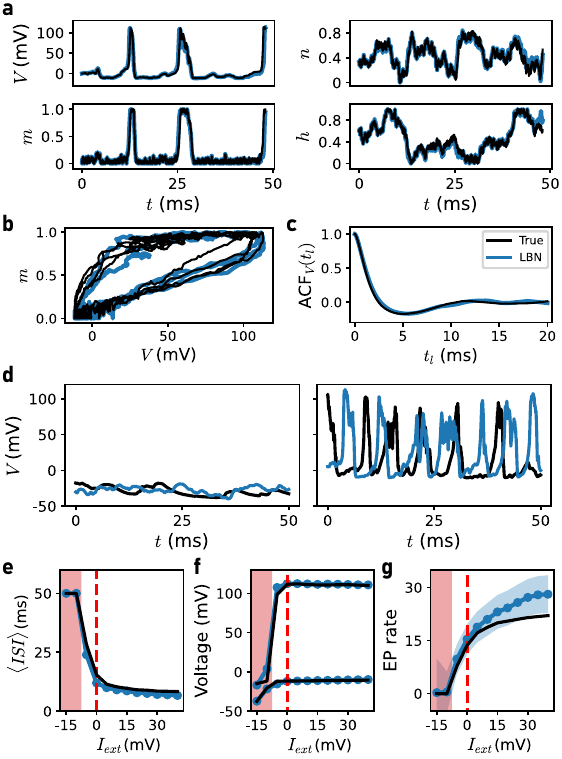}
    \vskip -0.1in
    \caption{
    OLE inference for the spiking neuron model.
    (a) Trajectories of four state variables generated from both the true Hodgkin--Huxley (HH) model (black) and the trained LBN (blue) with the same random seed.
    (b) Trajectories in the phase space of $V$ and $m$.
    (c) Autocorrelation function of $V(t)$, denoted by $ACF_{\bm{V}}(t_l)$, for both trajectories with respect to time lag $t_l$.
    (d) Generated trajectories at $I_{\rm ext}=-15~\text{mV}$ (left) and $I_{\rm ext}=30~\text{mV}$ (right) from the LBN trained at $I_{\rm ext}=0~\text{mV}$ and the true HH model.
    (e) Mean inter-spike interval ($ISI$), (f) mean of the minimum and maximum values of $V$, and (g) entropy production rate ($\langle \dot{S} \rangle$) with varying external current $I_{\rm ext}$. The blue shaded area in (g) indicates the uncertainty of $\langle \dot{S} \rangle$.
    When there are no spikes in the generated trajectories, the $ISI$ is set to the length of the trajectory ($=50~\text{ms}$).
    The red dotted line indicates $I_{\rm ext}=0$, which is the value used for training, and the red shaded area indicates the non-spiking region.
    }\label{fig4}
\end{figure}

{\it Spiking neuron model} --- To demonstrate the broad applicability of our method with a highly challenging scenario, we apply LBN to a well-known biological neuron model called the Hodgkin--Huxley (HH) model~\cite{hodgkin1952quantitative}.
The HH model is a paradigmatic neuron model that describes the dynamics of action potentials in neurons using a set of four highly nonlinear, coupled, first-order differential equations with four state variables: $V(t)$ is the membrane voltage ($\rm{mV}$), and $n(t)$, $m(t)$, and $h(t)$ are dimensionless probabilities that represent the activation of potassium ion channels and the activation and deactivation of sodium ion channels, respectively (see \ref{sec:ApendixD4}).
By introducing a homogeneous stochastic force on each gating variable (i.e., $n(t)$, $m(t)$, and $h(t)$), the stochastic HH model can be mathematically described with an OLE.
This introduced stochasticity leads to random motion of the channel variables, resulting in a more intricate and unpredictable spike pattern for $V(t)$, as depicted in Fig.~\ref{fig4}(a).
We generate multiple trajectories ($M=50$) governed by this complex OLE and train LBN using these trajectories to infer the drift field and the diffusion matrix.

Remarkably, despite the highly complex OLE of the HH model, LBN successfully learns both the drift field and the diffusion matrix, achieving low total errors of $\mathcal{E}^2_{\bm{\Phi}}=0.015$ and $\mathcal{E}^2_{\bm{\mathsf{D}}}=0.0009$ (see \ref{sec:ApendixD4} for the convergence of total errors). 
Moreover, using these trained estimators, we verify that the trained LBN can generate trajectories that closely match the model simulations when using the same random seed as the model simulation [Fig.~\ref{fig4}(a)].
The generated trajectories trace precisely the same closed curve in the phase space [Fig.~\ref{fig4}(b)] and exhibit a similar autocorrelation function ${\rm ACF}_{\bm{V}}(t_l)$ with the time lag $t_l$ [Fig.~\ref{fig4}(c)].

To validate the practical utility of LBN as a proxy for the true system in this case, we conduct tests to assess its quantitative analysis capabilities in a variety of environments beyond the training environment.
One intriguing phenomenon in the HH model is the pronounced variation in the spike pattern of $V(t)$ in response to changes in $I_{\rm ext}$, known as a Hopf bifurcation~\cite{xie2008controlling}. 
This bifurcation refers to the transition from a non-spiking state to a spiking state as $I_{\rm ext}$ varies.
Spiking neuron dynamics are commonly characterized by the inter-spike interval ($ISI$) and the minimum and maximum of $V(t)$, where $ISI$ becomes finite and the discrepancy between the $V(t)$ minimum and maximum abruptly becomes significant after the critical point.
Notably, our models were only trained at $I_{\rm ext}=0$. 
In this experiment, we measure the $ISI$ and $V(t)$ using generated trajectories from the trained LBN, while applying a constant $I_{\rm ext}$ by adding the drift vector for $V(t)$ at each step, and compare with the true values measured from generated trajectories via the true model [Fig.~\ref{fig4}(d)]. Interestingly, LBN precisely distinguishes both non-spiking and spiking regions as well as captures the extrema of $V(t)$ with respect to $I_{\rm ext}$ [Fig.~\ref{fig4}(e) and~\ref{fig4}(f)], indicating that LBN successfully works beyond the training range.
We then calculate the entropy production rate $\langle \dot{S} \rangle$ from the generated trajectories and verify that LBN adeptly captures the abrupt increase of $\langle \dot{S} \rangle$ at the critical point [Fig.~\ref{fig4}(g)] caused by the occurrence of spikes (\ref{sec:ApendixD4}).
The results in Fig.~\ref{fig4}(e) and~\ref{fig4}(f) show that our approach can be used to understand both thermodynamic and dynamical properties of systems beyond the training environment as well as to predict the critical point of a nonequilibrium phase transition.
Furthermore, results also show that the uncertainty of measured $\langle \dot{S} \rangle$ becomes larger with increasing distance from the training point $I_{\rm ext} = 0$ because the number of unobserved states increases.
The increasing uncertainty with distance implies increasing errors of prediction, suggesting that uncertainty can play a role as an indicator of the reliability of the prediction results.
Accordingly, the ability of our model to accurately capture the spiking dynamics and bifurcation point, despite being trained at $I_{\rm ext}=0$, demonstrates its versatility and capability to provide reliable quantitative analysis in diverse settings of the HH model.

\subsection{Underdamped Langevin equation (ULE)}

In this section, we show that our method can be extended to infer the ULE with a few modifications.
While the OLE can efficiently describe many stochastic systems on molecular scales, various macroscale organisms or short-time movements of microscale systems should be addressed by the ULE on account of inertial effects~\cite{huang2011direct, attanasi2014information, lowen2020inertial, bruckner2021learning}.
A major problem in ULE inference, which is absent in the OLE case, is that the velocity $\bm{v}$ is not directly accessible and should be estimated by observed trajectories of $\bm{x}$ in experiments.
To tackle such practical situations, we estimate the velocity as $\hat{\bm{v}}(t) \equiv [\bm{x}(t+\Delta t) - \bm{x}(t)]/\Delta t$ and set a state variable $\bm{z} \equiv [\bm{x}^{\rm T}, \hat{\bm{v}}^{\rm T}]^{\rm T} \in \mathbb{R}^{2d}$ for constructing datasets.

The discretized version of the ULE is given by
\begin{equation}
\begin{aligned}
    \Delta \bm{x}(t_i)|_{\bm{x}(t_i), \bm{v}(t_i)} &= \bm{v}(t_i)\Delta t + \bm{R}_{\bm{x}}(t_i),\\
    \Delta \bm{v}(t_i)|_{\bm{x}(t_i), \bm{v}(t_i)} &= \bm{\Phi}(\bm{x}, \bm{v}, t_i)\Delta t \\ &+ \sqrt{2\bm{\mathsf{D}}(\bm{x}, \bm{v}, t_i)}\Delta \bm{W}(t_i) + \bm{R}_{\bm{v}}(t_i),
\end{aligned}
\end{equation}
where the inertial mass is set to 1 and the remainders satisfy $\langle \bm{R}_{\bm{x}} (t_i) \rangle, \langle \bm{R}_{\bm{v}} (t_i) \rangle \sim \mathcal{O}(\Delta t^2)$. Since $\hat{\bm{v}}(t_i)$ is used instead of $\bm{v}(t_i)$, we derive the expression of $\Delta \hat{\bm{v}}(t_i)|_{\bm{x}(t_i), \hat{\bm{v}}(t_i)}$ and show that $\Delta \hat{\bm{v}}(t_i)$ can be understood as a sample from a given Gaussian distribution:
\begin{equation}
\begin{aligned}
    &\Delta \hat{\bm{v}}(t_i)|_{\bm{x}(t_i), \hat{\bm{v}}(t_i)} \\&\sim \mathcal{N}\biggl(\left[\bm{\Phi}(\bm{x}, \hat{\bm{v}}, t_i)-\frac{1}{3}\bm{\mathsf{D}}(\bm{x}, \hat{\bm{v}}, t_i)\partial_{\hat{\bm{v}}} \ln P(\bm{x}, \hat{\bm{v}}, t_i) \right]\Delta t, \\ & \quad\qquad \frac{4}{3}\bm{\mathsf{D}}(\bm{x}, \hat{\bm{v}}, t_i) \Delta t \biggr),
\end{aligned}
\label{eq:ULE_OnsagerMachlup}
\end{equation}
where $P(\bm{x}, \hat{\bm{v}}, t_i)$ is the probability density function (PDF) of $\bm{x}$ and $\hat{\bm{v}}$ at $t_i$.
Compared with Eq.~\eqref{eq:OLE_OnsagerMachlup}, the resulting Gaussian distribution exhibits markedly different properties: the average of the Gaussian distribution is not solely determined by the drift field, and the prefactor of the distribution variance is reduced by multiplying by $2/3$.
Equation~\eqref{eq:ULE_OnsagerMachlup} indicates that the observed dynamics are described by a fundamentally different ULE from the original ULE of the system, featuring an additional drift term proportional to the diffusion matrix and smaller fluctuations by the smaller diffusion matrix.
Note that these discrepancies are not dependent on $\Delta t$, implying that we cannot remove them even if we sample trajectories with $\Delta t \rightarrow 0$.
As a consequence of the leading-order biases introduced in the straightforward extension of Eq.~\eqref{eq:OLE_OnsagerMachlup}, it is necessary to develop new unbiased estimators~\cite{lehle2015analyzing, pedersen2016how, ferretti2020building, bruckner2020inferring}.
Here, we obtain our unbiased ULE estimators to first order in $\Delta t$ as follows:
\begin{equation}
\begin{aligned}
    \hat{\bm{\Phi}}(\bm{x}, \bm{v}, t) &= \hat{\bm{\Psi}}_f(\bm{x}, \hat{\bm{v}}, t) + \frac{1}{4}\left[ \hat{\bm{\Psi}}_f(\bm{x}, \hat{\bm{v}}, t) - \hat{\bm{\Psi}}_b(\bm{x}, \hat{\bm{v}}, t) \right], \\
    \hat{\bm{\mathsf{D}}}(\bm{x}, \bm{v}, t) &= \frac{3}{2}\frac{\langle \Delta \hat{\bm{v}}(t)\Delta \hat{\bm{v}}(t)^{\rm T}\rangle|_{\bm{x}(t), \hat{\bm{v}}(t)}}{2\Delta t},
\end{aligned}
\label{eq:ULE_estimators}
\end{equation}
where
\begin{equation}
\begin{aligned}
    \hat{\bm{\Psi}}_f(\bm{x}, \hat{\bm{v}}, t) &\equiv \frac{\langle \Delta \hat{\bm{v}}(t) \rangle|_{\bm{x}(t), \hat{\bm{v}}(t)}}{\Delta t},\\
    \hat{\bm{\Psi}}_b(\bm{x}, \hat{\bm{v}}, t) &\equiv \frac{\langle \Delta \hat{\bm{v}}(t-\Delta t) \rangle|_{\bm{x}(t), \hat{\bm{v}}(t)}}{\Delta t}.
\end{aligned}
\end{equation}
\noindent
To implement $\hat{\bm{\Phi}}(\bm{x}, \hat{\bm{v}}, t)$, we combine two neural networks $\hat{\bm{\Psi}}_f(\bm{x}, \hat{\bm{v}}, t)$ and $\hat{\bm{\Psi}}_b(\bm{x}, \hat{\bm{v}}, t)$ trained by assigning $\tilde{\bm{y}}(t_i)=\Delta \hat{\bm{v}}(t_i)|_{\bm{x}(t), \hat{\bm{v}}(t)}/\Delta t$ and $\tilde{\bm{y}}(t_i)=\Delta \hat{\bm{v}}(t_{i-1})|_{\bm{x}(t), \hat{\bm{v}}(t)}/\Delta t$, respectively, for constructing datasets.
For the diffusion estimator, we assign $\tilde{\bm{y}}(t_i)=3\Delta \hat{\bm{v}}(t_i)\Delta \hat{\bm{v}}(t_i)^{\rm T}|_{\bm{x}(t), \hat{\bm{v}}(t)}/(4\Delta t)$, which corresponds to the straightforward generalization of the diffusion estimator in Eq.~\eqref{eq:OLE_estimators} multiplied by $3/2$.
Detailed derivations of Eqs.~\eqref{eq:ULE_OnsagerMachlup} and \eqref{eq:ULE_estimators} are given in \ref{sec:ApendixC1}.

\begin{figure}[!t]
    \includegraphics[width=\linewidth]{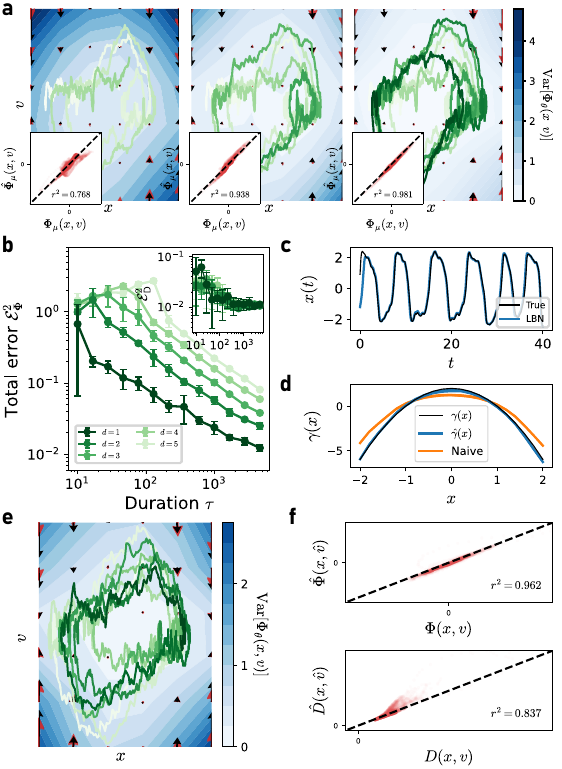}
    \vskip -0.1in
    \caption{
    Underdamped Langevin equation (ULE) inference for a stochastic van der Pol oscillator.
    (a) Example $xv$ trajectories and inferred drift fields (red arrows) for the $d=1$ case with increasing $\tau$ from left to right ($\tau \simeq 17$, $130$, and $1000$). The black arrows represent the exact fields and the colormap indicates the uncertainties of predictions $\rm{Var}[\bm{\Phi}_{\bm{\theta}}(x, v)]$. Insets: $\hat{\bm{\Phi}}(x, v)$ vs. ${\bm{\Phi}}(x, v)$.
    (b) Total errors of LBN for drift field ($\mathcal{E}^2_{\bm{\Phi}}$) and diffusion matrix ($\mathcal{E}^2_{\bm{\mathsf{D}}}$, inset) prediction with increasing $\tau$, respectively, for $d=1, \dots, 5$. Error bars indicate the standard deviation of estimates from five independent trajectories and estimators. 
    (c) Trajectories of $x(t)$ generated from both true model (black) and trained LBN (blue) with the same random seed.
    (d) True effective friction coefficient $\gamma(x)$ and that inferred by LBN $\hat{\gamma}(x)$ with varying $x$.
    Here, the `Naive' label indicates the inferred values by a straightforward extension of Eq.~\eqref{eq:OLE_estimators}, $\hat{\bm{\Psi}}_{f}$.
    (e) Example $xv$ trajectory and drift fields for the $d=1$ case with multiplicative noise.
    The meanings of the symbols and colormap are the same as in (a).
    (f) Scatter plots between ${{\Phi}}(x, v)$ and $\hat{{\Phi}}(x, v)$ (top) and ${\mathsf{D}}(x, v)$ and $\hat{\mathsf{D}}(x, v)$ (bottom) for the multiplicative noise case.
    }
\label{fig5}
\end{figure}

{\it Stochastic van der Pol oscillator} ---
We explore a stochastic $d$-dimensional van der Pol oscillator~\cite{van1926on, bruckner2020inferring}, characterized by a nonlinear drift field $\Phi_\mu(\bm{x}, \bm{v}) = k(1-x_\mu^2)v_\mu - x_\mu$ and a homogeneous diffusion matrix $\mathsf{D}_{\mu\nu}=T\delta_{\mu\nu}$.
First, we examine the impact of increasing $\tau$ on the accuracy and uncertainty of drift inference for $d=1$.
In Fig.~\ref{fig5}(a), as expected, we observe a trend where the accuracy increases (from $r^2 = 0.761$ to $r^2=0.978$) while the uncertainty diminishes (fading colormaps from left to right) with $\tau$. 
Notably, the contour plot representing uncertainties exhibits a shape akin to that of the trajectory, growing larger with distance from the center.
This observation reflects that LBN effectively learns the drift field based on $\mathcal{D}_{\rm tr}$ and manifests larger uncertainty in regions unseen during training.
Furthermore, we validate the effectiveness of our method in higher-dimensional systems with $d=1, \dots, 5$, noting a decrease in total errors with increasing $\tau$ as depicted in Fig.~\ref{fig5}(b).

We now compare the trajectories generated from the true model and trained LBN with the same random seed for $d=1$ case. Our observation reveals that LBN precisely predicts the system's evolution over time [Fig.~\ref{fig5}(b)].
Moving forward, we measure the effective friction coefficient $\gamma$, which is a prefactor of the linear term of $v$ in the drift field [Fig.~\ref{fig5}(c)]. 
In the van der Pol oscillator system, the effective friction coefficient is position-dependent, given by $\gamma(x) = k (1 - x^2)$.
To extract $\gamma(x)$ from the trained drift field at a given $x$, we obtain $\hat{\Phi}(x, v)$ at a fixed $x$ and within the range of $v$, specifically $v \in [-0.1, 0.1]$, to neglect higher-order terms of $v$. 
We then estimate $\hat{\gamma}(x)$ using a linear regression applied to $\hat{\Phi}(x, v)$ and $v$.
Remarkably, as depicted in Fig.~\ref{fig5}(d), LBN exactly recovers $\gamma(x)$ by $\hat{\gamma}(x)$ across varying $x$ and identifies regions where a sign switch of $\gamma(x)$ occurs without any corrections, additional estimates of other quantities, or assumptions about $\gamma(x)$.
These results can be compared to the case of using $\hat{\bm{\Psi}}_f$, a naive generalization of the OLE estimator, as the drift estimator, where the obtained effective friction coefficient is biased by a factor of $2/3$ as observed in linear systems~\cite{pedersen2016how, ferretti2020building}.
We thus verify that our estimators are unbiased and provide an accurate reconstruction of the underlying Langevin equation.
It is noteworthy that this estimation method for $\gamma(x)$ via LBN can be extended to any nonlinear system, even in cases with no exact linear term of $v$. 
In such situations, $\hat{\gamma}(x)$ is estimated by $\partial_{\hat v}\hat{\Phi}(x, \hat{v})|_{\hat{v}=0}$, representing a coefficient of the linear term in the first-order Taylor expansion.

We additionally evaluate the performance of LBN in the presence of multiplicative noise $\mathsf{D}(\bm{x}, \bm{v}) = (T_0 + T_x x_\mu^2 + T_v v_\mu^2)\delta_{\mu\nu}$ in the stochastic van der Pol oscillator system with $d=1$. 
LBN successfully infers both drift and diffusion fields as shown in Fig.~\ref{fig5}(e) and~\ref{fig5}(f), implying that our approach can be applied to a wide range of underdamped systems.
Results of higher-dimensional systems and convergence with respect to $\tau$ are illustrated in \ref{sec:ApendixD5}.

\begin{figure}[!t]
    \includegraphics[width=\linewidth]{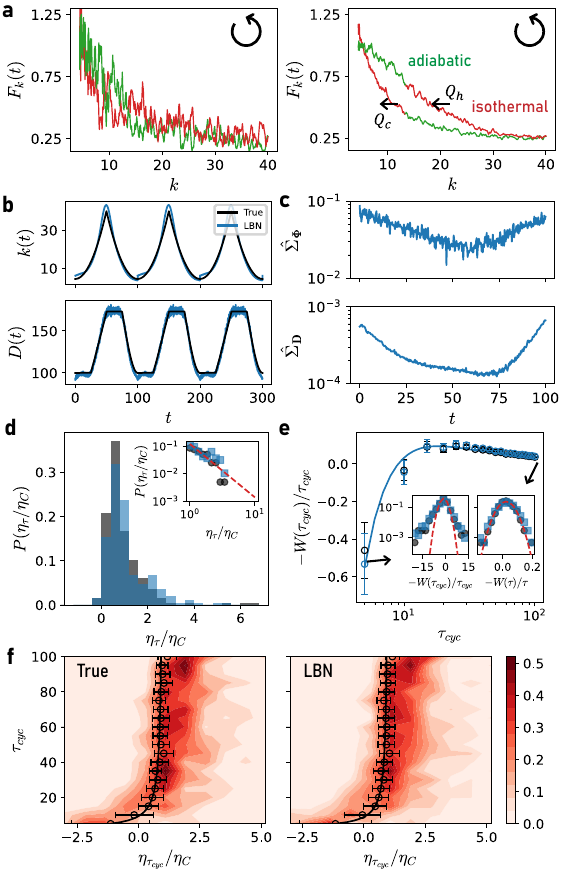}
    \vskip -0.1in
    \caption{
    ULE inference for the Brownian Carnot engine.
    (a) Mean Clapeyron diagrams obtained from trajectories of the true engine in $\mathcal{D}_{\rm tr}$ with $M=50$ (left) and our trained engine with $M=2000$ (right).
    (b) True and trained protocols of stiffness $k(t)$ and diffusion coefficient $D(t)$ with respect to time $t$.
    (c) Uncertainties of drift (top) and diffusion (bottom) inferences, $\hat{\Sigma}_{\bm{\Phi}}$ and $\hat{\Sigma}_{\bm{\mathsf{D}}}$, with respect to $t$.
    (d) PDFs of the efficiency, $P (\eta_{\tau}/\eta_C)$, of the true and trained engine with $\tau=100$, where $\eta_C$ is the Carnot efficiency.
    The inset shows the tails of the two distributions, where the red dotted line is a fitted line to a power law with an exponent of $-1.92$.
    (e) Work rate of the true and trained engine, $-W(\tau_{cyc}) / \tau_{cyc}$, with respect to cycle duration $\tau_{cyc}$. The average work rate is the power, i.e., $\mathcal{P}(\tau_{cyc})\equiv-\langle W(\tau_{cyc}) \rangle / \tau_{cyc}$.
    The two insets show PDFs of the work rate at $\tau_{cyc}=\tau$ (left) and $\tau_{cyc}=\tau/20$ (right), where the red dotted line is a fitted line to a Gaussian distribution.
    (f) Contour plots of the PDFs of $\eta_{\tau_{cyc}}/\eta_C$ obtained from the true (left) and trained (right) engines with respect to $\tau_{cyc}$. 
    Symbols and error bars indicate the average and the standard deviation of the values, respectively.
    Solid lines in (e) and (f) indicate fitted lines between values and the expected law, referred to from Ref.~\cite{martinez2016brownian}.
    In (d) and (f), $\eta_{\tau}$ and $\eta_{\tau_{cyc}}$ are obtained from summing over $n_{cyc}=10$ and $n_{cyc}=40$ cycles, respectively.
    Our trained engine is trained solely at $\tau_{cyc}=\tau=100$. 
    }\label{fig6}
    \vskip -0.1in
\end{figure}

{\it Brownian Carnot engine} ---
As the final challenging scenario for ULE inference, we consider the well-known Brownian Carnot engine, which was first realized in Ref.~\cite{martinez2016brownian}, to demonstrate the practical utility of LBN and validate its ability to handle non-stationary dynamics.
This system involves a single particle ($d=1$) trapped by a harmonic potential $U(x, t)=k(t)x(t)^2/2$ with a controllable stiffness $k(t)$ and in contact with a thermal bath with a constant friction coefficient $\gamma$ and controllable temperature $T(t)$.
Similar to the classical Carnot engine, the cycle of this engine, with a cycle duration $\tau_{cyc}$, also comprises two isothermal and adiabatic processes by tuning the stiffness $k(t)$ and the diffusion coefficient $D(t) \equiv \gamma T(t)$ with respect to time, simultaneously (see \ref{sec:ApendixD6} for detailed protocols).
We set all four thermodynamic processes to have the same length $\tau_{cyc}/4$ per cycle.
Here, the increase (decrease) of $k(t)$ corresponds to a compression (expansion), and the adiabatic process means that the average of exchanged heat $\dbar Q(t)$ is zero during this process, not that $\dbar Q(t)$ is always zero at every moment~\cite{martinez2016adiabatic}.
In the two isothermal processes, the environment temperature $T$ is kept constant at $T_c$ or $T_h$ ($> T_c$).
The conjugated force for $k(t)$ is given by $F_k(t) \equiv \partial U/\partial k = x(t)^2/2$, resulting in an increment of work applied to the particle, denoted by $\dbar W(t) = F_k(t) dk(t)$. 
The heat transferred from the thermal bath to the particle is then given by $\dbar Q(t) = dE(t) - \dbar W(t)$, where the total energy is defined as $E=kx^2/2 + v^2/2$.
For LBN to learn the time-varying drift and diffusion fields, $\mathcal{D}_{\rm tr}$ consists of multiple trajectories ($M=50$) of a cycle with $\tau_{cyc}=\tau=100$, and we assign an input vector to $[x, \hat{v}, t_i]^{\rm T}$ to encapsulate the temporal information.
The left panel in Fig.~\ref{fig6}(a) illustrates a mean Clapeyron diagram obtained from trajectories in $\mathcal{D}_{\rm tr}$, where pronounced fluctuations hinder the identification of a discernible pattern.

Despite the highly fluctuating trajectories of the Brownian Carnot engine [Fig.~\ref{fig6}(a)], we verify that LBN accurately learns the time-varying drift and diffusion fields with $r^2 = 0.952$ and $r^2=0.906$, respectively (\ref{sec:ApendixD6}).
Applying the trained LBN, we generate more trajectories ($M=2000$) and recover a more discernible mean Clapeyron diagram [right panel in Fig.~\ref{fig6}(a)] than the diagram obtained from the training trajectories.
This shows the potential applicability of our approach to clarify spatiotemporal patterns of trajectories that cannot be identified due to their fluctuations by generating multiple trajectories.
Fig.~\ref{fig6}(b) presents the true and inferred protocols of $k(t)$ and $D(t)$ with respect to $t$, showing that LBN not only exactly learns the time-varying fields but also can be used to infer the model parameters, as similarly demonstrated in Fig.~\ref{fig5}(d).
Here, $k(t)$ is estimated by linear regression between $\hat{\Phi}(x, t)$ and $x$ at a fixed $v$ with varying $t$.

It is noteworthy that the curves depicting the uncertainties in the drift and diffusion inferences exhibit a distinct trend change in the vicinity of transition points between thermodynamic processes [Fig.~\ref{fig6}(c)]: the slope of $\hat{\Sigma}_{\Phi}$ transforms from negative to positive around $t=\tau/2$, while the slope of $\hat{\Sigma}_{\mathsf{D}}$ becomes flatter near $t=\tau/4$ and changes its sign near $t=3\tau/4$.
These trend change points almost correspond to the magnitude change points of the drift and diffusion fields with respect to $t$ due to time-varying protocols.
That is, the larger the true field, the smaller the corresponding uncertainty. This relationship is natural from an information-theoretic perspective when viewing actual fields as transmitted signals and others as environmental influences.
In this way, we confirm that uncertainties effectively capture the magnitude changes in time-varying protocols, serving as a valuable proxy for errors in such processes.

Let us now examine whether the trained LBN actually works as a true Brownian Carnot engine.
To compare with the true one, thermodynamic quantities such as heat, work, and energy are measured from multiple trajectories ($M=2000$) generated by the true system and trained LBN, respectively (Appendices~\ref{sec:ApendixD1} and \ref{sec:ApendixD6}).
As can be seen in Fig.~\ref{fig6}(d), the PDFs of engine efficiencies $\eta_{\tau}$ of the true and trained models are almost identical, including their tails (inset).
Here, the efficiency $\eta_{\tau}$ is calculated by $-W^{(n_{cyc})}_{\tau}/Q^{(n_{cyc})}_{h, \tau}$, where $W^{(n_{cyc})}_{\tau}$ and $Q^{(n_{cyc})}_{h, \tau}$ are the sum of works and absorbed heats (during the hot isothermal process) over $n_{cyc}$ cycles, respectively, and $\eta_C \equiv 1- T_c/T_h$ is the Carnot efficiency.
Note that the tails of the two distributions follow a power law with exponent $-1.92$, the same as the result in Ref.~\cite{martinez2016brownian}.
This result directly demonstrates the applicability of LBN to imitate real complex stochastic engines with only a small amount of data.

Furthermore, we examine whether our LBN trained at $\tau_{cyc}=\tau=100$ effectively emulates the true engine even when confronted with varying $\tau_{cyc}$.
The relationship between power and efficiency in heat engines typically presents a trade-off because of the imposition of a quasistatic limit ($\tau_{cyc} \rightarrow \infty$) to achieve Carnot efficiency, a condition that coincides with diminishing power.
Finding the optimal $\tau_{cyc}$ for a specific engine is a formidable challenge, with the aim to maximize power while closely approaching Carnot efficiency~\cite{blickle2012realization, martinez2016brownian, pietzonka2018universal, krishnamurthy2023overcoming}. 
Therefore, if LBN trained solely at a specific $\tau_{cyc}$ can consistently yield comparable results with the true engine beyond the training region, the potential exists to substantially reduce the experimental costs required for parameter optimization and the analysis of statistical properties of thermodynamic quantities.

As apparent in Fig.~\ref{fig6}(e) and~\ref{fig6}(f), LBN remarkably reproduces the behavior of the true engine, accurately capturing powers and efficiencies across various $\tau_{cyc}$.
The powers of the true model and LBN, denoted as $\mathcal{P}(\tau_{cyc})\equiv -\langle W(\tau_{cyc}) \rangle /\tau_{cyc}$, exhibit inverted U-shaped curves with increasing $\tau_{cyc}$, suggesting that they reach maximum values at specific $\tau_{cyc} = \tau^*_{cyc}$.
To determine $\tau^*_{cyc}$, we fit the estimated $\mathcal{P}(\tau_{cyc})$ vs. $\tau_{cyc}$ to the known law $\mathcal{P}(\tau_{cyc})=-(\langle W_{\infty} \rangle + \Sigma_{ss}/\tau)/\tau$, where $\langle W(\tau_{cyc}) \rangle \rightarrow \langle W_{\infty} \rangle$ in the limit $\tau_{cyc} \rightarrow \infty$ and $\Sigma_{ss}$ is the dissipation per cycle~\cite{sekimoto2010stochastic, esposito2010efficiency, martinez2016brownian}.
Using this well-fitted line with parameters $\langle W({\infty}) \rangle \simeq -3.40$ and $\Sigma_{ss} \simeq 30.75$, we confirm that the maximum of $\mathcal{P}(\tau_{cyc})$ is obtained at $\tau^*_{cyc} \simeq 18.08$.
Regarding efficiency $\eta_{\tau_{cyc}}$, both efficiencies calculated from the true model and LBN monotonically decrease with decreasing $\tau_{cyc}$, even taking a negative average value at small $\tau_{cyc}$.
We then fit our LBN data to the known law given by $\eta_{\tau_{cyc}}=(\eta_C + \tau_W/\tau)/(1+\tau_Q/\tau)$, yielding $\tau_{W} \simeq -3.78$ and $\tau_Q \simeq -1.67$.
Substituting $\tau^*_{cyc}$ into the fitting line of $\eta_{\tau_{cyc}}$, the efficiency at maximum power is determined to be $\eta_{\tau^*_{cyc}} \simeq 0.24$, closely matching the Curzon--Ahlbon efficiency $\eta_{CA}=1-\sqrt{T_c/T_h} \simeq 0.24$ --- the expected efficiency of a Carnot engine at maximum power~\cite{curzon1975efficiency}.
Moreover, we can observe that LBN accurately replicates the distribution of observables beyond the training region [insets of Fig.~\ref{fig6}(e) and contour plots of Fig.~\ref{fig6}(f)].
These results suggest possibilities to analyze the statistics of observables such as the changes in the exponents of tail distributions or the different fitting curves of distributions from Gaussian to non-Gaussian with respect to $\tau_{cyc}$.

\section{Summary and discussions}

Decoding the dynamics governing a system's motion from data stands as an enduring puzzle in the realm of physics, and in particular,
deciphering the equation of motion in stochastic systems, with their erratic trajectories influenced by environmental fluctuations, presents a particularly formidable challenge.
In this paper, we introduce Langevin Bayesian networks, or LBN, a method designed to uncover the underlying Langevin equation governing observed stochastic systems.
By inputting observed trajectories into LBN in the absence of detailed system knowledge, our method adeptly discerns the unbiased drift field and diffusion matrix, based on rigorous derivations.
It then reconstructs the Langevin equation by combining these components for both overdamped and underdamped regimes, allowing us to emulate the underlying systems beyond the training region.
The incorporation of Bayesian neural networks further enhances our methodology as it grants the capability to estimate uncertainties associated with predictions, thereby establishing a lower bound of prediction errors.
To showcase the versatility of our method, we presented demonstrations across various scenarios, including challenging models such as the Hodgkin--Huxley neuron model and the Brownian Carnot engine.

Although we highlighted several advantages of our method in the results section, let us recap them here with some additional insights. 
First, unlike many previous approaches that hinge on specific binning methods or predetermined functional forms of fields, LBN breaks free from such dependencies as well as from the stationarity of the system.
This adaptability, derived from the application of neural network architecture, empowers our approach to navigate a myriad of unknown stochastic systems, especially when their underlying dynamics are intricately complex.
To address any concerns about the impact of network architecture on LBN's performance, we confirm in \ref{sec:ApendixB3} that its performance remains consistently robust despite changes in architecture.
Additionally, we compared LBN and a popular OLE inference method, called stochastic force inference, and found that LBN outperforms SFI across system dimensions in several examples.
Second, LBN offers quantitative uncertainty estimates for predictions, which it achieves through Bayesian inference and stochastic neural networks.
The capability to estimate uncertainty enables the inference of prediction errors with a lower bound.
By doing so, LBN mitigates the black box problem often associated with neural network-based approaches, and thus can provide insights into their reliability.
These uncertainties can also act as a proxy for prediction errors, serving as a cautionary signal for unreliable predictions and helping prevent misinterpretation of the system.
This becomes especially useful when applying the inferred Langevin equation beyond the training range, such as in unobserved or poorly sampled regions, where direct validation of predictions is difficult. In such cases, elevated uncertainty can help flag potential extrapolation errors.
However, it is essential to acknowledge that the lower bound of errors provided by the uncertainties may be poor when certain regions of the input space are sparsely populated with data or when the test data significantly differs from the training data, hence, it is crucial to be mindful of these potential limitations.
Third, the accessibility of output gradients effectively addresses issues in gradient calculation.
Many methods, especially binning-based methods, are inevitably weak in obtaining the gradient of the diffusion field.
In contrast, through the application of backpropagation, LBN can access output gradients, which allows to obtain deterministic forces even in inhomogeneous diffusion systems (see Fig.~\ref{fig3}).
Lastly, it is noteworthy that our neural network-based method offers computational efficiency, especially for large datasets.
Traditional kernel-based methods require expensive training computations, scaling as $\mathcal{O}(|\mathcal{D}|^3)$ due to kernel matrix inversions, with dataset size $|\mathcal{D}|$. Similarly, regression-based methods with predetermined basis functions suffer from combinationally growing numbers of fitting parameters with increasing system dimensionality, resulting in substantial training complexity. In contrast, neural networks avoid these scalability issues because their training complexity scales linearly with dataset size ($\mathcal{O}(|\mathcal{D}|)$ per epoch) without expensive kernel inversions or combinatorial growth in parameters~\cite{unke2021machine}.
This computational advantage is particularly valuable when dealing with large-scale simulations or experiments, where the computational burden can be significantly reduced.

Let us explore additional dimensions where our method can demonstrate its versatility.
Examining the Hodgkin--Huxley model (Fig.~\ref{fig4}), we highlight LBN's ability to replicate the critical behavior of the HH model across various $I_{\rm ext}$, even though the corresponding dynamics were not encountered during the training procedure.
Particularly, the critical behaviors of $\langle \dot{S} \rangle$ and other thermodynamic quantities in nonequilibrium systems have been investigated in previous works~\cite{ge2009thermodynamic, tome2012entropy, zhang2016critical, seara2021irreversibility}, and it has been suggested that $\langle \dot{S} \rangle$ can be used as a warning signal for predicting nonequilibrium phase transitions~\cite{yan2023thermodynamic}.
In this context, our method serves to statistically analyze thermodynamic and dynamical behaviors. 
This enables us to anticipate whether and when critical transitions occur beyond the training range while monitoring the uncertainties in their predictions.
An additional noteworthy facet is demonstrated in the Brownian Carnot engine (refer to Fig.~\ref{fig6}), where LBN showed an additional ability in terms of trajectory data augmentation. 
By generating trajectories via the trained Langevin equation, such augmentation can elucidate dynamical averaged patterns and thermodynamic statistics, including the tails of efficiency distributions.

We anticipate the broad use of LBN in modeling the complex behaviors of living organisms, climate dynamics, financial data trends, and the design of efficient nanodevices.
While this paper covers compelling aspects of our methodology, intriguing avenues remain for future exploration.
We assumed that the stochastic noise is a Gaussian white noise according to the central limit theorem; however, this theorem is often not fulfilled in reality, resulting in many biological systems exhibiting a power-law behavior of mean square displacement, known as anomalous diffusion.
Several mathematical models have been proposed to describe these systems~\cite{yonkeu2020can, baral2024stochastically, goswami2024anomalous}, and efforts have been made to classify which models best describe the given data~\cite{metzler2014anomalous, munozgil2021objective, seckler2022bayesian}.
Therefore, integrating these efforts on anomalous diffusion with our approach would be an interesting extension of our method.
Similarly, delving into the dynamics of non-Markovian systems through neural network-based methods poses another fascinating challenge.
Another practical consideration is the issue of variable sampling rates due to missing data;
this is commonly resolved by performing linear interpolation between two far points in time, which may potentially lead to incorrect predictions.
Extending LBN to address this issue by incorporating time-step information into the input state could offer a valuable improvement.
As in this case, by assigning different data labels or providing more information to the input state, various extensions of our method can be readily developed.
Note that one can also extend our method with other Bayesian neural network architectures or different implementations such as MC-dropout~\cite{gal2016dropout}, SWAG~\cite{maddox2019simple}, etc., which may further improve performance and obtain more reliable uncertainty estimates (plese refer to \ref{sec:ApendixB1}).

\section*{Code availability}
The Python codes generating the dataset and for the LBN are available at \url{https://github.com/qodudrud/LBN}.

\section*{Acknowledgments}
This work was supported by the Global-LAMP Program of the National Research Foundation of Korea (NRF) grant funded by the Ministry of Education (No. RS-2023-00301976, Y.B.) and the Basic Science Research Program through the National Research Foundation of Korea (NRF Grant No. 2022R1A2B5B02001752, S.H.; NRF Grant No. RS-2025-00514776, H.J.)

\newpage
\appendix

\section{Background on Bayesian Neural Networks}
\label{sec:ApendixA}

A Bayesian neural network (BNN) is a type of stochastic neural network that learns through Bayesian inference~\cite{mackay1992practical, hinton1993keeping, barber1998ensemble, jospin2022hands-on, arbel2023primer}.
While conventional artificial neural networks (ANNs) have driven a revolution in machine learning, they tend to be overconfident in their predictions and fail to capture model uncertainty; in other words, they cannot say ``I don't know''.
This incompetence can lead to serious problems, such as leading users to make erroneous decisions based on unreliable model predictions.
Several methods have been proposed to relieve this issue, with BNNs being a major breakthrough in many fields including physics~\cite{yang2019adversarial, lin2021detection, mancarella2022seeking, seckler2022bayesian}, autonomous robotics~\cite{michelmore2020uncertainty}, medical diagnosis~\cite{reijnen2020Preoperative}, and financial forecasting~\cite{chandra2021bayesian}.
This section provides prior background on the BNNs and algorithms used in our framework.

ANNs are powerful universal continuous function approximators, trained using a backpropagation algorithm~\cite{goodfellow2016deep}.
Feedforward networks are the quintessential example of ANNs, which consist of multiple layers of functions $\{\bm{l}^{(0)}, \dots, \bm{l}^{(n)}\}$:  
\begin{equation}
\begin{aligned}
    &\bm{l}^{(0)} = \bm{z}, \\
    &\bm{l}^{(i)} = \bm{s}^{(i)} \left(\bm{W}^{(i)} \bm{l}^{(i-1)} + \bm{b}^{(i)} \right) \;\;\; \forall i \in [1, n], \\
    &\bm{y}_{\bm{\theta}} = \bm{l}^{(n)},
\end{aligned}
\end{equation}
where $\bm{z}$ denotes the input data, $\bm{W}^{(i)}$ and $\bm{b}^{(i)}$ are the weight and bias of the $i$-th layer $\bm{l}^{(i)}$, $\bm{\theta} \equiv [\bm{\theta}^{(1)}, \bm{\theta}^{(2)}, \dots, \bm{\theta}^{(n)}]^{\rm T}$ with $\bm{\theta}^{(i)} \equiv [\bm{W}^{(i)}, \bm{b}^{(i)}]$ denotes the trainable parameters of the neural network, and $\bm{s}^{(i)}$ is an activation function that introduces nonlinearity to the network.
Feedforward networks can be summarized as $\bm{y}_{\bm{\theta}}(\bm{z}) \equiv \bm{l}^{(n)}(\bm{l}^{(n-1)} ( \ldots \bm{l}^{(2)} (\bm{l}^{(1)} (\bm{z})) \ldots )$, which maps $\bm{z}$ to corresponding output $\bm{y}_{\bm{\theta}}(\bm{z})$ with respect to trainable parameters $\bm{\theta}$.
$\bm{\theta}$ are trained by minimizing the objective function chosen by the users according to their purpose.
Since each parameter has a single value in conventional ANNs, it always returns a fixed prediction for a given input.
When $\bm{\theta}$ is not a single value but instead is randomly sampled from a particular distribution, the network is called a \textit{stochastic neural network}, which can provide an ensemble of predictions by multiple sampled $\bm{\theta}$.
BNNs are basically built using stochastic neural networks, where the distribution of their parameters $\bm{\theta}$ is updated by applying Bayes' theorem, i.e., $P(\bm{\theta}|\mathcal{D}) = P(\mathcal{D}|\bm{\theta})P(\bm{\theta})/P(\mathcal{D})$. Here, $P(\bm{\theta}|\mathcal{D})$ is the posterior, $P(\bm{\theta})$ is the prior, $P(\mathcal{D}|\bm{\theta})$ is the likelihood, and $P(\mathcal{D})$ is the marginal likelihood (or evidence).
Comparing the predictions by multiple sampled $\bm{\theta} \sim P(\bm{\theta}|\mathcal{D})$ allows us to estimate uncertainty; the smaller the variance of the predictions, the more reliable the trained model is.

While Bayesian inference methods have broad applicability, their implementation is not straightforward due to the intractability of the posterior distribution.
Calculating the marginal likelihood can be exceedingly difficult, and even if achieved, sampling from the posterior becomes a formidable task due to the high dimensionality of the sampling space.
To tackle these issues, the most widely employed methods include variational inference and Markov chain Monte Carlo methods~\cite{jospin2022hands-on, arbel2023primer}.
In our framework, we use a variational inference method and the Bayes-by-backprop algorithm because they are computationally cheap and easily adapted to many deep learning methods. 
The subsequent subsections introduce variational inference and the associated backpropagation method.

\begin{figure}[!t]
    \includegraphics[width=\linewidth]{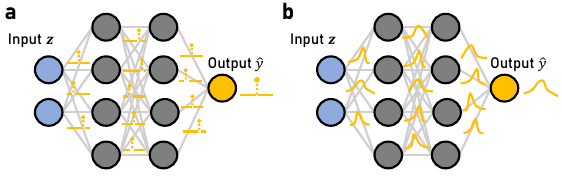}
    \vskip -0.1in
    \caption{
    Schematic of (a) point estimate neural networks and (b) stochastic neural networks. Ordinary neural networks are built using point estimate neural networks with single-valued weights, whereas Bayesian neural networks are built using stochastic neural networks with probability distributions over the weights.
    }\label{figS1}
\end{figure}

\subsection{Variational inference}
\label{sec:ApendixA1}

Variational inference stands as an efficient method for approximating probability distributions~\cite{hinton1993keeping, blei2017variational}.
It is particularly useful in Bayesian models to approximate the true posterior distribution $P(\bm{\theta} | \mathcal{D})$ due to its intractability.
The core concept of variational inference involves approximating $P(\bm{\theta} | \mathcal{D})$ with a more tractable distribution $Q_{\bm{\phi}}(\bm{\theta})$, called the variational distribution, parameterized by $\bm{\phi}$.
The closeness between $P(\bm{\theta} | \mathcal{D})$ and $Q_{\bm{\phi}}(\bm{\theta})$ is assessed through the Kullback--Leibler divergence (KL divergence), defined by
\begin{equation}
\begin{aligned}
    D_{\rm KL}\left[Q_{\bm{\phi}} (\bm{\theta}) || P(\bm{\theta} | \mathcal{D})\right] = \int Q_{\bm{\phi}}(\bm{\theta}) \ln\left[ \frac{Q_{\bm{\phi}}(\bm{\theta})}{P(\bm{\theta} | \mathcal{D})} \right] d\bm{\theta}.
\end{aligned}
\end{equation}
Unfortunately, computing the KL divergence requires knowledge of $P(\bm{\theta} | \mathcal{D})$, which is infeasible.
To circumvent this, we expand the KL divergence as:
\begin{equation}
\begin{aligned}
    D_{\rm KL}\left[Q_{\bm{\phi}} || P(\cdot | \mathcal{D})\right] &= \ln P(\mathcal{D}) + \mathcal{F}\left[ Q_{\bm{\phi}} \right],
\end{aligned}
\end{equation}
where $\mathcal{F}\left[ Q_{\bm{\phi}} \right]$, known as the variational free energy or negative evidence lower bound (ELBO), is defined by 
\begin{equation}
\begin{aligned}
    \mathcal{F}\left[ Q_{\bm{\phi}} \right] &\equiv \mathcal{L}^E(\bm{\phi}, \mathcal{D}) + \mathcal{L}^C(\bm{\phi}),
\end{aligned}
\label{app_eq:free_energy0}
\end{equation}
with
\begin{equation}
\begin{aligned}
    \mathcal{L}^E(\bm{\phi}, \mathcal{D}) &\equiv -\int Q_{\bm{\phi}} (\bm{\theta}) \ln P(\mathcal{D} | \bm{\theta}) d\bm{\theta},\\
    \mathcal{L}^C(\bm{\phi}) &\equiv D_{\rm KL}\left[Q_{\bm{\phi}} (\bm{\theta}) || P(\bm{\theta}) \right].
\end{aligned}
\end{equation}
As $P(\mathcal{D})$ is constant, minimizing $D_{\rm KL}\left[Q_{\bm{\phi}} || P(\cdot | \mathcal{D})\right]$ is equivalent to minimizing $\mathcal{F}\left[ Q_{\bm{\phi}} \right]$, which becomes a function independent of computing $P(\bm{\theta}|\mathcal{D})$.
Note that $\mathcal{F}\left[ Q_{\bm{\phi}} \right]$ can also be conceptualized from an information-theoretic perspective as a cost of minimum description length~\cite{hinton1993keeping}.

As shown in Eq.~\eqref{app_eq:free_energy0}, $\mathcal{F}\left[ Q_{\bm{\phi}} \right]$ consists of two cost functions: the error loss $\mathcal{L}^{E}$ and the complexity loss $\mathcal{L}^{C}$.
The error loss $\mathcal{L}^{E}$ represents the average negative log-likelihood under $Q_{\bm{\phi}}(\bm{\theta})$, diminishing as prediction accuracy improves.
Conversely, the complexity loss $\mathcal{L}^{C}$ quantifies the dissimilarity between $Q_{\bm{\phi}}(\bm{\theta})$ and $P(\bm{\theta})$, operating as a regularizer for $\bm{\phi}$ by enforcing constraints on $Q_{\bm{\phi}}(\bm{\theta})$ to be close to $P(\bm{\theta})$.
Thus, minimizing $\mathcal{F}[Q_{\bm{\phi}}]$ can be interpreted as finding minima for $\mathcal{L}^E$ with the constraint of reducing $\mathcal{L}^{C}$.
In practice, however, $\mathcal{L}^{C}$ can exhibit disproportionate magnitudes for large models or small datasets in practice~\cite{liu2018advbnn}, potentially overshadowing our primary objective of minimizing $\mathcal{L}^E$. To address this, we introduce a small prefactor $\kappa$ to moderate the impact of $\mathcal{L}^{C}$, yielding the modified variational free energy:
\begin{equation}
\begin{aligned}
    \mathcal{F}\left[ Q_{\bm{\phi}} \right] \equiv \mathcal{L}^{E} + \kappa \mathcal{L}^{C},
\end{aligned}
\label{app_eq:free_energy}
\end{equation}
with $0 \leq \kappa \leq 1$.
This approach, known as employing \textit{cold posteriors}~\cite{wenzel2020how}, effectively reduces the regualrization strength associated with the prior $P(\bm{\theta})$, thereby enhancing predictive performance.
Therefore, the training of our models on $\mathcal{D}_{\rm tr}$ involves minimizing the loss function $\mathcal{F}\left[ Q_{\bm{\phi}} \right]$ in Eq.~\eqref{app_eq:free_energy} with respect to $\bm{\phi}$ and finding minima for $\mathcal{L}^E$ on the validation dataset.

Let us derive calculable forms of $\mathcal{L}^E$ and $\mathcal{L}^C$ of our method.
For simplicity, we skip the layer index and denote each entry in $\bm{\theta}$ as $\bm{\theta}_i$.
Following the mean-field variational inference technique~\cite{blei2017variational}, we assume that entries in $\bm{\theta}$ are mutually independent and each $\bm{\theta}_i$ is governed by a distinct factor $\bm{\phi}_i$.
If we set the prior $P(\bm{\theta})$ as the isotropic multivariate Gaussian $\mathcal{N}(\bm{0}, \sigma_{\rm pri}\bm{I})$, and if variational distributions for $\bm{\theta}_i$, denoted by $Q_{\bm{\phi}_i}(\bm{\theta}_i)$, are also Gaussian, then $Q_{\bm{\phi}}$ is expressed as:
\begin{equation}
\begin{aligned}
    Q_{\bm{\phi}}(\bm{\theta}) = \prod_{\bm{\theta}_i \in \bm{\theta}} Q_{\bm{\phi}_i}(\bm{\theta}_i) \text{ where } Q_{\bm{\phi}_i} = \mathcal{N}(\mu_i, \sigma_i^2).
\end{aligned}
\label{app_eq:Q_eq}
\end{equation}
In this case, $\bm{\phi}_i$ should contain the parameters for $({\mu}_i, {\sigma}_i^2)$ (the exact form of $\bm{\phi}$ is given in the next subsection).
Note that if we assume $Q_{\bm{\phi}_i}(\bm{\theta}_i)$ as a delta distribution that assigns probability $1$ to a particular parameter and $0$ to all others, the networks become ANNs [see Fig.~\ref{figS1}(a)].
Using the fact that both $P(\bm{\theta})$ and $Q_{\bm{\phi}_i}(\bm{\theta}_i)$ are Gaussian and that the KL divergence can be computed between two Gaussian distributions, $\mathcal{L}^c$ can be obtained by
\begin{equation}
\begin{aligned}
    \mathcal{L}^C(\bm{\phi}) = \sum_{\bm{\theta}_i \in \bm{\theta}} \left( \ln{\frac{\sigma_{\rm pri}}{\sigma_i}} + \frac{\mu_i^2 + \sigma_i^2 - \sigma_{\rm pri}^2}{2\sigma_{\rm pri}^2} \right).
\end{aligned}
\label{app_eq:L^C}
\end{equation}
Thus, $\mathcal{L}^C(\bm{\phi})$ can be analytically calculated during the training.
To derive the form of $\mathcal{L}^E$, let the likelihood $P(\mathcal{D}|\bm{\theta})$ be Gaussian with unit variance, as is common.
Then the negative log-likelihood is expressed as
\begin{equation}
\begin{aligned}
    \mathcal{L}^E(\bm{\phi}, \mathcal{D}) &= \frac{1}{2}\left\langle \left\langle \left( \tilde{\bm{y}}- \bm{y}_{\bm{\theta}}\left(\bm{z}\right) \right)^2 \right\rangle_{\bm{\theta}} \right\rangle_{\mathcal{D}} + C,
\end{aligned}
\label{app_eq:L^E0}
\end{equation}
where $C$ is a constant resulting from the normalization factor.
Here, $\langle \cdot \rangle_{\mathcal{D}}$ denotes the average over $\mathcal{D}$ with $(\bm{z}, \tilde{\bm{y}}) \in \mathcal{D}$, where $\bm{z}$ is an input and $\tilde{\bm{y}}$ is the corresponding target in $\mathcal{D}$, $\langle \cdot \rangle_{\bm{\theta}} \equiv \int Q_{\bm{\phi}} (\bm{\theta}) d\bm{\theta}$ is the marginalization over $\bm{\theta}$, and $\bm{y}_{\bm{\theta}}(\bm{z})$ is the BNN output for a given input $\bm{z}$ and a sampled $\bm{\theta}$ from $Q_{\bm{\phi}}(\bm{\theta})$.
Since we cannot analytically compute $\mathcal{L}^E$, we apply Monte Carlo integration,
\begin{equation}
\begin{aligned}
    \mathcal{L}^E (\bm{\phi}, \mathcal{D}) \simeq \frac{1}{N_{\rm mc}} \sum_{k=1}^{N_{\rm mc}} \frac{1}{2} \left\langle \left( \tilde{\bm{y}}- \bm{y}_{\bm{\theta}^k}\left(\bm{z}\right) \right)^2 \right\rangle_{\mathcal{D}} + C,
\end{aligned}
\label{app_eq:L^E}
\end{equation}
where $\bm{\theta}^k$ is independently sampled from $Q_{\bm{\phi}}(\bm{\theta})$ at each computation. 

While the mean-field approximation and the Gaussian assumption on both the prior and variational distribution are commonly adopted for their computational simplicity and efficiency, they are known to have limitations.
In particular, the independence assumption in Eq.~\eqref{app_eq:Q_eq} can lead to underestimated posterior variance and limited expressiveness of the variational distribution.
However, recent work has shown that this limitation can be alleviated: mean-field approximations with at least two Bayesian hidden layers can, in principle, recover the true posterior~\cite{farquhar2020liberty}, suggesting sufficient expressiveness in deep architectures despite their simplicity.
Furthermore, although we mitigate the sensitivity to the prior by employing cold posteriors, an improper prior can still hinder predictive performance, and the optimal choice of $\kappa$ remains an open question~\cite{wenzel2020how, chaudhuri2022how}.
We acknowledge these inherent trade-offs--- advantages and limitations that are actively debated~\cite{arbel2023primer}---and view the search for optimal model structures and prior distributions for learning stochastic dynamics as an important direction for future work.

\subsection{Bayes-by-backprop (BBB) algorithm}

Here we introduce how we update the parameters of neural networks via backpropagation, a method called the Bayes-by-backprop (BBB) algorithm~\cite{blundell2015weight}.
The stochasticity of $\bm{\theta}$ opens the advantages of BNNs but also the disadvantage of hard-to-apply deep learning due to the difficulty of performing backpropagation.
The BBB algorithm resolves this problem using a re-parameterization trick:
let $\bm{\epsilon}$ be a random variable following a probability distribution $q(\bm{\epsilon})$ and let $\bm{\theta}=t(\bm{\epsilon}, \bm{\phi}) \sim q_{\bm{\phi}}(\bm{\theta})$, where $t(\bm{\epsilon}, \bm{\phi})$ is a deterministic transformation.
Using the fact that $q_{\bm{\phi}}(\bm{\theta})d\bm{\theta}=q(\epsilon)d\epsilon$, we can show that
\begin{equation}
\begin{aligned}
    &\partial_{\bm{\phi}} \int_{\bm{\phi}} q_{\bm{\phi}} (\bm{\theta}') f(\bm{\theta}', \bm{\phi}) d\bm{\theta}' \\
    &= \int_{\bm{\epsilon} }q(\bm{\epsilon}) [ \partial_{\bm{\theta}} f(\bm{\theta}, \bm{\phi}) \partial_{\bm{\phi}} \bm{\theta} + \partial_{\bm{\phi}}f(\bm{\theta}, \bm{\phi}) ] d\bm{\epsilon},
\end{aligned}
\label{app_eq:bbb}
\end{equation}
i.e., the derivative of an expectation is equal to the expectation of a derivative.
Let us set $q_{\bm{\phi}}(\bm{\theta}) = Q_{\bm{\phi}}(\bm{\theta})$ and $f(\bm{\theta}, \bm{\phi}) = - \ln P(\mathcal{D}|\bm{\theta}) + \kappa \ln [Q_{\bm{\phi}}(\bm{\theta})/P(\bm{\theta})]$.
Then $\mathcal{F}\left[ Q_{\bm{\phi}} \right]$ can be rewritten as
\begin{equation}
\begin{aligned}
    \mathcal{F}\left[ Q_{\bm{\phi}} \right] = \int Q_{\bm{\phi}}(\bm{\theta}) f(\bm{\theta}, \bm{\phi}) d\bm{\theta} \propto \sum_k f(\bm{\theta}^k, \bm{\phi}),
\end{aligned}
\end{equation}
where $\bm{\theta}^k$ is the $k$-th parameter sampled from $Q_{\bm{\phi}}(\bm{\theta})$.
Applying Eq.~\eqref{app_eq:bbb}, therefore, we can substitute $\bm{\nabla}_{\bm{\phi}} \mathcal{F}[Q_{\bm{\phi}}]$ into $\sum_k \bm{\nabla}_{\bm{\phi}} f(\bm{\theta}^k, \bm{\phi})$ with $\bm{\theta} \sim Q_{\bm{\phi}}(\bm{\theta})$ for performing backpropagation.
If we suppose $Q_{\bm{\phi}}(\bm{\theta})$ is a Gaussian distribution $\mathcal{N}(\bm{\mu}, \bm{\sigma}^2)$ and trainable parameter $\bm{\phi} \equiv (\bm{\mu}, \bm{\rho})$ with $\bm{\sigma}=\ln[1+\exp(\bm{\rho})]$, the update procedure of $\bm{\phi}$ is as follows:
\begin{algorithm}[H]
\begin{algorithmic}[1]
\REQUIRE{Trainable variational parameter $\bm{\phi} \equiv (\bm{\mu}, \bm{\rho})$, optimizer, training dataset $\mathcal{D}_{\rm tr}=\{\mathcal{D}_{{\rm tr}, i}\}_{i=1}^{\mathcal{B}}$.}
\LOOP
    \STATE Draw $\bm{\epsilon} \sim \mathcal{N}(0, \bm{\mathsf{I}})$.
    \State $\bm{\theta} = \bm{\mu} + \ln[1 + \exp(\bm{\rho})] \odot \bm{\epsilon}$ where $\odot$ is pointwise multiplication.
    \State $f[Q_{\bm{\phi}}]= - \ln P(\mathcal{D}_{{\rm tr}, i}|\bm{\theta}) + \kappa \ln [Q_{\bm{\phi}}(\bm{\theta})/P(\bm{\theta})]/\mathcal{B}$.
    \State Compute the gradient $\bm{\nabla}_{\bm{\phi}}f = {\rm backprop}_{\bm{\phi}}(f)$
    \State Update $\bm{\phi}$ with the optimizer.
\ENDLOOP
\end{algorithmic}
\caption{Update rule of the variational parameter}
\label{app_alg:BBB_alg}
\end{algorithm}
\noindent In Algorithm~\ref{app_alg:BBB_alg}, we divide the second term in $f[Q_{\bm{\phi}}]$ by the number of batches $\mathcal{B}$ due to our use of the minibatch optimization method~\cite{graves2011practical}.

\section{Training details of LBN}
\label{sec:ApendixB}

We provide clarity on the notations before delving into the following sections.
Our main goal is to reconstruct the Langevin equation from observed trajectories, and to this end, we construct a dataset $\mathcal{D}$ consisting of $M$ independent stochastic trajectories, each with a length $L$.
This can be expressed by $\mathcal{D} \equiv \left\{ \left\{ \left(\bm{z}^{j}(t_i), \tilde{\bm{y}}^j(t_i)\right) \right\}_{i=1}^{L} \right\}_{j=1}^{M}$.
The elements $\left(\bm{z}^{j}(t_i), \tilde{\bm{y}}^j(t_i)\right)$ within $\mathcal{D}$ consist of pairs, where ${\bm{z}}^j(t_i)$ represents the state vector of the system, which serves as an input, and $\tilde{\bm{y}}^j(t_i)$ represents a target vector (label).
The sampling times are denoted as $t_i \equiv i\Delta t$, and the duration of a trajectory is $\tau \equiv L\Delta t$.
Note that the target vector $\tilde{\bm{y}}^j(t_i)$ is not the true value we aim to obtain, such as a drift vector or diffusion matrix.
The training and validation datasets used during the training are denoted by $\mathcal{D}_{\rm tr}$ and $\mathcal{D}_{\rm val}$, respectively, and the test dataset used to assess the performance on unseen data is denoted by $D_{\rm te}$.
For simplicity, we will omit the superscript $j$ indicating the trajectory index in the state and target vectors.

\subsection{Error metrics}
\label{sec:ApendixB1}

We employ two error metrics to evaluate the performance of LBN: pointwise error $e^2_{\bm{y}}(\bm{z}, t)$ and total error $\mathcal{E}_{\bm{y}}^2$.
The pointwise error $e^2_{\bm{y}}(\bm{z}, t)$ at a given state $\bm{z}(t)$ assesses local performance at individual points in trajectories and is defined by:
\begin{equation}
\begin{aligned}
    e^2_{\bm{y}}(\bm{z}, t) \equiv \frac{ \left\langle \left[ \bm{y} (\bm{z}, t) - \bm{y}_{\bm{\theta}}(\bm{z}, t) \right]^2 \right\rangle_{\bm{\theta}}}{ \left\langle \bm{y}_{\bm{\theta}}^2 (\bm{z}, t) \right\rangle_{\bm{\theta}}},
\end{aligned}
\end{equation}
where $\bm{y}(\bm{z}, t)$ is the true drift field or diffusion matrix for a given state vector $\bm{z}(t)$.
As discussed in the main text, $e^2_{\bm{y}}(\bm{z}, t)$ has a lower bound on the relative uncertainty of predictions, denoted by $\hat{\Sigma}_{\bm{y}}$, and becomes highly correlated with it after the training. 
The derivation is shown in \ref{sec:ApendixC2}.
The total error $\mathcal{E}_{\bm{y}}^2$ assesses global performance on the dataset $\mathcal{D}$ and, similar to $\mathcal{L}^E$, is defined as follows:
\begin{equation}
\begin{aligned}
    \mathcal{E}_{\bm{y}}^2 \equiv \frac{ \left\langle \left\langle \left[ \bm{y} (\bm{z}, t) - \bm{y}_{\bm{\theta}} (\bm{z}, t) \right]^2 \right\rangle_{\bm{\theta}} \right\rangle_{\mathcal{D}}}{\left\langle \left\langle {\bm{y}}_{\bm{\theta}}^2 (\bm{z}, t) \right\rangle_{\bm{\theta}}\right\rangle_{\mathcal{D}}}.
\end{aligned}
\label{app_eq:def_E2}
\end{equation}
Thus, $\mathcal{E}_{\bm{y}}^2$ can be obtained by taking the average of both the denominator and numerator of $e^2_{\bm{y}}(\bm{z}, t)$ over $\mathcal{D}$.
This metric is also employed for comparison with another inference method in the main text, known as stochastic force inference (SFI)~\cite{frishman2020learning}.
The SFI method, well-established for OLE inference, relies on linear regression into a set of predetermined basis functions.
We utilize the publicly available implementation of this method from the authors' GitHub repository.
To compare with LBN, we choose the best results of SFI (i.e., the lowest errors) by varying the order of polynomial basis from $0$ to $5$ for all plots in the main text.
As the SFI method provides deterministic predictions [i.e., $\bm{\theta} \sim \delta(\bm{\theta}_0 - \bm{\theta})$] for given inputs, $\mathcal{E}_{\bm{y}}^2$ is computed based on a single series of drift predictions along a trajectory in $\mathcal{D}$.

\begin{figure*}[!ht]
    \includegraphics[width=\linewidth]{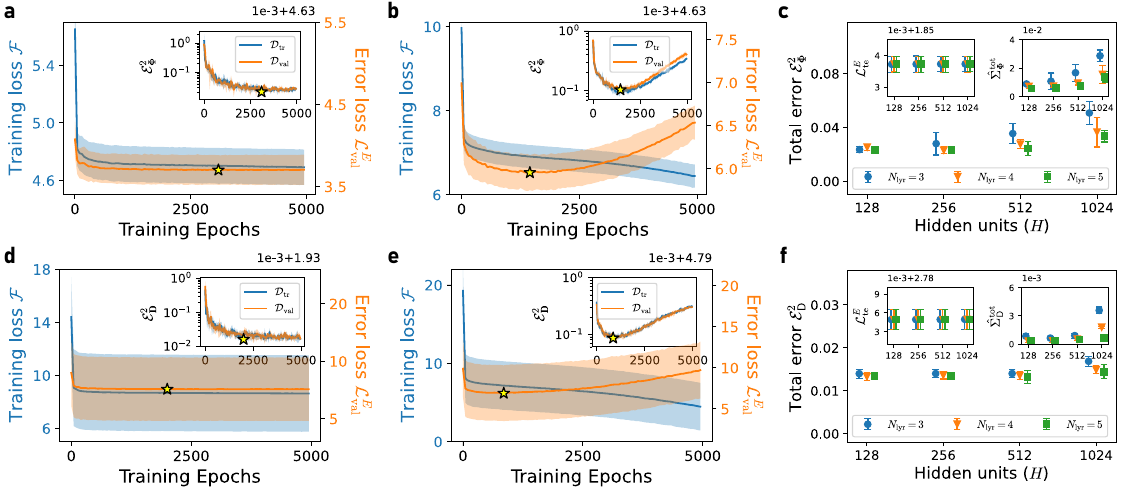}
    \vskip -0.1in
    \caption{
    Loss curves throughout training epochs and architecture robustness for drift estimators in the nonlinear force model (a--c) and diffusion estimators in the inhomogeneous diffusion force model (d--f).
    In figures of loss curves (a), (b), (d), (e), the training loss and error loss are denoted by $\mathcal{F}$ on $\mathcal{D}_{\rm tr}$ (blue line, left axis) and $\mathcal{L}^E_{\rm val} \equiv \mathcal{L}^E(\bm{\phi}, \mathcal{D}_{\rm val})$ (orange line, right axis), respectively, with yellow stars marking the minima of $\mathcal{L}^E_{\rm val}$.
    Insets: total errors of corresponding estimators over $\mathcal{D}_{\rm tr}$ and $\mathcal{D}_{\rm val}$ throughout training epochs with yellow stars marking the values at the epoch that minimizes $\mathcal{L}^E_{\rm val}$.
    (c) and (f) depict the total error on $\mathcal{D}_{\rm te}$ of corresponding estimators with varying LBN configurations. Here, $N_{\rm lyr}$ represents the number of layers composing neural networks and $H$ represents the number of hidden units (see Table~\ref{table:LBN_architecture}). Insets: error loss on $\mathcal{D}_{\rm te}$ and total relative uncertainty, denoted by $\mathcal{L}^E_{\rm te} \equiv \mathcal{L}^E(\bm{\phi}, \mathcal{D}_{\rm te})$ and $\hat{\Sigma}^{\rm tot}_{\bm{y}}$, respectively, with varying configurations.
    }\label{figS2}
\end{figure*}

\subsection{Training algorithm}
\label{sec:ApendixB2}

The training procedure of LBN on training dataset $D_{\rm tr}$ is as follows:
\begin{algorithm}[H]
\begin{algorithmic}[1]
\REQUIRE{LBN, optimizer, training and validation datasets $\mathcal{D}_{\rm tr}$ and $\mathcal{D}_{\rm val}$.}
\STATE Initialize $\mathcal{L}^E_{\rm best} \gets \infty$ and $\bm{\phi}^{\star} \gets \bm{\phi}_0$.
\LOOP
    \STATE Compute $\mathcal{F}$ over $\mathcal{D}_{\rm tr}$ by \begin{align}\label{apdx_eq:J_train}
            \mathcal{F} = \mathcal{L}^E(\bm{\phi}, \mathcal{D}_{\rm tr})+ \kappa \mathcal{L}^{C}(\bm{\phi}).
    \end{align}
    Here, $\mathcal{L}^C(\bm{\phi})$ and $\mathcal{L}^E(\bm{\phi}, \mathcal{D}_{\rm tr})$ are computed by Eqs.~\eqref{app_eq:L^C} and \eqref{app_eq:L^E}, respectively.
    \STATE Compute gradients $\nabla_{\bm{\phi}} \mathcal{F}$.
    \STATE Update $\bm{\phi}$ via the optimizer with Algorithm~\ref{app_alg:BBB_alg}.
    \STATE Compute the error loss on $\mathcal{D}_{\rm val}$, $\mathcal{L}^E(\bm{\phi}, \mathcal{D}_{\rm val})$.
    \IF{$\mathcal{L}^E(\bm{\phi}, \mathcal{D}_{\rm val}) < \mathcal{L}^E_{\rm best}$}
        \STATE $\mathcal{L}^E_{\rm best} \gets \mathcal{L}^E(\bm{\phi}, \mathcal{D}_{\rm val})$ and $\bm{\phi}^{\star} \gets \bm{\phi}$
    \ENDIF
\ENDLOOP
\end{algorithmic}
\caption{Training procedure of LBN}
\label{app_alg:training}
\end{algorithm}
\noindent 
As explained in \ref{sec:ApendixA1}, LBN is trained by minimizing $\mathcal{F}$ on $\mathcal{D}_{\rm tr}$.
The validation dataset $\mathcal{D}_{\rm val}$ is utilized to prevent the selection of an overfitted model and ensure the choice of the best model with minimal error by monitoring $\mathcal{L}^E(\bm{\phi}, \mathcal{D}_{\rm val})$.
During training, we identify and save the model parameter $\bm{\phi}^*$ corresponding to the minimum $\mathcal{L}^E(\bm{\phi}, \mathcal{D}_{\rm val})$ and designate it for the final trained model (steps 6--9 in Algorithm~\ref{app_alg:training}). 
This strategy minimally impacts the performance of our method in the absence of overfitting, as demonstrated in Fig.~\ref{figS2}(a) and~\ref{figS2}(d) with their insets.
However, in the presence of overfitting, the strategy proves effective in capturing the best model and minimizing errors, as evidenced in Fig.~\ref{figS2}(b) and~\ref{figS2}(e) with their insets.

\subsection{LBN architecture and hyperparameters}
\label{sec:ApendixB3}

It is important to emphasize that the diffusion matrix must be positive semi-definite and symmetric by definition~\cite{gardiner2004handbook}.
To ensure that the inferred diffusion matrix satisfies these conditions, we employ Cholesky decomposition to decompose $\bm{\mathsf{D}}_{\bm{\theta}}$ into $\bm{\mathsf{L}}_{\bm{\theta}} \bm{\mathsf{L}}^{\rm T}_{\bm{\theta}}$, where $\bm{\mathsf{L}}_{\bm{\theta}}$ is a real lower triangular matrix with positive diagonal entries:
\begin{equation}
\begin{aligned}
    {{\mathsf{L}}}_{\bm{\theta}, \mu \nu} = 
    \begin{cases}
        l_{\bm{\theta}, \mu \nu} & \text{if $\mu < \nu$}, \\
        \ln\left[1 + \exp(l_{\bm{\theta}, \mu \mu})\right] & \text{if $\mu = \nu$}, \\
        0 & \text{otherwise},
    \end{cases}
\end{aligned}
\label{eq:choleskyL}
\end{equation}
with network outputs $l_{\bm{\theta}, \mu \nu}$. 
Note that the diagonal entries of $\bm{{\mathsf{L}}}_{\bm{\theta}}$ are parameterized as in Eq.~\eqref{eq:choleskyL} to always be positive.
Thus, by inferring $\bm{\mathsf{L}}_{\bm{\theta}}$ and reconstructing $\bm{\mathsf{D}}_{\bm{\theta}}$ accordingly, we infer the diffusion matrix satisfying the above conditions while reducing the parameter search space.

\begin{figure}[t]
    \includegraphics[width=\linewidth]{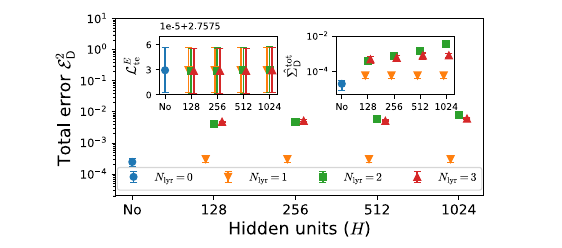}
    \vskip -0.1in
    \caption{
    Total error on $\mathcal{D}_{\rm te}$ of diffusion estimators with varying configurations. Here, $N_{\rm lyr}$ represents the number of layers composing the neural networks and $H$ represents the number of hidden units (see Table~\ref{table:LBN_architecture}). Insets: error loss on $\mathcal{D}_{\rm te}$ and total relative uncertainty, denoted by $\mathcal{L}^E_{\rm te} \equiv \mathcal{L}^E(\bm{\phi}, \mathcal{D}_{\rm te})$ and $\hat{\Sigma}^{\rm tot}_{\bm{y}}$, respectively, with varying configurations.
    }\label{figS3}
\end{figure}

Table~\ref{table:LBN_architecture} outlines the network architecture of LBN.
To speed up training and reduce the number of parameters, the last two network layers are configured as stochastic layers (denoted as `BayesFC layer'), while the others are configured as fully connected layers like ordinary neural networks (denoted as `FC layer').
The input dimension $d_{\rm in}$ is varied depending on the input state vector $\bm{z}$ of the target systems; for instance, $d_{\rm in}=d$ for the OLE systems in the main text, $d_{\rm in}=2d$ for the stochastic van der Pol oscillator, and $d_{\rm in}=2d+1$ ($= 3$ since $d=1$) for the Brownian Carnot engine.
In the case of the diffusion estimator, `BayesFC layer $2$' produces $d (d+1)/2$-dimensional outputs, which correspond to the entries of $\bm{\mathsf{L}}_{\bm{\theta}}$.
Performing $\bm{\mathsf{L}}_{\bm{\theta}} \bm{\mathsf{L}}^{\rm T}_{\bm{\theta}}$ in the `Cholesky layer', we can obtain the diffusion matrix $\bm{\mathsf{D}}_{\bm{\theta}}$.
In the `Output layer', we correct bias induced by techniques such as data normalization (refer to \ref{sec:ApendixC3}).
Note that when inferring the homogeneous diffusion matrix, the input and function approximators become unnecessary, rendering the `Input' layer and network layers entirely redundant.
Instead, we establish an inputless structure by eliminating these redundant layers and replacing $l_{\bm{\theta}, \mu\nu}$ with a parameter sampled from a Gaussian distribution $\mathcal{N}(\mu_i, \sigma_i^2)$, akin to $\bm{\theta}$ in the network layers.
To evaluate the performance of various diffusion estimator configurations, including the inputless structure denoted by $N_{\rm lyr}=0$, we examine total error $\mathcal{E}^2_{\bm{\mathsf{D}}}$, error loss $\mathcal{L}^E(\bm{\phi}, \mathcal{D}_{\rm te})$, and total relative uncertainty $\hat{\Sigma}_{\bm{\mathsf{D}}}^{\rm tot}$  (Fig.~\ref{figS3}).
Here, $\hat{\Sigma}^{\rm tot}_{\bm{y}}$ is defined by $\hat{\Sigma}^{\rm tot}_{\bm{y}} \equiv \left\langle {\rm Var}[\bm{y}_{\bm{\theta}}(\bm{z}, t)]\right\rangle_{\mathcal{D}}/\left\langle \left\langle {\bm{y}}_{\bm{\theta}}^2 (\bm{z}, t) \right\rangle_{\bm{\theta}}\right\rangle_{\mathcal{D}}$, similarly to the definition of $\mathcal{E}^2_{\bm{y}}$ in Eq.~\eqref{app_eq:def_E2}.
Our verification reveals that, while  $\mathcal{E}^2_{\bm{\mathsf{D}}}$ are consistently low across all configurations, the inputless structure exhibits the lowest total errors.
The suitability of the architecture is less evident in $\mathcal{L}^E(\bm{\phi}, \mathcal{D}_{\rm te})$; however, the magnitudes of $\hat{\Sigma}_{\bm{\mathsf{D}}}^{\rm tot}$ align with those of $\mathcal{E}^2_{\bm{\mathsf{D}}}$, suggesting that $\hat{\Sigma}_{\bm{\mathsf{D}}}^{\rm tot}$ can be employed to select the most appropriate architecture.
Thus, in practical scenarios where knowledge or an assumption that the diffusion matrix of the system is homogeneous, the inputless structure is a more suitable configuration for the diffusion estimator.
In cases where this knowledge or assumption is absent, training both configurations (normal and inputless structure) and subsequently selecting the configuration with lower error loss and uncertainty is a prudent approach.

To evaluate the architecture robustness, we examine the total errors $\mathcal{E}^2_{\bm{y}}$, error losses $\mathcal{L}^E(\bm{\phi}, \mathcal{D}_{\rm te})$, and total relative uncertainties $\hat{\Sigma}^{\rm tot}_{\bm{y}}$ for the test dataset $\mathcal{D}_{\rm te}$ with varying model configurations, maintaining the number of stochastic layers at $2$ while adjusting the number of fully connected layers.
As shown in Fig.~\ref{figS2}(c) and~\ref{figS2}(f), neither $\mathcal{E}^2_{\bm{y}}$ nor $\mathcal{L}^E(\bm{\phi}, \mathcal{D}_{\rm te})$ are significantly varied and consistently maintain low values across different model configurations.
Notably, in Fig.~\ref{figS3}, we confirm that the magnitudes of $\hat{\Sigma}_{\bm{y}}^{\rm tot}$ align with those of $\mathcal{E}^2_{\bm{y}}$.
This alignment supports the high correlation between them and underscores the utility of $\hat{\Sigma}_{\bm{y}}^{\rm tot}$ for selecting a more appropriate architecture.
Based on these observations, we opt for $N_{\rm lyr}=4$ and $H=256$ for all examples, except for the homogeneous diffusion estimator.
Note that our approach could potentially be extended with other Bayesian deep learning techniques to further enhance performance, such as MC-dropout~\cite{gal2016dropout}, SWAG~\cite{maddox2019simple}, etc.
Exploring these avenues presents an intriguing prospect for future work.

The hyperparameters used in our experiments are listed in Table~\ref{table:LBN_hyperparameters}.
We used ELU as the activation function and the Adam optimizer~\cite{kingma2014adam} to minimize $\mathcal{F}[Q_{\bm{\phi}}]$.
Learning rates were commonly set to $10^{-4}$ but were adjusted in a few examples to achieve improved performance.
We did not strive to optimally tune the hyperparameters, so more appropriate ones for each system may be found.
All runs were conducted on a single NVIDIA TITAN V GPU.

\begin{table}[!t]
\caption{Architecture of drift (top) and diffusion (bottom) estimators with $N_{\rm lyr}=4$: layer name, input dimension, output dimension, and activation function. $d$ and $d_{\rm in}$ are the system and input dimension, respectively, $H$ is the number of hidden units, `FC layer' indicates a fully connected layer, and `BayesFC layer' indicates a fully connected stochastic layer, as explained in \ref{sec:ApendixA}.}
\centering
\begin{tabular}{c|c|c|c}
\hline
\multicolumn{4}{c}{
\centering
    \textbf{Drift estimator}} \\
\hline
Layer name & Input dim. & Output dim. & Act. func.\\
\hline
    Input $\bm{z}$ & $d_{\rm in}$ & $d_{\rm in}$ &  None \\
    FC layer $1$ & $d_{\rm in}$ & $H$ &  ELU \\
    FC layer $2$ & $H$ & $H$ &  ELU \\
    BayesFC layer $1$ & $H$ & $H$ &  ELU \\
    BayesFC layer $2$ & $H$ & $d$ & None \\
    Output layer & $d$ & $d$ &  None\\
\hline
\multicolumn{4}{c}{\vspace{-0.1 in}}\\
\hline
\multicolumn{4}{c}{
\centering
    \textbf{Diffusion estimator}} \\
\hline
Layer name & Input dim. & Output dim. & Act. func.\\
\hline
    Input $\bm{z}$ & $d_{\rm in}$ & $d_{\rm in}$ &  None \\
    FC layer $1$ & $d_{\rm in}$ & $H$ &  ELU \\
    FC layer $2$ & $H$ & $H$ &  ELU \\
    BayesFC layer $1$ & $H$ & $H$ &  ELU \\
    BayesFC layer $2$ & $H$ & $d(d+1)/2$ & None \\
    Cholesky layer & $d(d+1)/2$ & $d(d+1)/2$ &  None\\
    Output layer & $d(d+1)/2$ & $d(d+1)/2$ &  None\\
\hline
\end{tabular}
\label{table:LBN_architecture}
\end{table}

\begin{table}[!t]
\caption{Hyperparameters of LBN}
\centering
\begin{tabular}{c|c|c}
\hline
   Name & Value & Notes \\
\hline
    $\mathcal{B}$ & $32$ & Number of equally sized batches \\
    $\sigma_{\rm pri}$ & $0.1$ & Stdv. of the prior distribution \\
    $\rho_{0}$ & $-5$ & Initial $\rho$ of the posterior distribution \\
    $\kappa$ & $0.2$ & Weight of the complexity cost \\
    $N_{\bm{\theta}}$ & $200$ & Cardinality of $\{ \bm{\theta} \}$ sampled from $P\left( \bm{\theta} | \mathcal{D}_{tr}\right)$ \\
    $N_{\rm mc}$ & $3$ & Number of sampled $\bm{\theta}$ to calculate $\mathcal{L}^{E}$ \\
\hline
\end{tabular}
\label{table:LBN_hyperparameters}
\end{table}

\section{Analytic calculations}
\label{sec:ApendixC}

In this section, we carry out the analytic calculations used in the main text. 
Here, we consider a general stochastic system whose coordinates $\bm{z}(t)$ evolve in time according to the Langevin equation as written by
\begin{equation}
\begin{aligned}
    \dot{\bm{z}}(t) = \bm{\Phi}(\bm{z}, t) + \sqrt{2\bm{\mathsf{D}}(\bm{z}, t)}\bm{\xi}(t),
\end{aligned}
\label{app_eq:GeneralLangevinEq}
\end{equation}
in the Itô convention where $\bm{\xi}(t)$ is Gaussian white noise with $\langle \xi_\mu (t) \rangle = 0$ and $\langle \xi_\mu (t) \xi_\nu (t') \rangle = \delta(t-t')\delta_{\mu\nu}$.
$\bm{\Phi}(\bm{z}, t)$ and $\bm{\mathsf{D}}(\bm{z}, t)$ are the drift vector and diffusion matrix, respectively, defined by
\begin{equation}
\begin{aligned}
    \bm{\Phi}(\bm{z}, t) &= \lim_{\Delta t \rightarrow 0} \frac{1}{\Delta t} \langle \bm{z}(t+\Delta t)-\bm{z}(t) \rangle|_{\bm{z}(t)}, \\
    \bm{\mathsf{D}}(\bm{z}, t) &= \lim_{\Delta t \rightarrow 0} \frac{1}{2\Delta t} \left\langle \left( \bm{z}(t+\Delta t)-\bm{z}(t) \right)^2 \right\rangle \Big|_{\bm{z}(t)}.
\end{aligned}
\end{equation}
These two functions uniquely define the stochastic process and are commonly referred to as Kramers--Moyal coefficients of the Fokker--Planck equation~\cite{risken1996fokker}.

\subsection{Derivation of estimators}
\label{sec:ApendixC1}

Here, we consider a $d=1$ case for simplicity.
In this case, It\^{o}'s formula in integral form, which is mainly used to derive the estimators, is given by
\begin{equation}
\begin{aligned}
    g(z, s) &= g(z, t_0) + \int_{t_0}^{s} dW(s') \sqrt{2D(z, s')}\partial_z g(z, s')\\
    & + \int_{t_0}^{s} ds' \big[\Phi(z, s') \partial_z g(z, s') + 
 D(z, s')\partial^2_z g(z, s') \big] 
\end{aligned}
\end{equation}
for an arbitrary function $g(z, t)$ and $s>t_0$.

Throughout this section, we use several It\^{o} integrals~\cite{kloeden1992numerical, gardiner2004handbook, lehle2015analyzing}:
\begin{equation}
\begin{aligned}
    \mathcal{I}^{(n)}_{w} &\equiv \int_{t}^{t+n\Delta t} dW(s) \sim \mathcal{O}(\Delta t^{1/2}), \\
    \mathcal{I}^{(n)}_{ww} &\equiv \int_{t}^{t+n\Delta t} dW(s) \int_{t}^{s} dW(s') \sim \mathcal{O}(\Delta t), \\
    \mathcal{I}^{(n)}_{0} &\equiv \int_{t}^{t+n\Delta t} ds = n\Delta t, \\
    \mathcal{I}^{(n)}_{0w} &\equiv \int_{t}^{t+n\Delta t} ds \int_{t}^{s} dW(s') \sim \mathcal{O}(\Delta t^{3/2}), \\
    \mathcal{I}^{(n)}_{0ww} &\equiv \int_{t}^{t+n\Delta t} ds \int_{t}^{s} dW(s') \int_{t}^{s'} dW(s'') \sim \mathcal{O}(\Delta t^{2}), \\
    \mathcal{I}^{(n)}_{00} &\equiv \int_{t}^{t+n\Delta t} ds \int_{t}^{s} ds' = \frac{1}{2} (n\Delta t)^2, \qquad \text{etc.,}\\
\end{aligned}
\end{equation}
with $dW(s) = \xi(s)ds$ and $\mathcal{I}^{(1)}_w \equiv \Delta W(t)$, where $W(t)$ is the Wiener process.
Additionally, we use the average values of the It\^{o} integrals and their multiple products: 
the average of an It\^{o} integral is zero if it contains any integration with respect to a Wiener process, e.g., the averages of the It\^{o} integrals above are all zero except $\mathcal{I}^{(n)}_{0}$ and $\mathcal{I}^{(n)}_{00}$, and
\begin{equation}
\begin{aligned}
    \left\langle \mathcal{I}^{(n)}_{0w} \mathcal{I}^{(m)}_{0w} \right\rangle &= \left(\frac{1}{2} n^2 m - \frac{1}{6}n^3 \right) \Delta t^3,
\end{aligned}
\label{app_eq:Ito_multi_integral}
\end{equation}
with $m \geq n$. In addition, $\left\langle \prod_k \mathcal{I}_{\bm{b}_k}^{(n_k)} \right\rangle = 0$ if the total number of nonzero entries in the subindex $\bm{b}_k$ is odd, e.g., $\left\langle \mathcal{I}^{(n)}_{0w} \mathcal{I}^{(m)}_{0ww} \right\rangle=0$.

\subsubsection{OLE estimators}

Now, let us start to derive the OLE estimators.
The OLE can be represented as the integral equation shown below:
\begin{equation}
\begin{aligned}
    \Delta x(t)|_{x(t)} &= \int_{t}^{t + \Delta t} ds \;\Phi(x, s) \\&+ \int_{t}^{t + \Delta t} dW(s)\sqrt{2{D}(x, s)},
\end{aligned}
\label{app_eq:OLE_integral}
\end{equation}
where $\Delta x(t) \equiv x(t+\Delta t) - x(t)$.
By applying It\^{o}'s formula into the first integrand in Eq.~\eqref{app_eq:OLE_integral}, the first integral can be written as 
\begin{equation}
\begin{aligned}
    \int_{t}^{t + \Delta t} ds\; \Phi(x, s)  = \Phi(x, t) \Delta t + R_1(t),
\end{aligned}
\end{equation}
where the remainder $R_1(t) \sim \mathcal{O}(\Delta t^{3/2})$.
Considering the fact that $dW(t)$ is $\mathcal{O}(\Delta t^{1/2})$, the second integrand can also be given by
\begin{equation}
\begin{aligned}
    &\int_{t}^{t + \Delta t} dW(s) \sqrt{2{D}(x, s)} \\& = \sqrt{2{D}(x, t)}\Delta W(t)
    + \partial_{x} {D}(x, t) \mathcal{I}^{(1)}_{ww} + R_2(t),
\end{aligned}
\end{equation}
where the remainder $R_2(t) \sim \mathcal{O}(\Delta t^{3/2})$.
Thus, Eq.~\eqref{app_eq:OLE_integral} can be rewritten to
\begin{equation}
\begin{aligned}
    \Delta x(t)|_{x(t)} &= \Phi(x, t) \Delta t + \sqrt{2D(x, t)}\Delta W(t)
    \\& + \partial_{x} {D}(x, t) \mathcal{I}^{(1)}_{ww} + R(t),
\end{aligned}
\label{app_eq:OLE_integral2}
\end{equation}
where the remainder $R(t) \equiv R_1(t) + R_2(t)$ satisfies $R(t) \sim \mathcal{O}(\Delta t^{3/2})$ and $\langle R(t) \rangle \sim \mathcal{O}(\Delta t^2)$~\cite{gardiner2004handbook}.
The elimination of the $\mathcal{O}(\Delta t^{3/2})$ term in $R(t)$ when taking the ensemble average holds true in multiple dimensions.
Taking the ensemble average of $\Delta x(t)$ and $\Delta x(t)^2$ given $x(t)$, we can obtain the drift field and diffusion matrix as follows:
\begin{equation}
\begin{aligned}
    \Phi(x, t) &= \frac{\langle \Delta x (t) \rangle|_{x(t)}}{\Delta t} + \mathcal{O}(\Delta t), \\
    D(x, t) &= \frac{\langle \Delta x(t)^2 \rangle|_{x(t)}}{2\Delta t} \\ &- \frac{1}{2} \left( \Phi(x, t)^2
    + \frac{1}{2} \partial_x D(x, t) \right) \Delta t + \mathcal{O}(\Delta t^2),
\end{aligned}
\label{app_eq:OLE_estimator}
\end{equation}
with $\left\langle \mathcal{I}_{ww}^{(1)} \mathcal{I}_{ww}^{(1)} \right\rangle = \left\langle \mathcal{I}_{00}^{(1)} \right\rangle = (1/2)\Delta t^2$.
Neglecting $\mathcal{O}(\Delta t)$, we utilize the corresponding first term on the RHS in Eq.~\eqref{app_eq:OLE_estimator} as the drift and diffusion estimators, respectively.
However, in cases where we have exact knowledge of the drift field or when the drift field is sufficiently large to warrant consideration, inclusion of the second term order of $\mathcal{O}(\Delta t)$ in the diffusion estimator may lead to improved accuracy.

\subsubsection{ULE estimators}

Let us derive the ULE estimators in a manner similar to the OLE estimators.
We follow similar notations and procedures as in Refs.~\cite{lehle2015analyzing, bruckner2020inferring}.
If we have direct access to both $x(t)$ and $v(t)$, we can naturally expect the ULE estimators to take the same form as Eq.~\eqref{app_eq:OLE_estimator}, with the replacement of $\Delta x(t)$ by $\Delta v(t)$. However, in practical scenarios, we can only observe $x(t)$, necessitating the estimation of $v(t)$ from the observed trajectories of $x(t)$.
Considering these more realistic scenarios, we first expand $\Delta x^{(n)}(t) \equiv x(t+n\Delta t) - x(t)$ by applying the ULE as follows:
\begin{equation}
\begin{aligned}
    \Delta x^{(n)}(t)|_{x(t), v(t)} &= \int_{t}^{t+n\Delta t} ds \, v(s) \\
            &= v(t) n\Delta t + \int_{t}^{t+n\Delta t} ds [v(s) - v(t) ] \\
            &= v(t)n\Delta t + \int_{t}^{t+n\Delta t} ds \biggl[ \int_{t}^{s} ds' \Phi(s') \\ 
            &\qquad\qquad\qquad + \int_{t}^{s} dW(s') \sqrt{2D(s')} \biggr].
\end{aligned}
\label{app_eq:ULE_dx}
\end{equation}
Here, for simplicity, we use $\Phi(s') \equiv \Phi(x, v, s')$ and $D(s') \equiv D(x, v, s')$.
Applying It\^{o}'s formula in Eq.~\eqref{app_eq:ULE_dx}, we obtain
\begin{equation}
\begin{aligned}
    \int_{t}^{t+n\Delta t} ds  \int_{t}^{s} ds' \Phi(s') &= \Phi(t) \mathcal{I}^{(n)}_{00} + R_1(t),
\end{aligned}
\label{app_eq:ULE_Ito1}
\end{equation}
and
\begin{equation}
\begin{aligned}
&\int_{t}^{t+n\Delta t} ds \int_{t}^{s} dW(s') \sqrt{2D(s')} \\
&= \sqrt{2D(t)} \mathcal{I}^{(n)}_{0w} + \partial_v D(t) \mathcal{I}^{(n)}_{0ww} + R_2(t),
\end{aligned}
\label{app_eq:ULE_Ito2}
\end{equation}
where the remainders satisfy $R_1(t), R_2(t) \sim \mathcal{O}(\Delta t^{5/2})$ and $ \langle R_1(t) \rangle, \langle R_2(t) \rangle \sim \mathcal{O}(\Delta t^{3})$.
Substituting Eqs.~\eqref{app_eq:ULE_Ito1} and \eqref{app_eq:ULE_Ito2} into Eq.~\eqref{app_eq:ULE_dx}, $\Delta x^{(n)}(t)$ is given by
\begin{equation}
\begin{aligned}
    \Delta x^{(n)}(t)|_{x(t), v(t)} &= v(t) n\Delta t + \sqrt{2D(t)} \mathcal{I}^{(n)}_{0w} \\
    &+ \Phi(t)\mathcal{I}^{(n)}_{00} + \partial_v D(t)\mathcal{I}^{(n)}_{0ww} + R_3(t),
\end{aligned}
\label{app_eq:ULE_dx^n}
\end{equation}
where $R_3(t) \equiv R_1(t) + R_2(t)$.
When we define the velocity estimator $\hat{v}(t)$ as $\Delta x^{(1)}(t) / \Delta t$, the difference between $\hat{v}(t)$ and $v(t)$ can be obtained from Eq.~\eqref{app_eq:ULE_dx^n}:
\begin{equation}
\begin{aligned}
    \hat{v}(t) - v(t) = \frac{1}{\Delta t} \sqrt{2D(t)} \mathcal{I}^{(1)}_{0w} + \mathcal{O}(\Delta t),
\end{aligned}
\label{app_eq:ULE_vv}
\end{equation}
and the velocity increment estimator $\Delta \hat{v}(t)$ is given by
\begin{equation}
\begin{aligned}
    \Delta \hat{v}(t)|_{x(t), v(t)} &= \frac{ x(t+2\Delta t) - 2 x(t + \Delta t) + x(t)}{\Delta t} \\
    &= \frac{ \Delta x^{(2)}(t) - 2 \Delta x^{(1)}(t)}{\Delta t} \\
    &= \frac{1}{\Delta t} \biggl[ \sqrt{2D(t)} \left(\mathcal{I}^{(2)}_{0w} - 2\mathcal{I}^{(1)}_{0w} \right) + \Phi(t) \Delta t^2 \\
    &+ \partial_v D(t) \left( \mathcal{I}^{(2)}_{0ww} - 2\mathcal{I}^{(1)}_{0ww} \right) \biggr] + R_4 (t),
\end{aligned}
\label{app_eq:ULE_dv}
\end{equation}
where $R_4(t) \sim \mathcal{O}(\Delta t^{3/2})$ and $\langle R_4(t) \rangle = \mathcal{O}(\Delta t^2)$, and we use $\mathcal{I}^{(2)}_{00} - 2\mathcal{I}^{(1)}_{00} = \Delta t^2$.

One more step is needed to describe the observed dynamics of $\hat{v}(t)$. As can be seen in Eq.~\eqref{app_eq:ULE_dv}, the increment of $\hat{v}(t)$ is conditioned on the given $x(t)$ and $v(t)$; we have to replace $v(t)$ with $\hat{v}(t)$ by multiplying $P(x, v, t)/P(x, \hat{v}, t)$, where $P(x, v, t)$ denotes the PDF of $x$ and $v$ at $t$.
By expanding $P(x, v, t)/P(x, \hat{v}, t)$, we can obtain
\begin{equation}
\begin{aligned}
    \frac{P(x, v, t)}{P(x, \hat{v}, t)} = 1-\left(\hat{v}(t)-v(t) \right)\partial_{\hat{v}} \ln P(x, \hat{v}, t) + \mathcal{O}(\Delta t).
\end{aligned}
\label{app_eq:Pv}
\end{equation}
Thus, multiplying $\Delta \hat{v}(t)|_{x(t), v(t)}$ by Eq.~\eqref{app_eq:Pv}, we can obtain the observed dynamics of $\hat{v}(t)$ as follows:

\begin{equation}
\begin{aligned}
    \Delta \hat{v}&(t)|_{x(t), \hat{v}(t)} = \frac{1}{\Delta t} \biggl[ \sqrt{2D(t)} \left(\mathcal{I}^{(2)}_{0w} - 2\mathcal{I}^{(1)}_{0w} \right) \\& + \Phi(t) \Delta t^2
     -2D(t) \partial_{\hat{v}} \ln P(x, \hat{v}, t) \frac{1}{\Delta t} \mathcal{I}^{(1)}_{0w}\left(\mathcal{I}^{(2)}_{0w} - 2\mathcal{I}^{(1)}_{0w} \right) \\ &+ \partial_v D(t) \left( \mathcal{I}^{(2)}_{0ww} - 2\mathcal{I}^{(1)}_{0ww} \right) \biggr] + R_4 (t).
\end{aligned}
\label{app_eq:ULE_dv2}
\end{equation}

Note that the leading term of the fluctuation in $\Delta \hat{v}(t)$ is $\sqrt{2D(t)}\left(\mathcal{I}^{(2)}_{0w} - 2\mathcal{I}^{(1)}_{0w} \right)/\Delta t$, which is of order $\mathcal{O}(\Delta t^{1/2})$ (zero average), similar to the fluctuations of $\Delta x(t)$ and $\Delta v(t)$. However, the magnitude of this fluctuation is smaller due to the fact that $\left\langle \left( \mathcal{I}^{(2)}_{0w} - 2\mathcal{I}^{(1)}_{0w} \right)^2 \right\rangle/(\Delta t^2) = 2\Delta t/3$, in contrast to $\left\langle \Delta W(t)^2 \right\rangle = \Delta t$ [refer to Eq.~\eqref{app_eq:Ito_multi_integral}]. This difference prompts a modification in the diffusion estimator for ULE inference.
Another notable point is that the noise term of $\Delta \hat{v}(t+\Delta t)$, denoted by $\left(\mathcal{I}^{(2)}_{0w}(t+\Delta t) - 2\mathcal{I}^{(1)}_{0w}(t+\Delta t) \right)/\Delta t$, is not independent of the noise term of $\Delta \hat{v}(t+\Delta t)$, denoted by $\left(\mathcal{I}^{(2)}_{0w}(t) - 2\mathcal{I}^{(1)}_{0w}(t) \right)/\Delta t$. 
If we consider the noises at $t$ and $t+\Delta t$, the autocorrelation between them is calculated by
\begin{equation}
\begin{aligned}
    &\left\langle \frac{\left(\mathcal{I}^{(2)}_{0w}(t) - 2\mathcal{I}^{(1)}_{0w}(t) \right)}{\Delta t} \frac{\left(\mathcal{I}^{(2)}_{0w}(t+\Delta t) - 2\mathcal{I}^{(1)}_{0w}(t+\Delta t) \right)}{\Delta t} \right\rangle \\ &= \frac{\left\langle \mathcal{I}^{(2)}_{0w}(t) \mathcal{I}^{(2)}_{0w}(t+\Delta t) \right\rangle}{\Delta t^2} - 2 \frac{\left\langle \mathcal{I}^{(2)}_{0w}(t) \mathcal{I}^{(1)}_{0w}(t+\Delta t) \right\rangle}{\Delta t^2} \\
    &= \frac{1}{6}\Delta t,
\end{aligned}
\end{equation}
for a nonzero $\Delta t$ [refer to Eq.~\eqref{app_eq:Ito_multi_integral}]. 
This nonzero autocorrelation indicates that the dynamics of $\hat{v}(t)$ is not Markovian and fluctuates with colored noise rather than white noise.
The discrepancy between observed and underlying dynamics can lead to a misguided understanding of the system; therefore, we must handle the observed dynamics with caution.

Let us go back to the derivation of ULE estimators.
By taking the ensemble average of $\Delta \hat{v}(t)$ and $\Delta \hat{v}(t)^2$ at given $x(t)$ and $\hat{v}(t)$, we can obtain
\begin{equation}
\begin{aligned}
    \frac{\langle \Delta \hat{v}(t) \rangle|_{x(t), \hat{v}(t)}}{\Delta t} &= \Phi(t)-\frac{1}{3}D(t)\partial_{\hat{v}} \ln P(x, \hat{v}, t) + \mathcal{O}(\Delta t),  \\
    \frac{\langle \Delta \hat{v}(t)^2 \rangle|_{x(t), \hat{v}(t)}}{2 \Delta t} &= \frac{2}{3} D(x, v, t) + \mathcal{O}(\Delta t),
\end{aligned}
\label{app_eq:ULE_estimator}
\end{equation}
where we use the properties $\left\langle \mathcal{I}^{(n)}_{0w} \right\rangle = \left\langle \mathcal{I}^{(n)}_{0ww} \right\rangle = 0$ and $\left\langle \left( \mathcal{I}^{(2)}_{0w} - 2\mathcal{I}^{(1)}_{0w} \right)^2 \right\rangle = 2\Delta t^3/3$.
Unlike Eq.~\eqref{app_eq:OLE_integral2}, we cannot retrieve the drift field and diffusion matrix solely through the ensemble average.
These biased results have been similarly reported in Refs.~\cite{lehle2015analyzing, pedersen2016how, ferretti2020building, bruckner2020inferring}.
Consequently, to retrieve the drift field, we need to calculate an additional term, $(1/3)\partial_{\hat v} \ln P(x, \hat{v}, t)$.

To obtain $(1/3)\partial_{\hat v} \ln P(x, \hat{v}, t)$, we calculate the ensemble average of $\Delta \hat{v}(t-\Delta t)$ at given $x(t)$ and $\hat{v}(t)$:
\begin{equation}
\begin{aligned}
    &\left\langle \Delta \hat{v}(t-\Delta t) \right\rangle|_{x(t), \hat{v}(t)} \\ &= \left\langle \Delta \hat{v}(t-\Delta t) \frac{P\left(x, \hat{v}, t\right)}{P\left(x, \hat{v}, t-\Delta t \right)} \right\rangle \biggr|_{x(t-\Delta t), \hat{v}(t-\Delta t)} \\
    &\simeq  \langle \Delta \hat{v}(t-\Delta t)  \rangle|_{x(t-\Delta t), \hat{v}(t-\Delta t)} \\
    & - \langle \partial_{\hat{v}} \ln P(x, \hat{v}, t-\Delta t) \Delta \hat{v}(t-\Delta t)^2 \rangle|_{x(t-\Delta t), \hat{v}(t-\Delta t)} \\
    &\simeq \left( \Phi(t) - \frac{5}{3} D(t)\partial_{\hat{v}} \ln P(x, \hat{v}, t) \right)\Delta t,
\end{aligned}
\label{app_eq:dv_back}
\end{equation}
where we use
\begin{equation}
\begin{aligned}
    P(x, \hat{v}, t) &= P(x, \hat{v}, t-\Delta t) - \Delta \hat{v}(t-\Delta t) 
    \\ &\times \partial_{\hat{v}} P(x, \hat{v}, t-\Delta t) + \mathcal{O}(\Delta t).
\end{aligned}
\end{equation}
Therefore, using Eqs.~\eqref{app_eq:ULE_estimator} and \eqref{app_eq:dv_back}, we can obtain the drift field and diffusion matrix of the underlying underdamped Langevin dynamics:
\begin{equation}
\begin{aligned}
    \Phi(x, v, t) &\simeq \Psi_f(x, \hat{v}, t) + \frac{1}{4} \left[ \Psi_f(x, \hat{v}, t) - \Psi_b(x, \hat{v}, t) \right], \\
    D(x, v, t) &\simeq \frac{3}{2} \frac{\langle \Delta \hat{v}(t)^2 \rangle|_{x(t), \hat{v}(t)}}{2 \Delta t},
\end{aligned}
\label{app_eq:ULE_estimator2}
\end{equation}
where
\begin{equation}
\begin{aligned}
    \Psi_f(x, \hat{v}, t) &\equiv \frac{\langle \Delta \hat{v}(t) \rangle|_{x(t), \hat{v}(t)}}{\Delta t},\\
    \Psi_b(x, \hat{v}, t) &\equiv \frac{\langle \Delta \hat{v}(t-\Delta t) \rangle|_{x(t), \hat{v}(t)}}{\Delta t}.
\end{aligned}
\end{equation}
We utilize Eq.~\eqref{app_eq:ULE_estimator2} in the drift and diffusion estimators; to obtain the drift field, two neural networks are used to train $\Psi_f(x, \hat{v}, t)$ and $\Psi_b(x, \hat{v}, t)$, separately.

\subsection{Error-uncertainty relation}
\label{sec:ApendixC2}

\begin{figure}[t]
    \includegraphics[width=\linewidth]{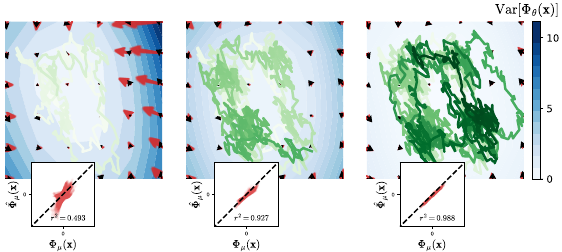}
    \vskip -0.1in
    \caption{
    Example trajectories and inferred drift fields (red arrows) for the $d=2$ case with increasing $\tau$ from left to right ($\tau \simeq 17$, $130$, and $1000$). The black arrows represent the exact fields and the colormap indicates the uncertainties of predictions $\rm{Var}[\bm{\Phi}_{\bm{\theta}}]$.
    Insets: $\hat{\bm{\Phi}}(\bm{x})$ vs. ${\bm{\Phi}}(\bm{x})$.
    }\label{figS4}
\end{figure}

We now derive the error-uncertainty relation, which has the same form as bias-variance decomposition in machine learning~\cite{goodfellow2016deep}.
For convenience, we omit parentheses in all notations, such as $\bm{y}_{\bm{\theta}}=\bm{y}_{\bm{\theta}}(\bm{z}, t)$ in this section.
We remind that the pointwise error $e^2_{\bm{y}}$ is defined by $\langle (\bm{y} - \bm{y}_{\bm{\theta}})^2  \rangle_{\bm{\theta}}/\langle \bm{y}_{\bm{\theta}}^2 \rangle_{\bm{\theta}}$.
Let us decompose $e^2_{\bm{y}}$ as follows:
\begin{equation}
\begin{aligned}
    e^2_{\bm{y}} \left\langle \bm{y}_{\bm{\theta}}^2 \right\rangle_{\bm{\theta}} &= \left\langle \left[ (\bm{y} - \hat{\bm{y}}) - (\bm{y}_{\bm{\theta}} - \hat{\bm{y}}) \right]^2 \right\rangle_{\bm{\theta}} \\
        &= \left\langle (\bm{y} - \hat{\bm{y}})^2 + (\bm{y}_{\bm{\theta}} - \hat{\bm{y}})^2 - 2 (\bm{y} - \hat{\bm{y}})(\bm{y}_{\bm{\theta}} - \hat{\bm{y}}) \right\rangle_{\bm{\theta}} \\
        &= {\rm Bias}^2[\bm{y}_{\bm{\theta}}]^2 + {\rm Var}[\bm{y}_{\bm{\theta}}] - 2(\bm{y} - \hat{\bm{y}}) \langle \bm{y}_{\bm{\theta}} - \hat{\bm{y}} \rangle_{\bm{\theta}},
\end{aligned}
\label{app_eq:err_uncert_proof}
\end{equation}
where ${\rm Bias}^2[\bm{y}_{\bm{\theta}}] \equiv (\bm{y} - \hat{\bm{y}})^2$ represents the error caused by a lack of training data or model assumptions such as the network architecture and optimizers.
In Eq.~\eqref{app_eq:err_uncert_proof}, the last term on the RHS is $0$, leading to the following relation:
\begin{equation}
\begin{aligned}
    e^2_{\bm{y}} = \hat{{\Sigma}}_{\bm{y}} + \frac{{\rm Bias}^2[\bm{y}_{\bm{\theta}}]}{\langle \bm{y}_{\bm{\theta}}^2 \rangle_{\bm{\theta}}},
\end{aligned}
\end{equation}
with the relative uncertainty of predictions $\hat{\Sigma}_{\bm{y}} \equiv {\rm Var}[\bm{y}_{\bm{\theta}}]/\langle \bm{y}_{\bm{\theta}}^2 \rangle_{\bm{\theta}}$.
The fact that ${\rm Bias}^2[\bm{y}_{\bm{\theta}}] \geq 0$ implies $e^2_{\bm{y}} \geq \hat{{\Sigma}}_{\bm{y}}$, signifying that $\hat{{\Sigma}}_{\bm{y}}$ provides a lower bound of $e^2_{\bm{y}}$.
Moreover, ${\rm Bias}^2[\bm{y}_{\bm{\theta}}]$ is minimized during the training by employing unbiased estimators, resulting in a high correlation between $e^2_{\bm{y}}$ and $\hat{{\Sigma}}_{\bm{y}}$.
Based on these two relations, $\hat{{\Sigma}}_{\bm{y}}$ can function as a proxy for $e^2_{\bm{y}}$, allowing us to assess whether the model is reliably trained.
Fig.~\ref{fig5}(a) in the main text and Fig.~\ref{figS4} qualitatively illustrate that the accuracy and the uncertainty of drift prediction, denoted by ${\rm Var}[\bm{\Phi}_{\bm{\theta}}(\bm{x})]$, increases and decreases as the duration of training trajectory $\tau$ increases, respectively, implying that the provided uncertainty can be used to reveal the reliability of predictions.

\subsection{Inference from normalized trajectories}
\label{sec:ApendixC3}

During the training of neural networks, it is a common practice to normalize the input data for various practical reasons, such as accelerating learning and reducing the risk of getting stuck in local optima~\cite{kim2020learning, seckler2022bayesian}.
We also employ data normalization to ensure stable learning, which necessitates the reconstruction of the inferred Langevin equation from the normalized data to obtain the original Langevin equation.
Let $\tilde{\bm{z}} \equiv \bm{\mathsf{B}}(\bm{z} - \bm{m})$ denote a normalized state of $\bm{z}$ where $\bm{m}$ is a $d$-dimensional constant vector and $\bm{\mathsf{B}}$ is a $d \times d$ real-valued matrix.
The Langevin equation for the normalized state can be written by 
\begin{equation}
\begin{aligned}
    \dot{\tilde{\bm{z}}}(t) = \tilde{\bm{\Phi}}(\tilde{\bm{z}}(t), t) + \sqrt{2\tilde{\bm{\mathsf{D}}}(\tilde{\bm{z}}(t), t)}\bm{\xi}(t),
\end{aligned}
\label{app_eq:NormalizedLangevinEq}
\end{equation}
where $\tilde{\bm{\Phi}}$ and $\tilde{\bm{\mathsf{D}}}$ are the drift vector and diffusion matrix for $\tilde{\bm{z}}$.
Using the fact that $\dot{\tilde{\bm{z}}} = \bm{\mathsf{B}}\dot{\bm{z}}$, Eq.~\eqref{app_eq:NormalizedLangevinEq} can be rewritten by
\begin{equation}
\begin{aligned}
    \dot{\bm{z}}(t) = \bm{\mathsf{B}}^{-1}\tilde{\Phi}(\tilde{\bm{z}}(t), t) + \bm{\mathsf{B}}^{-1}\sqrt{2\tilde{\bm{\mathsf{D}}}(\tilde{\bm{z}}(t), t)}\bm{\xi}(t).
\end{aligned}
\label{app_eq:NormalizedLangevinEq2}
\end{equation}
Comparing Eq.~\eqref{app_eq:NormalizedLangevinEq2} to the original Langevin equation~\eqref{app_eq:GeneralLangevinEq}, we can obtain $\bm{\Phi} = \bm{\mathsf{B}}^{-1}\bm{\Phi}$ and $\bm{\mathsf{D}} = \bm{\mathsf{B}}^{-1}\tilde{\bm{\mathsf{D}}}\bm{\mathsf{B}}^{-{\rm T}}$.
For example, if we set $\bm{\mathsf{B}} = {\rm diag}\left( \sqrt{\langle z_1^2 \rangle}, \sqrt{\langle z_2^2 \rangle}, \dots, \sqrt{\langle z_d^2 \rangle} \right)$ and $\bm{m} = \langle \bm{z} \rangle$ to be $\langle \tilde{\bm{z}}\rangle = 0$ and $\langle \tilde{\bm{z}}^2\rangle = 1$, the original drift vector and diffusion matrix can be easily recovered by
\begin{equation}
\begin{aligned}
    \Phi_\mu(\bm{z}, t) &= \frac{1}{\sqrt{\langle z_\mu^2 \rangle}} \tilde{\Phi}_\mu(\tilde{\bm{z}}(t), t), \\
    \mathsf{D}_{\mu\nu}(\bm{z}, t) &= \frac{1}{\sqrt{\langle z_\mu^2 \rangle \langle z_\nu^2 \rangle}} \tilde{\mathsf{D}}_{\mu\nu}(\tilde{\bm{z}}(t), t).
\end{aligned}
\end{equation}
This reconstruction process is conducted in the LBN `Output layer' (see Table~\ref{table:LBN_architecture}).
While this particular normalization method is applied to the input states in our approach, there are many other normalization techniques available. 
Depending on the specific purpose and requirements, other appropriate normalization methods can be chosen accordingly.

\section{Numerical simulation details with additional results}
\label{sec:ApendixD}

To generate stochastic trajectories, we apply the Euler--Maruyama method, which can be written as
\begin{equation}
\begin{aligned}
    \delta \bm{x}(t) &\equiv \bm{x}(t+\delta t) - \bm{x}(t) 
    \\ &= \bm{\Phi}(\bm{x}, t)\delta t + \sqrt{2\bm{\mathsf{D}}(\bm{x}, t)}\delta \bm{W}(t),
\end{aligned}
\end{equation}
for the overdamped regime with $\bm{\Phi}(\bm{x}, t) = \bm{F}(\bm{x}, t) + \bm{\nabla}_{\bm{x}} \cdot \bm{\mathsf{D}}(\bm{x}, t)$, and 
\begin{equation}
\begin{aligned}
    \delta \bm{x}(t) &\equiv \bm{v}(t)\delta t, 
    \\
    \delta \bm{v}(t) &\equiv \bm{v}(t+\delta t) - \bm{v}(t) 
    \\ &= \bm{\Phi}(\bm{x}, \bm{v}, t)\delta t + \sqrt{2\bm{\mathsf{D}}(\bm{x}, \bm{v}, t)}\delta \bm{W}(t),
\end{aligned}
\end{equation}
for the underdamped regime with $\bm{x}, \bm{v} \in \mathbb{R}^d$. 
Here, $\delta t$ is a simulation time-step and $\delta \bm{W}(t) \equiv \bm{W}(t+\delta t) - \bm{W}(t) = \bm{\xi}(t) \delta t$ where $\bm{W}(t)$ is the Wiener process, i.e., $\delta \bm{W}(t)$ is distributed with mean $0$ and variance $\delta t$.
Thus, in simulations, we set $\delta \bm{W}(t) = \sqrt{\delta t} \bm{\zeta}$ where $\bm{\zeta}$ is a $d$-dimensional vector independently sampled from a standard normal distribution.
Note that all simulations for constructing datasets are performed with $\delta t$, and the generated trajectories are sampled with a time-step $\Delta t$ ($> \delta t$) to imitate more realistic situations.
We generate $M$ trajectories with length $L \equiv \tau/\Delta t$ and divide them into training and validation sets, denoted by $\mathcal{D}_{\rm tr}$ and $\mathcal{D}_{\rm val}$, at a ratio of $5:1$. 
The ratio between sets may vary depending on the user's choice, taking into account the size of the datasets.
To demonstrate that our method can be generalized to infer the Langevin equation for unseen data, all the results in the main text are calculated from the test set $\mathcal{D}_{\rm te}$, which is independently generated and was not used in training.
We use $\mathcal{D}_{\rm te}$ having the same number of trajectories with $\tau = 100$ in all scenarios.

\subsection{Numerical calculations via trained LBN}
\label{sec:ApendixD1}

To generate trajectories via trained LBN, as illustrated in Figs.~\ref{fig2}(b), \ref{fig4}, \ref{fig5}(c), and \ref{fig6}, we calculate the average of the predictions, $\hat{\bm{\Phi}}(\bm{z}, t)$ and $\hat{\bm{\mathsf{D}}}(\bm{z}, t)$, by collecting the ensemble of predictions, $\left\{ \bm{\Phi}_{\bm{\theta}^k}(\bm{z}, t) \right\}_{k=1}^{N_{\bm{\theta}}}$ and $\left\{ \bm{\mathsf{D}}_{\bm{\theta}^k}(\bm{z}, t) \right\}_{k=1}^{N_{\bm{\theta}}}$, at each state $\bm{z}(t)$ and determine $\Delta \bm{z}(t)$ via the Euler--Maruyama method.
The algorithm is given as follows:
\begin{algorithm}[H]
\begin{algorithmic}[1]
\REQUIRE{Trained LBN, initial state $\bm{z}(t_0)$}
\State $i \leftarrow 0$
\LOOP
    \LOOP
        \State Draw $\bm{\theta}_{\bm{\Phi}}^k \sim Q_{\bm{\phi}_{\bm{\Phi}}}(\bm{\theta})$ and $\bm{\theta}_{\bm{\mathsf{D}}}^k \sim Q_{\bm{\phi}_{\bm{\mathsf{D}}}}(\bm{\theta})$, separately.
        \State Compute $\bm{\Phi}_{\bm{\theta}_{\bm{\Phi}}^k}(\bm{z}, t_i)$ and $\bm{\mathsf{D}}_{\bm{\theta}_{\bm{\mathsf{D}}}^k}(\bm{z}, t_i)$.
    \ENDLOOP
    \STATE  Obtain $\hat{\bm{\Phi}}(\bm{z}, t_i)$ and $\hat{\bm{\mathsf{D}}}(\bm{z}, t_i)$ from $\left\{ \bm{\Phi}_{\bm{\theta}_{\bm{\Phi}}^k}(\bm{z}, t_i) \right\}_{k=1}^{N_{\bm{\theta}}}$ and $\left\{ \bm{\mathsf{D}}_{\bm{\theta}_{\bm{\mathsf{D}}}^k}(\bm{z}, t_i) \right\}_{k=1}^{N_{\bm{\theta}}}$ by taking their average.
    \STATE Calculate $\Delta \bm{z}(t_i) = \hat{\bm{\Phi}}(\bm{z}, t_i)\Delta t + \sqrt{2 \hat{\bm{\mathsf{D}}}(\bm{z}, t_i)} \Delta \bm{W}(t_i)$.
    \State $\bm{z}(t_{i+1}) = \bm{z}(t_i) + \Delta z(t_i)$ and then $i \leftarrow i+1$.
\ENDLOOP
\end{algorithmic}
\caption{Generating trajectories via LBN}
\end{algorithm}
\noindent
To avoid confusion, we use subscripts $\bm{\Phi}$ and $\bm{\mathsf{D}}$ to indicate that the parameters of the drift and diffusion estimators are different.
Using obtained $\left\{ \bm{y}_{\bm{\theta}^k}(\bm{z}, t) \right\}_{k=1}^{N_{\bm{\theta}}}$ at each step, it is also feasible to calculate the uncertainties for generating the next state.

Calculation of stochastic thermodynamic quantities such as entropy production in Figs.~\ref{fig2}(b) and \ref{fig4}(g) is conducted similarly to the reconstruction of $\bm{F}(\bm{x})$ in the inhomogeneous diffusion system.
Taking entropy production as an example, the entropy production at each step, denoted by $\Delta S(t)$, can be calculated (in an OLE system) by 
\begin{equation}
\begin{aligned}
    \Delta S(t) = \bm{\mathsf{D}^{-1}} \bm{F}(\bm{x}, t) \circ \Delta \bm{x}(t),
\end{aligned}
\label{app_eq:ep}
\end{equation}
where $\circ$ denotes the Stratonovich product~\cite{sekimoto2010stochastic}.
Converting Eq.~\eqref{app_eq:ep} into It\^o convention gives
\begin{equation}
\begin{aligned}
    \Delta S(t) = \bm{\mathsf{D}^{-1}} \bm{F}\left(\bm{x}^{m}, t \right) \cdot \Delta \bm{x}(t),
\end{aligned}
\label{app_eq:ep_Ito}
\end{equation}
with $\bm{x}^{m}(t) \equiv [\bm{x}(t) + \bm{x}(t+\Delta t)]/2$.
Therefore, collecting the ensembles of force estimators and diffusion estimators, $\left\{\bm{F}_{\bm{\theta}}(\bm{x}^{m}, t)\right\}_{k=1}^{N_{\bm{\theta}}}$ and $\left\{\bm{\mathsf{D}}_{\bm{\theta}}\right\}_{k=1}^{N_{\bm{\theta}}}$, the corresponding entropy production estimator can be constructed by
\begin{equation}
\begin{aligned}
    \Delta S_{\bm{\theta}}(t) = \bm{\mathsf{D}_{\bm{\theta}}^{-1}} \bm{F}_{\bm{\theta}}(\bm{x}^m, t) \cdot \Delta \bm{x}(t),
\end{aligned}
\label{app_eq:inferred_ep}
\end{equation}
constructing an ensemble $\left\{ \Delta S_{\bm{\theta}}(t)\right\}_{k=1}^{N_{\bm{\theta}}}$.
This ensemble of estimations can be employed to obtain the average and uncertainties of the estimations.
The cumulative entropy production $\Delta S^t \equiv \sum_{i=0}^{t/\Delta t} \Delta S(t_i)$ in Fig.~\ref{fig2}(b) and entropy production rate in Fig.~\ref{fig4}(g) are obtained by this method, including their uncertainties.

\subsection{(OLE) Nonlinear force model (Fig.~\ref{fig2})}
\label{sec:ApendixD2}

\begin{figure}[t]
    \includegraphics[width=\linewidth]{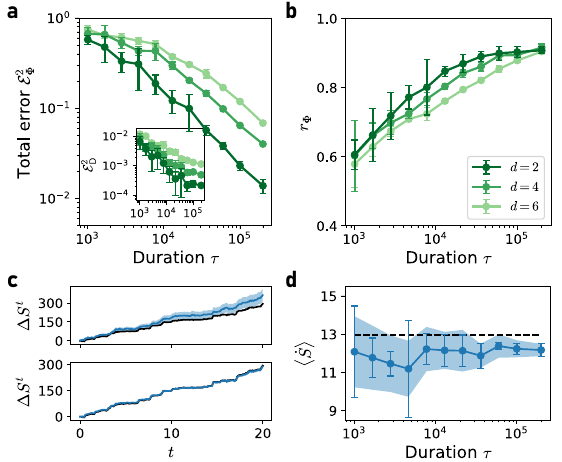}
    \vskip -0.1in
    \caption{ Additional results for a nonlinear force model.
    (a) Total errors of LBN for the drift field ($\mathcal{E}^2_{\bm{\Phi}}$) and diffusion matrix ($\mathcal{E}^2_{\bm{\mathsf{D}}}$, inset) and (b) Pearson correlation coefficient $r_{\bm{\Phi}}$ between $e^2_{\Phi}$ and $\hat{\Sigma}_{\Phi}$ with increasing the duration $\tau$ of the training trajectory for $d=2$, $4$, and $6$. 
    (c) Cumulative entropy production $\Delta S^t$ along a trajectory.
    The black solid line indicates the true values and the blue solid line and shaded area indicate the average and standard deviation of the ensemble of values $\left\{ \Delta S_{\bm{\theta}}(t)\right\}_{k=1}^{N_{\bm{\theta}}}$ obtained by LBN with $\tau \simeq 17$ (top) and $\tau=2000$.
    (d) Inferred entropy production rate $\langle \dot{S} \rangle$ with increasing $\tau$.
    The black dotted line indicates the true value and the blue solid line and shaded area indicate the average and standard deviation of the ensemble of values $\left\{ \dot{S}_{\bm{\theta}}\right\}_{k=1}^{N_{\bm{\theta}}}$.
    Error bars indicate the standard deviation of estimates from five independent trajectories and estimators.
    }\label{figS5}
\end{figure}

Our first model for testing the applicability of our approach to OLE inference is a stochastic process characterized by nonlinear force fields:
\begin{equation}
\begin{aligned}
    F_\mu(x) = -\sum_\nu \mathsf{A}_{\mu\nu}x_\nu + \alpha x_\mu e^{-x_\mu^2},
\end{aligned}
\end{equation}
with $\alpha = 10$ and $\mathsf{A}_{\mu \nu} = 2\delta_{\mu\nu} + 2\delta_{\mu, {\nu+1}} - 2\delta_{\mu, \nu-1}$.
The introduction of the nonlinear force accounts for the presence of a Gaussian-shaped obstacle at the center and is motivated by Ref.~\cite{frishman2020learning}. 
In this model, we specify an anisotropic diffusion matrix, i.e., $\mathsf{D}_{\mu \nu} = T\delta_{\mu\nu} - \sqrt{T}\delta_{\mu, {\nu+1}} - \sqrt{T}\delta_{\mu, \nu-1}$ with $T=3$.
The drift vector $\bm{\Phi}(\bm{x})$ is the same with $\bm{F}(\bm{x})$ because $\bm{\mathsf{D}}$ is homogeneous.
In Fig.~\ref{fig2} of the main text, we use a single trajectory ($M=1$) with $\Delta t = 0.01$, $\delta t = \Delta t / 10$, and $\tau = 2000$.
The convergence of total errors of the drift and diffusion estimators ($\mathcal{E}^2_{\bm{\Phi}}$ and $\mathcal{E}^2_{\bm{\mathsf{D}}}$) and Pearson correlation coefficient between pointwise errors of the drift estimator and relative uncertainties, denoted by $r_{\bm{\Phi}}$, with respect to the duration $\tau$ of training trajectory are illustrated in Fig.~\ref{figS5}(a) and~\ref{figS5}(b), respectively.
Moreover, we additionally infer the cumulative entropy production $\Delta S^t$ and entropy production rate $\langle \dot{S} \rangle$ by varying the duration $\tau$ of a training trajectory in Fig.~\ref{figS5}(c) and~\ref{figS5}(d), qualitatively illustrating the decreasing trend of both uncertainty and standard deviations between estimators.
This agreement of the trends supports that the uncertainty can provide the reliability of estimations.

\subsection{(OLE) Inhomogeneous diffusion model (Fig.~\ref{fig3})}
\label{sec:ApendixD3}

\begin{figure}[t]
    \includegraphics[width=\linewidth]{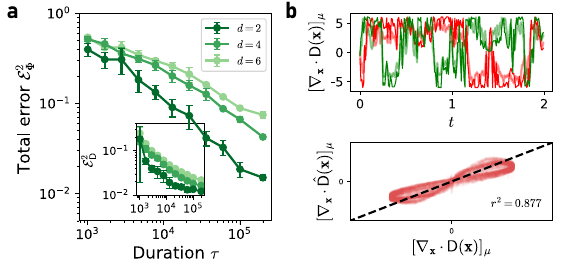}
    \vskip -0.1in
    \caption{
    Additional results for an inhomogeneous diffusion model.
    (a) Total errors of LBN for the drift field ($\mathcal{E}^2_{\bm{\Phi}}$) and diffusion matrix ($\mathcal{E}^2_{\bm{\mathsf{D}}}$, inset) with increasing the duration $\tau$ of a training trajectory for $d=2$, $4$, and $6$. 
    (b) Exact and inferred spurious drift, $[\bm{\nabla}_{\bm{x}}\cdot \bm{\mathsf{D}}(\bm{x}) ]_{\mu}$ and $[\bm{\nabla}_{\bm{x}}\cdot \hat{\bm{\mathsf{D}}}(\bm{x}) ]_{\mu}$, along a trajectory (top) and a scatter plot between them (bottom).
    }\label{figS6}
\end{figure}

To demonstrate the ability of LBN to infer an inhomogeneous diffusion matrix in the OLE, we opt for a deterministic force and diffusion matrix described by
\begin{equation}
\begin{aligned}
    \bm{F}(\bm{x}) &= -\bm{\mathsf{A}}\bm{x},\\
    \mathsf{D}_{\mu\nu} (\bm{x}) &= T ( 1 + \alpha e^{-x_{\mu}^2/2}) \delta_{\mu \nu},
\end{aligned}
\end{equation}
with $\alpha = 5$ and $\mathsf{A}_{\mu \nu} = 4\delta_{\mu\nu} - 2\delta_{\mu, {\nu+1}} - 2\delta_{\mu, \nu-1}$.
This model features a symmetric $\bm{\mathsf{A}}$ in contrast to the nonlinear force model, and thus the deterministic force does not induce circulation but rather directs a particle toward the center.
Here, due to the inhomogeneity of the diffusion matrix, a spurious drift emerges, resulting in the following drift vector:
\begin{equation}
\begin{aligned}
    \bm{\Phi}(\bm{x}) &= -\bm{\mathsf{A}}\bm{x} - \alpha T \bm{x} e^{-\bm{x}^2/2}.
\end{aligned}
\end{equation}
In the main text, we use a single trajectory ($M=1$) with $\Delta t = 0.01$, $\delta t = \Delta t / 10$, and $\tau = 2000$ [Fig.~\ref{fig3}(a--d)] for constructing $\mathcal{D}_{\rm tr}$.
Fig.~\ref{figS6} shows the convergence of total errors of the drift and diffusion estimators ($\mathcal{E}^2_{\bm{\Phi}}$ and $\mathcal{E}^2_{\bm{\mathsf{D}}}$) with increasing $\tau$ [Fig.~\ref{figS6}(a)] and the coincidence between the true and inferred spurious drift vector, $[\bm{\nabla}_{\bm{x}} \cdot \hat{\bm{\mathsf{D}}}(\bm{x})]_\mu$ and $[\bm{\nabla}_{\bm{x}} \cdot \bm{\mathsf{D}}(\bm{x})]_\mu$[Fig.~\ref{figS6}(b)].

\subsection{(OLE) Hodgkin--Huxley model (Fig.~\ref{fig4})}
\label{sec:ApendixD4}

The Hodgkin--Huxley (HH) model stands as a widely used mathematical neuron model that describes the dynamics of action potentials in neurons~\cite{hodgkin1952quantitative}.
This model is governed by four nonlinear first-order ordinary differential equations of four variables $x_i$ with $i \in [1, 4]$, where $x_1 \equiv V$ is the membrane potential in mV, $x_2 \equiv n$ is the activation gating variable of the potassium ion channels, and $x_3 \equiv m$, $x_4 \equiv h$ are gating variables that represent the activation and inactivation of the sodium ion channels, respectively. 
Gating variables $x_2$, $x_3$, and $x_4$ are dimensionless probabilities, confined to the range $[0, 1]$. 
Thus, we clamp $x_2$, $x_3$, and $x_4$ into the range $[0, 1]$ in simulations.

\begin{figure}[t]
    \includegraphics[width=\linewidth]{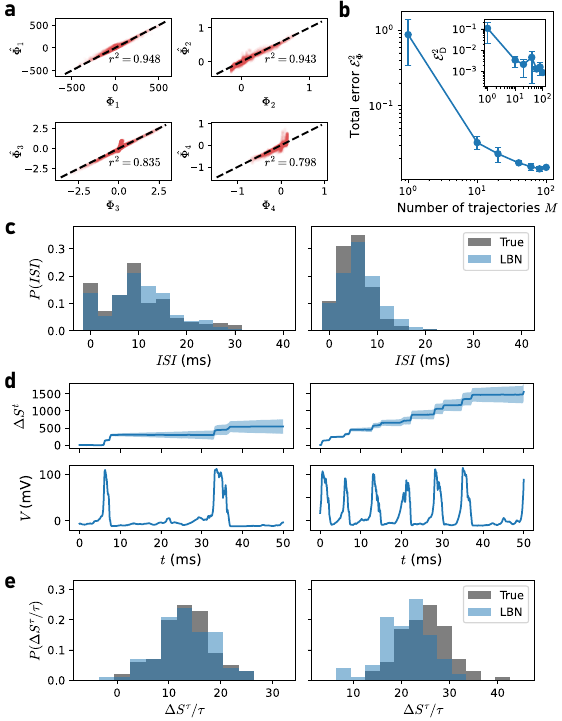}
    \vskip -0.1in
    \caption{
    Additional results for the spiking neuron model.
    (a) Scatter plots between inferred drift field $\hat{\Phi}_{\mu}$ vs. true drift field $\Phi_{\mu}$ with $\mu=1,\dots,4$.
    (b) Total errors of LBN for the drift field ($\mathcal{E}^2_{\bm{\Phi}}$) and diffusion matrix ($\mathcal{E}^2_{\bm{\mathsf{D}}}$) with increasing the number of training trajectories $M$.
    (c) PDF of the inter-spike interval ($ISI$) measured from trajectories generated by the true model and LBN trained at $I_{\rm ext}=0$.
    (d) Cumulative entropy production $\Delta S^t$ (top) and corresponding membrane potential $V(t)$ (bottom) with respect to time $t$.
    Shaded areas indicate the cumulative standard deviation of $\Delta S^t$.
    (e) PDF of $\Delta S^\tau/\tau$ measured from trajectories generated by the true model and LBN trained at $I_{\rm ext}=0$.
    In (c--e), the left (right) panel indicates the results from generated trajectories of $\tau=50$ with $I_{\rm ext}=0$~mV ($I_{\rm ext}=30$~mV).
    }\label{figS7}
\end{figure}

There exist numerous stochastic versions of the HH model, incorporating various fluctuations in neurons caused by synaptic noise and the inherent randomness of ion channel gating~\cite{goldwyn2011what}. 
In this paper, we consider a stochastic version of the HH model by adding a small Gaussian white noise into the gating variables.
The resulting system is expressed as an OLE as follows:
\begin{equation}
\begin{aligned}
    \dot{x}_1 &= F_1(\bm{x}) + \frac{I_{\rm ext}}{C_M}    \\
    \dot{x}_\mu &= F_{\mu} + \sqrt{2D}\xi_\mu(t) \text{  with  } \mu=2, 3, 4,
\end{aligned}
\label{app_eq:HH_Langevin}
\end{equation}
where 
\begin{equation}
\begin{aligned}
    F_1(\bm{x}) &= -\frac{1}{C_M}[ g_{\rm Na} x_3^3 x_4 (x_1 - E_{\rm Na}) + g_{\rm K} x_2^4 (x_1 - E_{\rm K}) \\ &\quad + g_{L} (x_1 - E_{L}) ],
    \\
    F_\mu &= \alpha_\mu (x_1) (1-x_\mu)-\beta_\mu(x_1) x_\mu \text{  with  } \mu=2, 3, 4,
\end{aligned}
\end{equation}
and $D={\rm diag}(0, T, T, T)$ with $T=0.01~{\rm ms}^{-1}$. The externally applied current is denoted as $I_{\rm ext}$ in Eq.~\eqref{app_eq:HH_Langevin}, serving as a bifurcation parameter of the HH model.
The complex nonlinear functions $\alpha_\mu(x_1)$ and $\beta_\mu(x_1)$ are given by~\cite{xie2008controlling}
\begin{equation}
\begin{aligned}
    \alpha_2(x_1) &= 0.01(10 - x_1)/(\exp[-(x_1-10)/10]-1),\\
    \beta_2(x_1) &= 0.125 \exp[-x_1/80],\\
    \alpha_3(x_1) &= 0.1(25 - x_1)/(\exp[-(x_1-25)/10]-1),\\
    \beta_3(x_1) &= 4 \exp[-x_1/18],\\
    \alpha_4(x_1) &= 0.07 \exp[-x_1/20],\\
    \beta_4(x_1) &= 1/(\exp[-(x_1-30)/10] + 1).
\end{aligned}
\end{equation}
Here, $C_M=1.0~{\rm \mu F/cm^2}$ denotes the membrane capacitance, $g_{\rm Na}=120~{\rm mS/cm^2}$, $g_K=36.0~{\rm mS/cm^2}$, and $g_L=0.3~{\rm mS/cm^2}$ denote the maximum conductance of the corresponding ionic currents, and $E_{\rm Na} = 115.0~{\rm mV}$, $E_{\rm K} = -12.0~{\rm mV}$, and $E_{\rm L} = 10.6~{\rm mV}$ denote the equilibrium potentials of the sodium, potassium, and leak currents.
We employ $100$ trajectories ($M=100$) with $\Delta t = 0.01~{\rm ms}$, $\delta t = \Delta t / 10$, and $\tau = 48~{\rm ms}$ for constructing $\mathcal{D}_{\rm tr}$.
Note that all trajectories in $\mathcal{D}_{\rm tr}$ are generated by the stochastic HH model with $I_{\rm ext} = 0$.
In Fig.~\ref{fig4}(d--g), we generate trajectories with $M=50$ and $\tau = 50~{\rm ms}$ with varying $I_{\rm ext}$ via the true model and LBN, and both run $10~{\rm ms}$ before recording trajectories to reach a stationary state.
The trained LBN achieves a performance of $\mathcal{E}^2_{\bm{\Phi}}=0.015$ and $\mathcal{E}^2_{\bm{\mathsf{D}}}=0.0009$; scatter plots of $\hat{\Phi}_{\mu}$ vs. ${\Phi}_{\mu}$ with $\mu=1,\dots,4$ are shown in Fig.~\ref{figS7}(a).
The convergence of the total errors with the increasing number of training trajectories $M$ is depicted in Fig.~\ref{figS7}(b).
The inter-spike interval ($ISI$) in Fig.~\ref{fig4}(e) is obtained by measuring the time between two spikes in the trajectories.
The PDFs of $ISI$ at $I_{\rm ext}=0$ and $I_{\rm ext}=30$ are illustrated in Fig.~\ref{figS7}(c), forming qualitatively similar PDFs from the true model and LBN.
The entropy production of the HH model in Fig.~\ref{fig4}(g) is computed by Eq.~\eqref{app_eq:ep} for the gating variables.
Fig.~\ref{figS7}(d) shows the cumulative entropy production $\Delta S^t$ and corresponding membrane voltage $V(t)$ with respect to $t$, and we confirm that the abrupt increase in entropy production occurs when $V(t)$ has a spike pattern, implying that the spike patterns are the main source of the increase in entropy production.
The $\Delta S^\tau / \Delta t$ PDFs at $I_{\rm ext}=0$~mV and $I_{\rm ext} = 30$~mV are shown in Fig.~\ref{figS7}(e); the two PDFs from the true model and LBN at $I_{\rm ext} = 0$~mV are nearly overlapped, while being rather different at $I_{\rm ext} = 30$~mV.

\subsection{(ULE) Stochastic van der Pol oscillator (Fig.~\ref{fig5})}
\label{sec:ApendixD5}

The van der Pol oscillator is a mathematical model that describes the behavior of a nonlinear oscillator with damping~\cite{van1926on}.
Introducing a Gaussian white noise into the governing equation of the system, originally formulated as a second-order differential equation, we consider a ULE with the following nonlinear drift vector and homogeneous diffusion matrix:
\begin{equation}
\begin{aligned}
    \Phi_\mu(\bm{x}, \bm{v}) &= k(1-x_\mu^2)v_\mu - x_i, \\
    \mathsf{D}_{\mu\nu} &= T \delta_{\mu\nu}
\end{aligned}
\end{equation}
with $k=2$ and $T=1$. Here, the effective friction coefficient can be expressed as $\gamma(x_{\mu}) = k(1-x_\mu^2)$.
We use a single trajectory ($M=1$) with $\Delta t = 0.01$ and $\delta t=\Delta t/5$ for training.

\begin{figure}[t]
    \includegraphics[width=\linewidth]{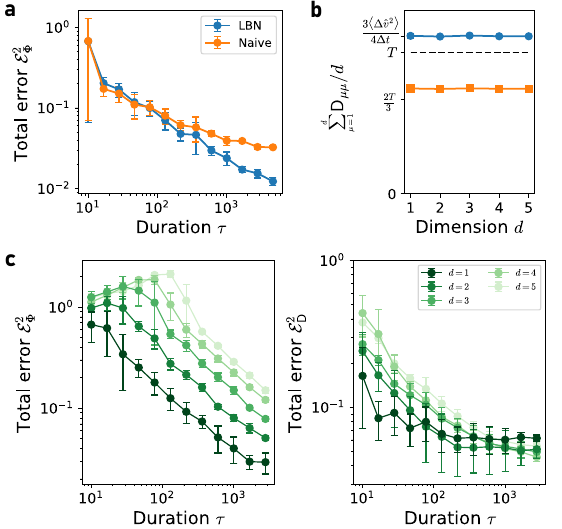}
    \vskip -0.1in
    \caption{ Additional results for a stochastic van der Pol oscillator.
    (a) Total errors for the drift field ($\mathcal{E}^2_{\bm{\Phi}}$) of LBN and a naive extension of the OLE estimator with increasing the duration $\tau$ of a training trajectory.
    (b) Average of diagonal entries ($\sum_{\mu=1}^d \mathsf{D}_{\mu\mu}/d$) of LBN and a naive extension of the OLE estimator with respect to the system dimension $d$.
    $T$ denotes the true value and $\frac{3\langle \Delta \hat{v}^2 \rangle}{4\Delta t}$ denotes the value obtained from a long trajectory.
    (c) Total errors for the drift field ($\mathcal{E}^2_{\bm{\Phi}}$, left) and diffusion matrix ($\mathcal{E}^2_{\bm{\mathsf{D}}}$) of LBN with increasing $\tau$ for $d=1, \dots, 5$ for the multiplicative noise case.
    }\label{figS8}
    \vskip -0.1in
\end{figure}

We use unbiased estimators for the drift field and diffusion matrix to infer the ULE of the system, deviating from the naive extension of OLE estimators.
To demonstrate the impact of these modifications, we compare the total errors and estimated diffusion matrix between the naive extended estimator and LBN.
Fig.~\ref{figS8}(a) illustrates that the total error of LBN with an unbiased estimator in Eq.~\eqref{eq:ULE_estimators} becomes smaller than the value of the naive extension with increasing $\tau$, indicating that our modification successfully corrects the leading-order bias.
The effect of modification in the diffusion estimator, the rescaling factor $3/2$, is depicted in Fig.~\ref{figS8}(b), resulting in more accurate estimates by LBN than the naive extension.
Note that $\sum_{\mu=1}^d \mathsf{D}_{\mu\mu}/d = T$ in the given system.
However, a slight discrepancy between the results of LBN and the true value $T$ is observed.
To investigate this discrepancy, we estimate $3\langle \Delta \hat{v}^2 \rangle/4\Delta t$ from an extended trajectory, and find it coincides with the results of LBN. This alignment implies that the discrepancy arises in the terms of $\mathcal{O}(\Delta t)$.

Fig.~\ref{fig5}(e) and~\ref{fig5}(f) show the results for the van der Pol system with an inhomogeneous diffusion matrix $\mathsf{D}_{\mu\nu} = (T_0 + T_x x_\mu^2 + T_v v_{\mu}^2)\delta_{\mu\nu}$ with $T_0 = 1$, $T_x = 0.3$, and $T_v = 0.1$.
Fig.~\ref{figS8}(c) shows the convergence of total errors for both the drift vector and diffusion matrix in the inhomogeneous diffusion case of the van der Pol oscillator.

\subsection{(ULE) Brownian Carnot engine (Fig.~\ref{fig6})}
\label{sec:ApendixD6}

One of the primary objectives in thermodynamics is the development of efficient thermal engines, a goal that has been extended to the micro- and nanoscales through the field of stochastic thermodynamics.
The Brownian Carnot engine stands out as a noteworthy endeavor in developing an efficient microscale engine, which was realized in Ref.~\cite{martinez2016brownian}.
In the experimental realization of this engine, precise control over a particle is achieved using an optical trap and a noisy electrostatic force, allowing for adjustments to the effective drift vector and diffusion coefficient, respectively.
We replicate the engine cycle with a cycle duration $\tau_{cyc}$ as closely as possible, varying time $t$, and defining each cycle to comprise four thermodynamic processes of equal length $\tau_{cyc}/4$.

\begin{figure}[t]
    \includegraphics[width=\linewidth]{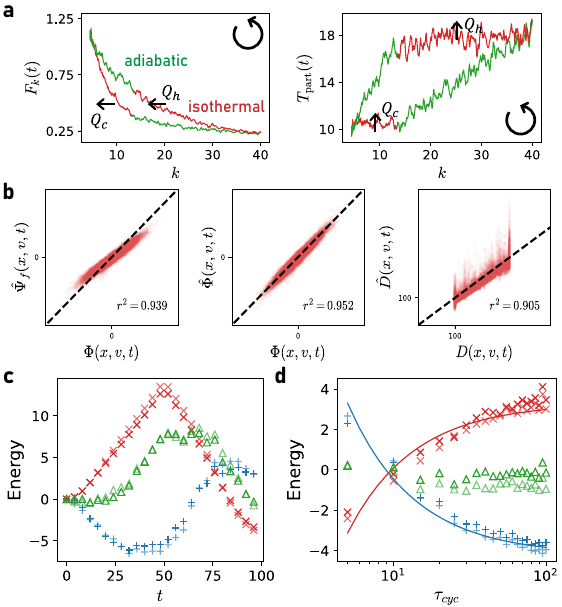}
    \vskip -0.1in
    \caption{
    Additional results for the Brownian Carnot engine.
    (a) Mean Clapeyron (left) and $T_{\rm part}-k$ (right) diagrams of the Brownian Carnot engine obtained from trajectories with $M=2000$.
    (b) Scatter plots between ${{\Phi}}(x, v, t)$ and $\hat{{\Psi}}(x, v, t)$ (left), ${{\Phi}}(x, v, t)$ and $\hat{{\Phi}}(x, v, t)$ (middle), and ${\mathsf{D}}(x, v, t)$ and $\hat{\mathsf{D}}(x, v, t)$ (right).
    (c) Ensemble averages of applied work ($\langle W(t) \rangle$, red crosses), absorbed heat ($\langle Q(t) \rangle$, blue pluses), and total energy ($\langle E(t) \rangle$, green triangles) with respect to time $t$.
    (d) Ensemble averages of total ($\langle W(\tau_{cyc}) \rangle$, red crosses), total absorbed heat ($\langle Q(\tau_{cyc}) \rangle$, blue pluses), and total energy ($\langle E(\tau_{cyc}) \rangle$, green triangles) with respect to cycle duration $\tau_{cyc}$.
    Solid lines are fits to $A+B/\tau_{cyc}$.
    In (c) and (d), the darker (lighter) symbols indicate the values obtained from the true engine (LBN).
    }\label{figS9}
\end{figure}

We consider a one-dimensional particle immersed in friction coefficient $\gamma$ and trapped in a harmonic potential $U(x, t)=k(t)x(t)^2/2$ with time-varying stiffness $k(t)$.
The drift field and stiffness are given by
\begin{equation}
\begin{aligned}
    \Phi(x, v, t) &= -\gamma v(t) -k(t) x(t) \\
    k(t) &= k_m + k_s t^2 \qquad t\in[0, \tau/2],
\end{aligned}
\end{equation}
where the stiffness is time-symmetric with $k(t)=k(\tau_{cyc}-t)$.
Note that the decrease (increase) of $k(t)$ corresponds to the expansion (compression) process.
Here, we fix the minimum and maximum values of $k(t)$ as $k_m = 40/9$ and $k_M=40$, respectively, and thus $k_s$ is calculated by $k_s = 32 k_m/\tau_{cyc}^2$.
The diffusion coefficient $D(t) \equiv \gamma T(t)$ remains constant in the isothermal steps and changes smoothly while $T^2(t)/k(t)$ remains constant to ensure that the average of transferred heat is zero in the adiabatic steps~\cite{martinez2016adiabatic}.
The change in $D(t)$ with respect to $t$ is summarized as follows:
\begin{equation}
\begin{aligned}
    D(t) = \begin{cases}
        \gamma T_c & \text{$0 \leq t < \frac{1}{4}\tau_{cyc}$ (isothermal)}, \\
        \gamma T_c \left[ k(t)/k_{qt} \right]^{1/2} & \text{$\frac{1}{4}\tau_{cyc} \leq t < \frac{1}{2} \tau_{cyc}$ (adiabatic)}, \\
        \gamma T_h & \text{$\frac{1}{2}\tau_{cyc} \leq t \leq \frac{3}{4}\tau_{cyc}$ (isothermal)}, \\
        \gamma T_h \left[ k(t)/k_{qt} \right]^{1/2} & \text{$\frac{3}{4}\tau_{cyc} \leq t < \tau_{cyc}$ (adiabatic)},
    \end{cases}
\end{aligned}
\end{equation}
with $\gamma=10$, $k_{qt} \equiv k(\tau_{cyc}/4)=40/3$, $T_c = 10$, and $T_h = 10\sqrt{3}$.
Fig.~\ref{figS9}(a) shows thermodynamic diagrams of the engine for $\tau_{cyc}=100$, where the conjugated force for $k(t)$ is $F_k(x, t) \equiv \partial U / \partial k = x(t)^2/2$ and the effective temperature of the particle is $T_{\rm part}(t) \equiv k(t)\langle x(t)^2 \rangle$.
Please see Fig.~\ref{fig6}(b) for the time evolution of $k(t)$ and $D(t)$.

The work $\Delta W(t)$ exerted on the particle, transferred heat $\Delta Q(t)$ from the thermal bath to the particle, and total energy change $\Delta E(t)$ in the intervals of time $[t, t+\Delta t]$ are calculated in a similar way as in Ref.~\cite{martinez2016brownian} as follows:
\begin{equation}
\begin{aligned}
    \Delta W(t) &= F_{k}(x, t) \circ \Delta k(t), \\
    \Delta Q(t) &= \Delta E(t) - \Delta W(t),  \\
    \Delta E(t) &= \frac{1}{2}[k(t+\Delta t)x(t+\Delta t)^2 - k(t)x(t)^2 ] \\ &+ \frac{1}{2}[v(t+\Delta t)^2 - v(t)^2],
\end{aligned}
\end{equation}
where the total energy is defined by $E(t) \equiv (1/2)k(t)x(t)^2 + (1/2)v(t)^2$.
Cumulative thermodynamic quantities at time $t$ along a trajectory are computed by the sum of these changes, e.g., $W(t) \equiv \sum_{i=0}^{t/\Delta t} \Delta W(t)$.
Fig.~\ref{figS9}(c) shows how the ensemble averages of $W(t)$, $Q(t)$, and $E(t)$ vary during a cycle with $\tau_{cyc}=100$: $E(t)$ ends near $0$ and $Q(t)$ does not vary in the ranges $[\tau_{cyc}/4, \tau_{cyc}/2]$ and $[3\tau_{cyc}/4, \tau_{cyc}]$, corresponding to the adiabatic processes. In addition, $W(t)$ increases (decreases) in the compression (expansion) process and ends with a negative value, implying that it performs as an engine.
The total thermodynamic quantities during a cycle, i.e., $W(\tau_{cyc})$, $Q(\tau_{cyc})$, and $E(\tau_{cyc})$, with varying $\tau_{cyc}$ are illustrated in Fig.~\ref{figS9}(d). Here,
$E(\tau_{cyc})$ remains consistently near $0$, while $W(\tau_{cyc})$ and $Q(\tau_{cyc})$ converge to their quasistatic averages.
In addition, we verify that $W(\tau_{cyc})$ becomes negative at small $\tau_{cyc}$, implying that the system performs as a refrigerator.

\begin{figure*}[t]
    \includegraphics[width=\linewidth]{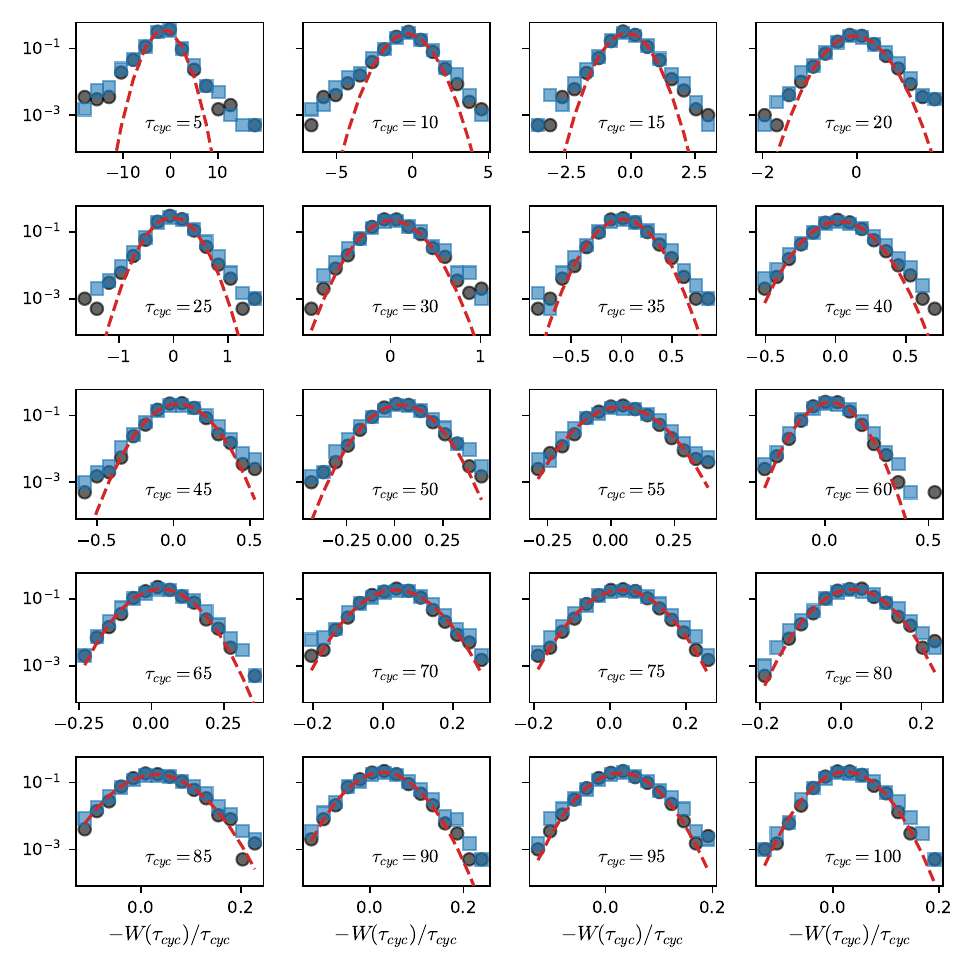}
    \vskip -0.1in
    \caption{
    The PDF of work rate, $-W(\tau_{cyc})/\tau_{cyc}$, with varying the duration $\tau_{cyc}$ of a cycle from $\tau_{cyc}=5$ to $100$. The red dotted lines are fitted lines to a Gaussian distribution. The black (blue) symbols are the values of the true engine (LBN).
    }\label{figS10}
\end{figure*}

\begin{figure*}[t]
    \includegraphics[width=\linewidth]{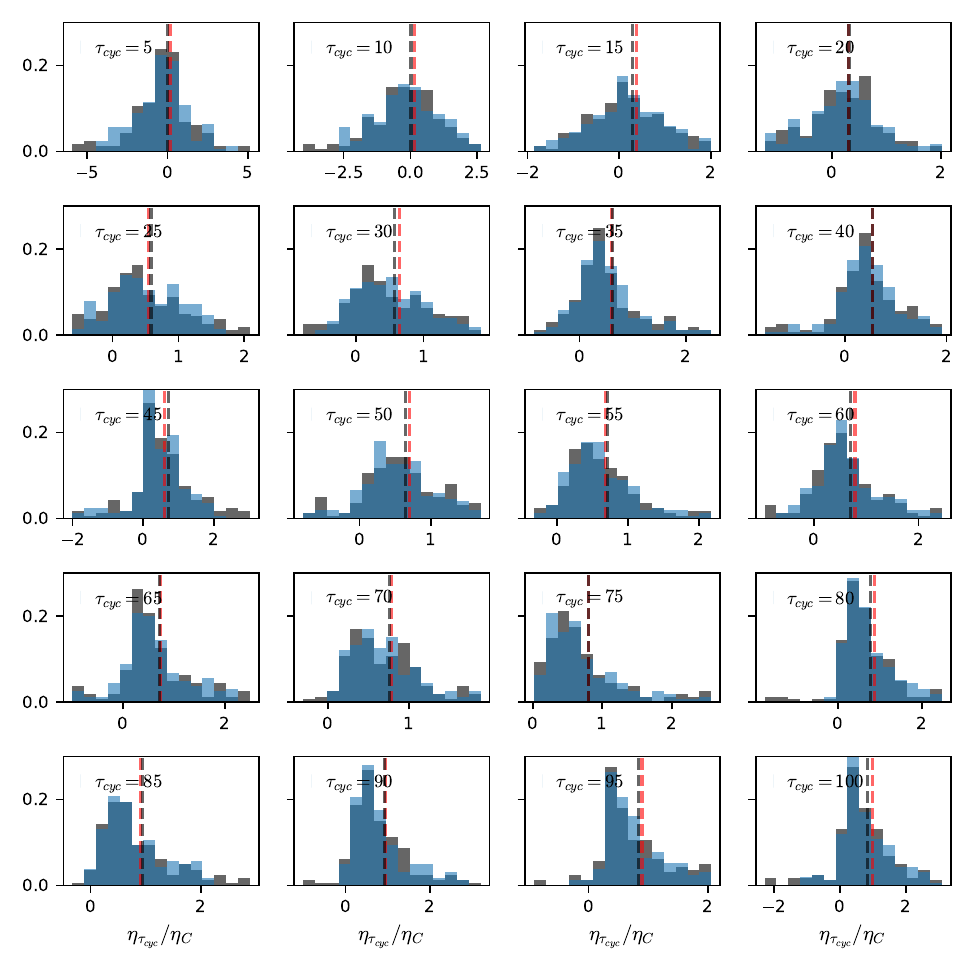}
    \vskip -0.1in
    \caption{
    The PDF of efficiency, $\eta_{\tau_{cyc}}$, with varying the duration $\tau_{cyc}$ of a cycle from $\tau_{cyc}=5$ to $100$ within the range from the $10$th percentile to the $90$th percentile. $\eta_{\tau_{cyc}}$ is obtained with $n_{cyc}=10$ and $\eta_C$ denotes the Carnot efficiency, calculated by $\eta_C \equiv 1-T_c/T_h \simeq 0.423$.
    The black (blue) bars are the values of the true engine (LBN).
    The black (red) dashed lines are the average of $\eta_{\tau_{cyc}}/\eta_C$ of the true engine (LBN).
    }\label{figS11}
\end{figure*}

For computing the efficiency $\eta_{\tau_{cyc}}$ of the engine, the traditional definition of efficiency is employed:
\begin{equation}
\begin{aligned}
    \eta_{\tau{cyc}} = \frac{-W^{(n_{cyc})}_{\tau_{cyc}}}{Q^{(n_{cyc})}_{h, \tau_{cyc}}},
\end{aligned}
\end{equation}
where $W^{(n_{cyc})}_{\tau_{cyc}}$ and $Q^{(n_{cyc})}_{h, \tau_{cyc}}$ are the sum of work and absorbed heat during all processes and the isothermal expansion process over $n_{cyc}$ cycles, respectively.
Note that as $n_{cyc} \rightarrow \infty$ increases, $\eta_{\tau{cyc}}$ becomes close to the long-term efficiency of the engine in the given duration of a cycle $\tau_{cyc}$, and the fluctuation of $\eta_{\tau{cyc}}$ shrinks.
$\eta_{\tau_{cyc}}$ in Fig.~\ref{fig6}(c) and~\ref{fig6}(f) are calculated with $n_{cyc}=10$ and $n_{cyc}=40$, respectively.
In Fig.~\ref{fig6}(f), the average of $\eta_{\tau_{cyc}}$ is calculated within the range from the $10$th percentile to the $90$th percentile due to the large fluctuation of $\eta_{\tau{cyc}}$.
The PDFs of the work rate $-W(\tau_{cyc})/\tau_{cyc}$ and efficiency $\eta_{\tau_{cyc}}$ with varying $\tau_{cyc}$ are illustrated in Figs.~\ref{figS10} and \ref{figS11}.
At short $\tau_{cyc}$, the PDF of the work rate deviates from a Gaussian distribution. For systems with harmonic potential, it is well-established that the work PDF is Gaussian when the driving duration is significantly longer than the system relaxation time~\cite{hendrix2001afast}. However, under fast driving conditions or with time-dependent stiffness, the terms regarding the relaxation time become significant, resulting in deviations from Gaussianity in the work PDF~\cite{speck2011work}.
As shown in Fig.~\ref{figS10}, the PDF of the work rate gradually converges toward a Gaussian distribution as $\tau_{cyc}$ increases.

We use $50$ trajectories ($M=50$) with $\Delta t = 0.01$, $\delta t=\Delta t/5$, and $\tau_{cyc} = \tau = 100$ for training, and the engine undergoes $100$ cycles before recording trajectories.
As depicted in Fig.~\ref{figS9}(b), LBN accurately infers both drift vector and diffusion matrix, achieving $\mathcal{E}^2_{\bm{\Phi}}=0.051$ and $0.00565$, respectively.
Moreover, it is worth noting that the naive extension of the OLE drift estimator $\hat{\Psi}(x, v, t)$ achieved a lower $r^2$ [left panel in Fig.~\ref{figS9}(b)] with $\mathcal{E}^2_{\bm{\Phi}}=0.253$.
In the process of generating trajectories via LBN with varying $\tau_{cyc} = \tau/n_{\tau}$ in Fig.~\ref{fig6}(d--f), the time data employed to construct the input vector are sampled from the original time data $[0, \Delta t, 2 \Delta t, \dots, L\Delta t]$ with the sampling rate $n_{\tau}$.
Thermodynamic quantities calculated from trajectories via LBN are compared with the true values in Figs.~\ref{figS9}, \ref{figS10}, and \ref{figS11}, demonstrating that LBN successfully learns the Brownian Carnot engine even beyond the training regime.

\newpage
\vfill
\newpage
\vfill
\clearpage








\bibliographystyle{elsarticle-num} 
\bibliography{CHAOS-D-24-06833-R1.bib}

\end{document}